# Programmable emergence through hierarchical self-organisation

Aladin Şura[1,✉] and F. Ömer Ilday[1,2,3,✉]

[1] Faculty of Electrical Engineering and Information Technology, Ruhr Universität Bochum, Universitätsstraße 150, 44801 Bochum, Germany

[2] Faculty of Physics and Astronomy, Ruhr Universität Bochum, Universitätsstraße 150, 44801 Bochum, Germany

[3] Center for Complex Laser-Matter Interactions, Ruhr Universität Bochum, Universitätsstraße 150, 44801 Bochum, Germany

✉ email: Aladin.Sura@ruhr-uni-bochum.de; Oemer.Ilday@ruhr-uni-bochum.de

Predictively steering self-organising systems with hierarchical structure toward intended outcomes across widely separated dynamical scales remains a fundamental challenge. Despite decades of progress, hierarchy remains a descriptive property rather than a mechanism for control. Here, we show how a high-dimensional stochastic system can be steered toward preselected target states by exploiting its internal hierarchy of timescales. Fast variables lock to slower collective variables; iterating this locking across scales progressively reduces the effective dynamics to a small set of externally controllable parameters. This enables directing the system while maintaining dynamical insulation from noise across scales, thereby turning hierarchy from an observation into a mechanism for programmable emergence. We experimentally demonstrate this approach in a fully tractable platform: a mode-locked laser, in which we assemble and stabilise more than 100 pulses into harmonic patterns over timescales spanning 10 orders of magnitude. Phenomena such as nucleation, quantum-noise-induced annihilation, long-range interactions and annealing-like ordering are predicted and exploited to transition the system to preselected states. Our results establish a route to programming emergent behaviour in driven-dissipative systems, with implications ranging from optics to materials.

A longstanding question in nonlinear science is whether self-organised systems[1] can be predictively programmed to yield predetermined emergent states. Hierarchical organisation, in which emergent structures at one scale become building blocks for larger-scale order, is ubiquitous across non-equilibrium systems. Although hierarchical systems have been extensively studied from the atomic scales to colloidal assembly and biological organisation[2-6], these efforts remain largely empirically driven, and hierarchy often remains a descriptive property[4], rather than a control mechanism.

This gap is addressed here by exploiting hierarchical timescale separation to enable predictive steering toward intended final states. We show that this becomes possible when the dynamics exhibit a hierarchy of timescales and each scale supports attractor states, such that the collective behaviour can be reduced to a few externally controlled parameters. The key is not the specific interactions, but the sequential locking of fast degrees of freedom to slow collective variables. Critically, the multi-timescale nature dynamically insulates emergent structures from noise across levels. We illustrate these principles first in a minimal theoretical model of hierarchical pattern formation, and then demonstrate experimentally in a mode-locked laser, chosen for its experimental accessibility.

Our approach builds on dimensional reduction via timescale separation, whereby rapidly relaxing internal dynamics lock to slower collective variables[8,9]. However, unlike conventional *post hoc* analysis at a single dynamical level, it is applied here iteratively across multiple emergent levels, enabling predictive steering of hierarchical self-organisation through a small set of external parameters. In systems with well-separated timescales, the dynamics can be organised into nested blocks in which fast degrees of freedom relax rapidly to attractors parametrised by progressively slower collective variables (Fig. 1a). Unlike in low-dimensional systems, these attractors are not merely fixed points or limit cycles of a few coordinates but typically high-dimensional vectors. They are therefore identified as the self-organised patterns or structures of the system, which we denote by $\mathscr{S}_{(1)}$, $\mathscr{S}_{(2)}$, …, $\mathscr{S}_{(m)}$. Each such organisational structure emerges, and is defined by the corresponding attractor of a given hierarchical level, rather than being attributed *post hoc*. We group the coordinates, both microscopic and emergent, according to the characteristic timescales of their evolution, $\mathbf{x}(t) = [\mathbf{x}_{(1)}, \mathbf{x}_{(2)}, \ldots, \mathbf{x}_{(n)}]$, where $\mathbf{x}_{(1)}$ evolves on the fastest timescale $\mathscr{T}_{(1)}$, $\mathbf{x}_{(2)}$ on the slower $\mathscr{T}_{(2)}$, and so on, up to the slowest block, $\mathbf{x}_{(n)}$ with $\mathscr{T}_{(n)}$ (Fig. 1a). The hierarchy requires $\mathscr{T}_{(1)} \ll \mathscr{T}_{(2)} \ll \cdots \ll \mathscr{T}_{(n)}$. Not every dynamical level necessarily gives rise to a spatial or temporal pattern. Consequently, there may be fewer organisational structures than hierarchical scales. Self-organised structures that serve as building blocks for higher-level organisation must also remain stable against the fluctuations present at other dynamical scales. Fortunately, distinct structures form on different timescales within the hierarchy. This separation is key to their stability: fluctuations on fast levels average out when viewed from slower ones.

The key physical insight is that iterative locking of fast dynamics to progressively slower collective variables collapses a high-dimensional stochastic system onto a small set of externally controllable parameters at the top of the hierarchy. This procedure, which we term hierarchical steering, enables programmable transitions between emergent states by varying the external parameters quasi-statically along a continuous path $\mathbf{\Lambda}(t)$. The entire set of coordinates tracks evolving attractors without breaking locking relations: at every infinitesimal step, fast coordinates relax rapidly, intermediate ones follow adiabatically, and the slowest coordinates remain quasi-static (Fig. 1b). Hierarchy is thus elevated from an observed feature to a mechanism for programmable emergence.

We illustrate hierarchical steering through a generalised Swift–Hohenberg model, one of the canonical descriptions of pattern formation across physics, highlighting the generality of the steering mechanism. In this toy system, fast pattern-forming dynamics are hierarchically coupled to slower collective fields with well-separated spatial and temporal scales, such that quasi-statically varied control parameters steer the formation of hierarchical spatial patterns. The final state depends not only on these parameters but on the order in which they are applied (Supplementary Section 2 and Supplementary Fig. 2).

Hierarchical organisation has been reported in biological systems[4], chemical[2,5] and colloidal[10] self-assembly, reaction-diffusion systems[11], laser-matter interactions[12,13], and lasers. They all involve multiple timescales, but vary greatly in their experimental accessibility. Among them, mode-locked lasers uniquely combine well-separated dynamical levels with full theoretical tractability and direct experimental access to every hierarchical scale. This makes lasers an ideal platform for testing the mechanistic description of the internal dynamics, dimensional reduction to a few external control parameters, and final-state steering without solving the full nonlinear dynamics.

**Hierarchical steering in a mode-locked laser**

Passively mode-locked lasers[14] support diverse nonlinear waveforms, including solitons[15,16], similaritons[17], dissipative solitons[18], and their combinations[19,20], which exhibit particle-like characteristics[21]. At sufficiently high powers, lasers[15-20] give rise to coexisting multiple pulses, which form a wide family of states with varying pulse numbers and spacings, including non-identical[22] and period-multiplied pulses[23]. In harmonic states[24], the pulses must be identical and equally spaced, and even then, many distinct states exist, differing only in pulse number. Transitions between them are triggered by intrinsic noise or external perturbations[25], often irreversibly[26]. Despite decades of research, passive harmonic mode-locking suffers from instabilities and poor reproducibility, with specific states achievable only sporadically and mainly by trial and error.

The mode-locking dynamics unfold across four hierarchical levels each characterised by a distinct temporal scale (Fig. 2). At the lowest level, longitudinal cavity modes self-lock into femtosecond pulses, forming the basic organisational structure, $\mathcal{S}_{(1)}$. The associated fast (picosecond to nanosecond scale) coordinates $\mathbf{x}_{(1)}$ rapidly relax to an attractor pulse shape parametrised by the pulse energy. The next level $\mathbf{x}_{(2)}$ is the set of pulse energies evolving over many roundtrips (over microseconds), which act as order parameters for $\mathbf{x}_{(1)}$. Above them (in the millisecond timescale), gain dynamics form $\mathbf{x}_{(3)}$, setting the operating point of the energy map. At the top level, the pulses assemble into ordered superstructures of pulse patterns $\mathcal{S}_{(2)}$, described by the slowest coordinates $\mathbf{x}_{(4)}$ (seconds). At this level, only the externally controlled parameters, which, in our laser, are the pump power $P_\mathrm{p}$, the filter offset $\Delta\lambda$ (the relative offset of the blue- and red-shifted spectral filters), and the discrete state $\mathbb{S}$ of the external pulse injection, taking on the values $\{\mathrm{off}, \mathrm{nuc}, \mathrm{ann}\}$. The state is typically off; when set to nuc or ann, it injects suprathreshold or sub-threshold pulses into the cavity for nucleation or annealing.

*Hierarchical Level 1: Pulse shape locked to a fast limit cycle*

At the fastest level of the hierarchy, the fast coordinates $\mathbf{x}_{(1)}$ correspond to the intra-cavity optical field $a\,(n;z,\tau)$. We adopt an operator-based description[27], where $a\,(n;z,\tau)$ evolves through a sequence of nonlinear operators, one for each part of the cavity: $a\left(n;z_1,\tau\right) = \hat{O}_1 a\,(n;0,\tau)$, $a\left(n;z_2,\tau\right) = \hat{O}_2 a\left(n;z_1,\tau\right)$, and so on, where $z$ denotes position along the cavity (periodic with cavity length $L_\mathrm{c}$), and $\tau$ is the temporal position with respect to the centre of the first pulse, and $n$ is the roundtrip number. The full cavity operator is the ordered product $\hat{O}_\mathrm{c} = \cdots\,\hat{O}_2 \cdot \hat{O}_1$. Only the existence and attractor nature of the limit cycle are required; explicit solution of the full evolution is not needed.

In the steady state, the optical field $a\,(n;z,\tau)$ collapses onto a limit cycle parametrised by the pulse energy $E(n)$, when the limit cycle's basin is broad compared to noise-induced excursions, a condition readily satisfied in lasers combining strong spectral filtering (wavelength-dependent loss) and nonlinear amplification[17-20,28,29],

$$\hat{O}_1 a\,(n;0,\tau) = a\left(n;z_1,\tau\right) \simeq \tilde{a}\left(E(n);\tau\right), \tag{1}$$

where $\tilde{a}\left(E(n);\tau\right)$ is the attractor pulse shape corresponding to the pulse energy $E(n)$.

This parametrisation is confirmed for our laser numerically and experimentally. Measured spectra collapse onto a single one-parameter family when plotted against pulse energy (Fig. 3 and Supplementary Fig. 4). This establishes pulse energy as the order parameter governing the fastest dynamical level. At the next hierarchical level, these energies evolve slowly from roundtrip to roundtrip and become the active dynamical variables.

*Hierarchical Level 2: Energy-map engineering for pulse nucleation and annihilation*

At the second level, the slower coordinates $\mathbf{x}_{(2)}$ emerge as the set of individual pulse energies $E_i(n)$, with $i = 1,\cdots,N$, for $N$ coexisting pulses. Their evolution from roundtrip to roundtrip is governed by a discrete stochastic energy map,

$$E_i(n+1) = \mathcal{F}\left(E_i(n), g, \Delta\lambda\right) + \mathcal{E}_i(\mathbb{S}) + \eta_E(n), \tag{2}$$

obtained by applying the full cavity operator $\hat{O}_{\mathrm{c}}\left(g, \Delta\lambda\right)$ to the parametrised pulse shape, which remains locked to the pulse energy. Here, a single-variable function, $\mathcal{F}$, captures the evolution of the pulse energy, $\mathcal{E}_i(\mathbb{S})$ is the energy of an injected pulse, present only when $\mathbb{S} = \mathrm{nuc}$, and $\eta_E(n)$ represents a weak noise that captures fluctuations in pulse energy per roundtrip. The noise term becomes significant only near bifurcations. The gain factor $g$ depends on pump power $P_{\mathrm{p}}$, varies slowly over about 100 roundtrips or more, and is therefore effectively constant during pulse energy evolution.

The fixed points, labelled as $E^{*,m}$ in order of increasing energy, define quantised energy levels for stable pulses. With $E^{*,0} = 0$ always present, they satisfy $E^{*,m} = \mathcal{F}\left(E^{*,m}, g, \Delta\lambda\right)$, stable if $|\partial\mathcal{F}/\partial E| < 1$. Their values and stability depend dynamically only on the slowly varying gain, which thus emerges as the order parameter for the pulse-energy evolution. They can be tuned through the filter offset $\Delta\lambda$ and pump power $P_{\mathrm{p}}$ providing a control protocol. Individual pulses can occupy different fixed points and therefore have different energies.

When $\partial\mathcal{F}/\partial E > 1$, a non-zero unstable fixed point, such as $E^{*,1}$, acts as a strict energy threshold for pulse nucleation. As predicted by the energy map and shown in Fig. 4a, an externally injected ($\mathbb{S} = \text{nuc}$) suprathreshold pulse with $E > E^{*,1}$ grows into a stable circulating pulse, while subthreshold pulses decay (Fig. 4b). This enables fully deterministic pulse nucleation.

Controlled pulse annihilation, in contrast, exploits proximity to a saddle-node bifurcation of the energy map. Near this bifurcation, the stable and unstable fixed points nearly merge. Rare quantum-noise-induced tunnelling events then allow individual pulses to cross, or tunnel across, the threshold and vanish one by one, while others remain unaffected. These fluctuations originate from spontaneous emission, since environmental and pump-induced classical noise sources act at much lower (up to kHz) frequencies and are too slow to affect individual pulses differently on the picosecond timescale separating them[30,31]. By tuning the distance to the bifurcation via pump power, the annihilation rate becomes fully controllable, with the likelihood of annihilation predicted by proximity to the bifurcation (Fig. 4c).

Together, creation and annihilation provide complete control over the discrete pulse number $N$ (Fig. 4c). Multi-pulse states, qualitatively analogous to multi-particle states (see Supplementary Section 13), can therefore be assembled and modified one excitation at a time, without destabilising the remaining pulses. This establishes pulse number as a controllable emergent degree of freedom.

*Hierarchical Level 3. Slow gain-depletion and recovery*

The even slower variables $\mathbf{x}_{(3)}$ correspond to the gain, which has a pump power-controlled baseline, $g_o$, and a much smaller intra-cavity-scale modulated component $g_m(\tau)$. The modulation occurs because the gain is depleted discretely by each pulse and recovers via pumping, therefore it is locked to the evolving pulse positions. As the gain baseline responds to changes in pulse number or pump power, it shifts the entire energy map quasi-statically over many roundtrips.

Creating a new pulse adds to the total depletion, gradually lowering the gain (Fig. 2b), shifting the energy map downward, and potentially driving the system to the saddle–node bifurcation where tunnelling annihilates a pulse. This allows accommodation of excess pulses at fixed pump power, verified experimentally by pulse creation and annihilation at the same pump power (Fig. 4a,b). The characteristic gain-adjustment time is $E_{\mathrm{sat}}T_{\mathrm{c}}/NE^{*,2} \approx 10^4 T_{\mathrm{c}}/N$, where $E_{\mathrm{sat}}$ is the gain saturation energy, *i.e.*, the pulse energy that would lower the gain to $1/e$ of its former value. In practice, pump power must be adjusted incrementally after each creation or annihilation to maintain the intended operating point. If the bifurcation is approached too rapidly, multiple pulses can be lost abruptly before the gain recovers, leading to uncontrolled collapse (Fig. 4d, black arrows).

Engineering the energy map (Supplementary Section 5) enables predictive assembly of non-identical pulses using only external control parameters. By fine-tuning the pump power $P_{\mathrm{p}}$, the energy map is brought close to a bifurcation, where quantum-noise-induced tunnelling transfers one pulse from the lower-energy fixed point $E^{*,2}$ to the higher-energy point $E^{*,4}$ (Fig. 4e). Moving away from the bifurcation stabilises the resulting non-identical pair (Fig. 4f). Deliberately frustrating this stabilisation produces a dynamic regime in which the gain evolves cyclically and the pulse energy switches periodically between $E^{*,2}$ and $E^{*,4}$. This results in breathing-pulse-like states[33] with ~0.1 ms oscillations, set by the slow gain timescale (Fig. 4g).

Finally, even after the baseline gain stabilises, weak intra-roundtrip gain modulations persist and encode the instantaneous pulse pattern,

$$g_{\mathrm{m}}(\tau) = g_0 \frac{E^{*,2}}{E_{\mathrm{sat}}} \left( N \frac{\tau}{T_{\mathrm{c}}} - \sum_{j=1}^{N} \Theta\left(\tau - \tau_j\right) \right), \quad (3)$$

where $\Theta$ is the Heaviside step function. Though their amplitude is only $\sim 10^{-4} g_0$, these modulations ($\mathscr{T}_{(4)}$ in Fig. 2b) encode the pulse pattern into the gain landscape, shifting the fixed-point energies parametrically according to pulse positions. Pulse positions therefore emerge as the natural order parameter at the next level. The coupling between hierarchical levels weakens and slows progressively.

*Hierarchical Level 4: Emergent pulse-pattern dynamics, long-range interactions and annealing*

We finally arrive at the top of the hierarchy. The slowest variables $\mathbf{x}_{(4)}$ comprise the pulse positions $\{\tau_i\}$, which give rise to the emergent pulse pattern. Building on a formal analogy between their stochastic temporal motion and weakly trapped Brownian particles[32], we describe the evolution of pulse positions via

$$\dot{\tau}_i(t) = -v_i\left(\tau_1(t), \cdots, \tau_j(t), \cdots, \tau_N(t), P_\mathrm{p}/N, \Delta\lambda, \mathbb{S}\right) + \eta_{\tau,i}(t, \mathbb{S}), \quad (4)$$

where $v_i$ denotes the relative speed of the $i^{\mathrm{th}}$ pulse and $\eta_{\tau,i}(t)$ is a stochastic term due to spontaneous emission, the variance of which defines an effective temperature in the trapped Brownian-particle analogy[31]. Hierarchical steering constraints the functional form of the relative speed terms $v_i$, with coefficients obtained from simulations (Methods).

In our laser, both acoustic[33] and gain-mediated[34] interactions contribute to pulse-to-pulse coupling. The gain-mediated interaction is approximately linear in the displacement from harmonic positions and largely independent of repetition rate, whereas the acoustic interaction is oscillatory in pulse separation and strongly repetition-rate dependent. For the harmonic states, we operate in the regime where the gain-mediated interaction prevails. Near the saddle-node bifurcation of the energy map, its strength scales as $\left(1 - \partial\mathcal{F}/\partial E\right)^{-1}$, acting as a stiff restoring force toward harmonic order.

Perfect harmonic states are not established immediately. The pulses can become trapped in shallow metastable, anharmonic temporal positions due to competing contributions from gain-mediated and acoustic interactions. Inspired by thermal annealing, we perturb the pattern by injecting a stream of subthreshold ($\mathbb{S} = \mathrm{ann}$, $E < E^{*,1}$) pulses. Because the injected pulses are asynchronous with the cavity pulse train, their deterministic action induces transient, effectively stochastic deformations of the gain landscape on timescales short compared to the slow pattern dynamics. For the pulse-position degrees of freedom, this is equivalent to a temporary increase in the effective noise strength, analogous to raising the effective temperature of the pulse pattern without perturbing faster hierarchical levels. This controlled agitation enables pulses trapped in shallow local gain maxima to escape and relax toward the global harmonic configuration (Fig. 5a), while pulses already at harmonic positions are less affected. This mechanism reproducibly leads to near-harmonic patterns with more than 100 intra-cavity pulses, inferred from the harmonic-to-fundamental repetition-rate ratio (Fig. 5b). We then fine-tune stiffness of the harmonic traps via the control parameters (Supplementary Section 7), and routinely reach fully harmonic states with equal spacing $T_{\mathrm{R}} = T_{\mathrm{c}}/N$ and supermode suppression (a measure of harmonic purity) exceeding 50 dB. The remaining deviations correspond to stochastic motion of pulses within their harmonic traps.

**Conclusions and Outlook**

Our results show that when a system with multiscale self-organisation exhibits strong timescale separation, emergence itself becomes programmable: external interventions applied at the slowest collective level propagate through the hierarchy while faster degrees of freedom remain locked and the corresponding organisational structures intact. Timescale separation further provides dynamical insulation, ensuring that emergent structures remain robust against fluctuations across scales, except when the system is deliberately positioned near bifurcations, where fluctuations become controllable tools for state transitions, as demonstrated by quantum-noise-induced pulse annihilation and annealing. Applied to harmonic mode-locking, this has yielded predictive protocols for pulse creation, annihilation, tailored pulse patterns, and annealing-like transitions to fully harmonic states, achievements that eluded decades of largely empirical approaches. This paves the way to increasing the number of harmonic pulses without destabilisation, thus opening the door to record-high repetition rates. Our approach should naturally extend to spatiotemporal mode-locking[27,35], enabling programmable control over larger numbers of coupled spatial modes than are currently accessible.

The requirement for well-separated dynamical timescales becomes demanding with increasing complexity: the total temporal span grows exponentially with the number of nested levels. Reassuringly, echoing Feynman's observation that "*there is plenty of room at the bottom*," there are many orders of magnitude of temporal dynamics at the fast end. In particular, laser–matter interactions provide direct access to processes as fast as attoseconds, the timescale of electronic responses[36]. Recent advances in laser-driven self-assembly and structure formation[37-41] demonstrate that such processes are not constrained by the diffraction limit, because they do not rely on point-by-point modification. Instead, distinct physical processes act on distinct timescales, from ultrafast heating and plasma formation to slower atomic structural changes, transport and reequilibration[42]. These can be selectively addressed, for example by lasers generating ultrashort pulses, grouped into nanosecond bursts[43], combined with microsecond-to-millisecond spatial beam shaping. In such systems, formation is localised to an interaction zone, where targeted steering proceeds through previously formed structures that persist and bias subsequent formation, resulting in permanent self-organised architectures. The combination of extreme temporal depth with programmable emergence may open a route for directed self-organisation in material systems[44].

# Methods

## General Requirements

Let $\mathbf{x}(t) = (x_1, x_2, \cdots)$ denote all dynamical coordinates, evolving according to a high-dimensional set of first-order equations, $\dot{\mathbf{x}} = \mathscr{F}(\mathbf{x}, \boldsymbol{\Lambda})$, with $\boldsymbol{\Lambda} = (\lambda_1, \cdots, \lambda_k)$ representing a small set of externally controlled parameters. The governing equations take a nested form after being grouped according to their time scales,

$$
\begin{aligned}
\dot{\mathbf{x}}_{(1)} &= \mathscr{F}_1(\mathbf{x}_{(1)}(t), \mathbf{x}_{(2)}, \cdots, \mathbf{x}_{(n)}, \boldsymbol{\Lambda}), \\
\dot{\mathbf{x}}_{(2)} &= \mathscr{F}_2(\mathbf{x}_{(2)}(t), \cdots, \mathbf{x}_{(n)}, \boldsymbol{\Lambda}), \\
&\ \vdots \\
\dot{\mathbf{x}}_{(n)} &= \mathscr{F}_n(\mathbf{x}_{(n)}(t), \boldsymbol{\Lambda}),
\end{aligned}
\tag{5}
$$

where, at each timescale, slower variables evolve so gradually that they are treated as effectively constant. $\mathbf{x}_{(i>1)}$ are the reduced coordinates obtained after applying the adiabatic elimination or related techniques. These slow coordinates act as order parameters for the faster coordinates, analogous in their role to the externally controlled $\boldsymbol{\Lambda}$. The fast variables rapidly relax to attractors, comprising stable fixed points or limit cycles, which are constant on their own timescale but drift slowly on longer times in response to the slow order parameters:

$$
\begin{aligned}
\mathbf{x}^*_{(1)} &= \boldsymbol{\Phi}^*_{(1)}(\mathbf{x}_{(2)}, \cdots, \mathbf{x}_{(n)}, \boldsymbol{\Lambda}), \\
\mathbf{x}^*_{(2)} &= \boldsymbol{\Phi}^*_{(2)}(\mathbf{x}_{(3)}, \cdots, \mathbf{x}_{(n)}, \boldsymbol{\Lambda}), \\
&\ \vdots \\
\mathbf{x}^*_{(n)} &= \boldsymbol{\Phi}^*_{(n)}(\boldsymbol{\Lambda})\,.
\end{aligned}
\tag{6}
$$

We use order parameter to denote slow variables that parametrise fast attractors, and collective variable more generally for emergent slow coordinates. These typically coincide: each $\mathbf{x}_{(>i)}$ serves as both an order parameter for $\mathbf{x}_{(i)}$ and a collective variable emerging from faster dynamics. At each level, large numbers of degrees of freedom are locked to progressively slower variables, yielding a nested sequence of reduced descriptions so that each attractor, or family of coexisting attractors, $\mathbf{\Phi}^*_{(k)}(\mathbf{x}_{(k>1)}, \cdots, \mathbf{x}_{(n)}, \mathbf{\Lambda})$ has fewer dependencies until only the externally controlled parameters remain at the top. At a given hierarchical level, multiple stable attractors may coexist, corresponding to distinct states whose selection is ultimately set by the externally controlled parameters. As $\mathbf{\Lambda}$ is varied, the attractor at each level deforms accordingly and may undergo bifurcations, and the system evolves from $\mathbf{x}^*$ to $\mathbf{x}^{*'}$. This progression need not be reversible, and the control-parameter trajectory required to steer the system back from $\mathbf{x}^{*'}$ back to $\mathbf{x}^*$ may differ entirely from the forward trajectory. Steering, therefore, reduces to identifying a continuous path $\mathbf{\Lambda}(t)$ that moves the top-level structure $\mathcal{S}_{(m)}$ toward the desired target (Fig. 1b).

The requirements are as follows: (i) Each block of coordinates $\mathbf{x}_{(i)}$ possesses a basin of attraction leading to a manifold (attractor) $M_i(\mathbf{x}_{(>i)}, \mathbf{\Lambda})$ on which the fast coordinates rapidly converge. This manifold may correspond to a fixed point or a limit cycle, and depends parametrically on the slower variables $\mathbf{x}_{(>i)}$ and the control parameters $\mathbf{\Lambda}$. The dynamics have no strong explicit dependence on faster coordinates $\mathbf{x}_{(<i)}$, which enter only through their fixed-point values, or for limit cycles, through time-averaged integrals of motion (*e.g.*, pulse energy). (ii) The linearised dynamics around this manifold exhibit well-separated relaxation rates. If $\nu_i^{\mathrm{fast}}$ denotes the real part of the dominant eigenvalue governing relaxation of $\mathbf{x}_{(i)}$, and $\nu_{i+1}^{\mathrm{slow}}$, the real part of the eigenvalue governing the evolution of $\mathbf{x}_{(i+1)}$, then the reduction[8,9] requires $|\nu_i^{\mathrm{fast}}| \gg |\nu_{i+1}^{\mathrm{slow}}|$. This condition must iteratively hold for all levels, and formalises the requirement $\mathcal{T}_{(1)} \ll \mathcal{T}_{(2)} \ll \cdots \ll \mathcal{T}_{(n)}$ or, more compactly, $\varepsilon_i \equiv \mathcal{T}_{(i)}/\mathcal{T}_{(i+1)} \ll 1$. This ensures that slower variables can be treated as quasi-static on the timescale of faster dynamics. The accuracy of adiabatic elimination improves with the degree of timescale separation increases[45]. (iii) Along slow drifts of $\mathbf{x}_{(>i)}$ and $\mathbf{\Lambda}$, the manifolds $\mathbf{\Phi}^*_{(i)}(\mathbf{x}_{(>i)}, \mathbf{\Lambda})$ vary smoothly except at isolated bifurcations. (iv) Stochastic fluctuations of fast variables must decorrelate rapidly enough that their cumulative influence on slower variables remains limited[10].

*Reduction of the degrees of freedom: deterministic treatment*

For clarity, we illustrate the reduction for a system with two timescales; the extension to multiple levels can be done iteratively. Suppose the dynamics can be written, after rescaling time so the slowest level evolves on a unit timescale, with an explicit small parameter $\varepsilon \ll 1$ separating the blocks of fast and slow coordinates:

$$\begin{aligned} \varepsilon\, \dot{\mathbf{x}}_{(1)} &= \mathscr{F}_1\left(\mathbf{x}_{(1)}, \mathbf{x}_{(2)}, \boldsymbol{\Lambda}\right), \\ \dot{\mathbf{x}}_{(2)} &= \mathscr{F}_2^{\text{full}}\left(\mathbf{x}_{(1)}, \mathbf{x}_{(2)}, \boldsymbol{\Lambda}\right), \end{aligned} \tag{7}$$

so that the time derivative of the fast block is $O(1/\varepsilon)$ larger than that of the slow block. (Different blocks can be separated by different $\varepsilon$; here, we consider only two levels so there is a single value.)

Under the required assumptions discussed above, the fast block of coordinates rapidly approaches its attracting manifold parametrised by the slower variables. We denote the attracting manifold by $\mathbf{x}^*_{(1)} = \boldsymbol{\Phi}_1\left(\mathbf{x}_{(2)}, \boldsymbol{\Lambda}\right)$, while noting that it could (as in the case of the laser) involve a stable limit cycle instead of a fixed point. Then, to leading order in $\varepsilon$ the slow coordinates are governed by the reduced equation $\dot{\mathbf{x}}_{(2)} = \mathscr{F}_2^{\text{full}}\left(\boldsymbol{\Phi}_1\left(\mathbf{x}_{(2)}, \boldsymbol{\Lambda}\right), \mathbf{x}_{(2)}, \boldsymbol{\Lambda}\right) = \mathscr{F}_2\left(\mathbf{x}_{(2)}, \boldsymbol{\Lambda}\right)$, whereby the dimensionality is reduced because $\mathbf{x}_{(1)}$ has been replaced by its locked values, while the slow drift of the eliminated variables is implicitly retained through their parametric dependence on $\mathbf{x}_{(2)}$ and $\boldsymbol{\Lambda}$. For systems with more than two timescales, this adiabatic elimination is applied iteratively from the lowest (fastest) to the top (slowest) hierarchy, thereby obtaining the reduced equations, $\mathscr{F}_2, \ldots, \mathscr{F}_n$. On any finite time interval that avoids bifurcations, the difference between solutions of the full and the reduced systems remains limited, $\left\| \mathbf{x}^{\text{full}}_{(>i)}(t) - \mathbf{x}^{\text{red}}_{(>i)}(t) \right\| \leq \eta\, \varepsilon$, for some finite $\eta$, which may depend on the time interval[46]. Practically, the error vanishes with greater separation of the timescales ($\varepsilon \to 0$).

*Reduction of the degrees of freedom: stochastic treatment*

The hierarchy acts as a bidirectional filter: fluctuations at a given level are either averaged out or effectively frozen when viewed from adjacent scales. To demonstrate the implications using stochastic averaging[46,9], we write:

$$\begin{aligned} \dot{\mathbf{x}}_{(1)}(t) &= \varepsilon^{-1}\mathcal{F}_1\left(\mathbf{x}_{(1)}, \mathbf{x}_{(2)}, \mathbf{\Lambda}\right) + \varepsilon^{-1/2}\sigma\left(\mathbf{x}_{(2)}, \mathbf{\Lambda}\right)\xi_t, \\ \dot{\mathbf{x}}_{(2)}(t) &= \mathcal{F}_2^{\text{full}}\left(\mathbf{x}_{(1)}, \mathbf{x}_{(2)}, \mathbf{\Lambda}\right). \end{aligned} \tag{8}$$

Here, we reintroduce $\mathcal{F}_2^{\text{full}}$ to capture the weak dependence of the slow coordinates on the fast ones (ignored for the deterministic treatment), $\xi_t$ represents a fast white noise, $\varepsilon = \mathcal{T}_{(1)}/\mathcal{T}_{(2)} \ll 1$ captures the timescale separation, and $\sigma(\mathbf{x}_{(2)}, \mathbf{\Lambda})$ sets the strength of the fast fluctuations. The scaling of the noise by $\varepsilon^{-1/2}$ represents the diffusive limit; if noise scaling were stronger, the fluctuations would not be averaged out and would likely destabilise the slower (higher) level. Conversely, if the noise scaled more weakly than the diffusive limit, it would be averaged out completely, exerting no influence on the slower level[9], which arises from the Central Limit Theorem[45].

As we move to the slow timescale of $\mathbf{x}_{(2)}$, the rapid oscillations average out. Under the assumption of ergodic fast process, the slow variable $\mathbf{x}_{(2)}$ converges in distribution to a diffusion process[45]

$$\dot{\mathbf{x}}_{(2)} = \mathcal{F}_2^{\text{det}}\left(\mathbf{x}_{(2)}, \mathbf{\Lambda}\right) + \sqrt{\Sigma(\mathbf{x}_{(2)}, \mathbf{\Lambda})}\,\zeta_t, \tag{9}$$

where $\mathcal{F}_2^{\text{det}}$ represents the deterministic reduced equations of motion for $\mathbf{x}_{(2)}$, representing how the system evolves without taking into account the effect of the fast noise acting on $\mathbf{x}_{(1)}$. The term $\Sigma(\mathbf{x}_{(2)}, \mathbf{\Lambda})$ is an effective diffusion tensor, which captures the residual impact of the fast fluctuations. This tensor is defined by a Green-Kubo-type covariance integral[45,9]

$$\Sigma(\mathbf{x}_{(2)}, \mathbf{\Lambda}) = \int_0^{\infty} \left\langle \Delta\mathcal{F}_2(\phi_t(\mathbf{x}_{(1)}), \mathbf{x}_{(2)}) \otimes \Delta\mathcal{F}_2(\mathbf{x}_{(1)}, \mathbf{x}_{(2)}) \right\rangle dt, \tag{10}$$

where $\Delta\mathscr{F}_2 = \mathscr{F}_2^{\text{full}} - \mathscr{F}_2^{\text{det}}$, and $\phi_t(\mathbf{x}_{(1)})$ denotes the fast stochastic trajectory of $\mathbf{x}_{(1)}$ under $\mathscr{F}_1$ with $\mathbf{x}_{(2)}$ having effectively fixed (frozen) values. The notation $\langle \cdot \rangle$ denotes an ensemble average with respect to the probability distribution of the fluctuations of the fast coordinates $\mathbf{x}_{(1)}$. Physically, this integral quantifies the cumulative effect of the correlations of the micro-kicks (such as spontaneous emission events) as they are experienced by the slow coordinates (*e.g.*, the pulse position). As the timescale separation grows ($\varepsilon \to 0$), the fast process decorrelates quickly in the timescale of the slow variables because the integral in Equation (10) is effective limited by $\mathscr{T}_{(1)}$. As a result, the magnitude of $\Sigma$ scales with $\varepsilon$[45]. However, as well-known in stochastic resonance and pattern selection[47], even a vanishingly small diffusion term can be physically vital. In our laser example, this residual stochasticity can act as a selection mechanism, such as governing the long-term pulse position ordering, and timing jitter of the pulses.

*Interplay between organisational hierarchy and hierarchy of timescales*

Organisational hierarchy denotes a nested series of emergent coherent structures, represented by $\mathscr{S}_{(1)}, \mathscr{S}_{(2)}, \ldots, \mathscr{S}_{(\mathrm{m})}$, where each $\mathscr{S}_{(\mathrm{k}+1)}$ comprises the simpler structures, $\mathscr{S}_{(\mathrm{k})}$, acting as building blocks (for example, patterns of pulses in the laser formed from individual pulses, which emerge from the interaction of longitudinal cavity modes). Timescale hierarchy refers to a separation in dynamic timescales for coordinate blocks, $\mathbf{x}_{(1)}, \mathbf{x}_{(2)}, \ldots, \mathbf{x}_{(n)}$, characterised by $\varepsilon_i \equiv \mathscr{T}_{(i)}/\mathscr{T}_{(i+1)} \ll 1$. These two hierarchies are conceptually distinct and need not be one-to-one; in general $m < n$. In the laser case studied here, $\mathscr{S}_{(1)}$ (the pulse shape) mainly arises from $\mathbf{x}_{(1)}$, whereas $\mathscr{S}_{(2)}$ (the pulse pattern) emerges from the slowest block of coordinates, $\mathbf{x}_{(4)}$. $\mathbf{x}_{(2)}$ and $\mathbf{x}_{(3)}$, govern energy and gain dynamics and do not represent distinct structures comparable to $\mathscr{S}_{(1)}$ and $\mathscr{S}_{(2)}$.

An implicit requirement is that each organisational structure $\mathscr{S}_{(\mathrm{k})}$ be sufficiently long-lived relative to the timescale on which they act as building blocks for a higher-level structure $\mathscr{S}_{(\mathrm{k}+1)}$. Unlike material particles, emergent structures are not inherently stable objects. When an organisational structure $\mathscr{S}_{(k)}$ arises from coordinates $\mathbf{x}_{(i)}$ and the higher structure $\mathscr{S}_{(k+1)}$ arises from coordinates $\mathbf{x}_{(j)}$ with $j \geq i+1$, a dynamic insulation mechanism provides stability, provided the assumptions of hierarchical steering (existence of attracting manifolds and timescale separation) hold. Fast deterministic dynamics relax to their attractor on times $\mathscr{T}_{(i)}$, while fast stochastic fluctuations acting on $\mathbf{x}_{(i)}$ are averaged out on the slower timescale of $\mathbf{x}_{(j)}$. Conversely, fluctuations acting on $\mathbf{x}_{(j)}$ on times $\tau_{(j)}$ appear effectively frozen or quasi-static over time intervals of order $\mathscr{T}_{(i)}$, and therefore do not act as order-disrupting noise for $\mathbf{x}_{(i)}$. This mutual decoupling is the dynamical insulation mechanism by which organisational structures at different levels are protected from destructive interscale perturbations. While fast noise averages to a small effective stochastic forcing on slow variables, rare large deviations (*e.g.*, tunnelling events near bifurcations) can still occur. Their rates depend exponentially on the appropriate action or barrier height and on the effective noise amplitude on the slow manifold. The interplay of slow bifurcations and weak noise underlies the controlled pulse annihilation events discussed in the main text.

In the laser example considered here, $\mathscr{S}_{(1)}$ (pulse shape) and $\mathscr{S}_{(2)}$ (pulse pattern) occupy widely separated timescales (over ten orders of magnitude), fulfilling the conditions above and yielding the dynamic insulation that makes hierarchical self-organisation (onset of mode-locking, followed by creation, tuning, annealing of pulses, and lastly onset of harmonic or anharmonic pulse patterns) controllable and repeatable.

**Mode-locking dynamics and operator formalism**

The pulse, with complex field $a(n; z, \tau)$ during the $n^{\text{th}}$ roundtrip, evolves through a sequence of nonlinear operators. Each operator $\hat{O}_m$, represents a distinct cavity section: $\hat{O}_1 a(n; 0, \tau) = a\left(n; z_1, \tau\right)$, $\hat{O}_2 a\left(n; z_1, \tau\right) = a\left(n; z_2, \tau\right)$, and so on, where $z$ denotes position along the cavity (periodic with cavity length $L_c$), and $\tau$ is the temporal position (also known as time delay) in a frame co-moving with the circulating optical field. Each section implements nonlinear propagation governed by a generalised nonlinear Schrödinger equation or discrete transformation, such as similariton or soliton propagation, or spectral filtering. The full cavity evolution is described by the concatenated operator, $\hat{O}_c = \cdots \hat{O}_2 \hat{O}_1$, such that $a_{n+1}(0,\tau) = \hat{O}_c\, a_n(0,\tau)$. The steady state, $a_{ss}(z,\tau)$, corresponds to a limit cycle of this system, satisfying $a_{ss}(0,\tau) = \hat{O}_c\, a_{ss}(0,\tau)$. For a broad range of initial conditions, $a(0;0,\tau)$, the laser converges after a small number of roundtrips, $a_{ss}(0,\tau) = \lim_{n\to\infty} \hat{O}_c^n a(0;0,\tau)$.

**Hierarchical Level 1**

We validate the hierarchical steering in a Mamyshev fibre laser, chosen for its easily adjustable saturable-absorption characteristics. A typical Mamyshev cavity consists of two nonlinear amplification arms[28], each preceded by a narrow spectral filter, one blue-shifted and one red-shifted. In many implementations, including ours, the amplification stages support similariton-like evolution[17,29]. The filters constrain the input spectrum entering each arm, such that the output pulse shape becomes uniquely determined, or parametrised, by its energy, $E(n)$. For a similariton, $|\tilde{a}|^2 = E(n)/\tau_p\left(E(n)\right) \exp\left(-\sum_{k=1}^{l} \tau^2/\tau_p^2\left(E(n)\right)\right)$, where higher $l$ denotes a nearly ideal parabolic similariton[49]. A second attractor follows the second filter, with analogous parametrisation. Pulse shaping mechanisms differ in the strength of the attraction to a single slow coordinate, and some require additional parameters or show slower convergence. As long as the basin of attraction of the limit cycle is broad compared to noise-induced excursions, stability is preserved. This description naturally extends to the multi-pulse regime with $N$ pulses, each with energy, $E_j(n)$, and temporal position, $\tau_j(n)$, given by $a_{tot}(z,\tau) = \sum_{j=1}^{N} a_j(z,\tau-\tau_j)$. Provided the pulses are well separated, their energies and positions can be treated as independent degrees of freedom, and each pulse converges to the same shape $a_j(z_1,\tau-\tau_j) = \tilde{a}\left(E_j(n);\tau\right)$.

A narrow filter effectively equalises all pulse shapes: the pulse shape in the frequency domain is given by $a(\omega) = |a(\omega)|e^{i\phi(\omega)}$, where $|a(\omega)|$ and $\phi(\omega)$ are the spectral amplitude and phase, respectively. After a filter with transmission $H(\omega)$, the spectrum becomes $a_{\text{out}}(\omega) = a(\omega)H(\omega)$. In the limit where the pulse bandwidth substantially exceeds the filter bandwidth, both the spectral amplitude and phase vary weakly across the transmitted band, $a_{\text{out}}(\omega) \approx |a(\omega_0)|\, e^{i\phi(\omega_0)}\, H(\omega)$, where $\omega_0$ is the filter centre frequency. Thus, the filter enforces a common spectral amplitude and phase, differing only by a multiplicative constant that sets the pulse energy. The remaining degree of freedom is a temporal shift that depends on the initial chirp. Transmission through the filter depends on the spectral overlap between the broadened pulse spectrum and the filter passband. The filter bandwidth is much narrower than the individual spectra, such that transmission is dominated by the partial overlap of the filter with a single one of the lobes generated by nonlinear broadening (Fig. 3). Approximating both the lobe and the filter as Gaussian, and that the spectrum shifts approximately linearly with pulse energy, the energy map can be expressed as $E(n+1) = E(n)\, C(1 - e^{-E^2(n)/\epsilon^2})e^{-(\Delta_0 - \kappa E(n)/\epsilon)^2/\Delta\lambda_{\text{eff}}^2}$, where $\Delta_0$ is the offset between the lobe and the filter, $\kappa$ is the spectral shift coefficient, $\epsilon$ scales the energy, and $\Delta\lambda_{\text{eff}}^2 = \Delta\lambda_\ell^2 + \Delta\lambda_f^2$ is the quadrature sum of the lobe and filter bandwidths. This minimal form captures the rise, maximum, and initial decay of the transmission curve at low energies (Fig. 4a,b). As pulse energy increases, the single-lobe approximation breaks down: once the first lobe shifts beyond optimal overlap, transmission can rise again due to partial overlap with the next lobe. Rather than modelling spectral shapes explicitly, we semi-phenomenologically represent the multi-lobed spectral structure sampled by the narrow filter. The resulting cavity energy map can then be written in the generic reduced form $E(n+1) = E(n)\, C(1 - e^{-E^2(n)/E_0^2})\, e^{-(\Delta_0 - \kappa E(n)/\epsilon)^2/\Delta\lambda_{\text{eff}}^2}\left(1 + m\cos(\Omega E(n)/\epsilon)\right)$, where the cosine term gives the oscillatory modulation, with scaling $\Omega$ and depth $m$. This parameterisation reproduces the energy map with two filters and two gain segments, including the emergence of the higher-energy stable fixed point $E^{*,4}$ (Fig. 4e). This generic form is not specific to Mamyshev cavities, but commonly observed in other lasers, such as those using nonlinear polarization rotation.

**Hierarchical Level 2**

Applying the cavity operator $\hat{O}_c\left(g_b, g_r, \Delta\lambda\right)$ to this pulse shape yields the shape at the following roundtrip, $\tilde{a}\left(E\left(n+1\right)\right)$, which determines $E\left(n+1\right)$. This defines a discrete stochastic energy map $E(n+1) = \mathscr{F}\left(E(n); g_b, g_r, \Delta\lambda\right) + \eta_E(n)$, where $\eta_E(n)$ represents weak roundtrip-roundtrip energy fluctuations. The map $\mathscr{F}$ is constructed directly from numerical pulse-propagation simulations as $\mathscr{F} \equiv \int \left|\hat{O}_c \tilde{a}\left(E\left(n\right)\right)\right|^2 d\tau$. The gains $g_b$ and $g_r$ of the blue and red arms are nearly constant during pulse energy evolution corresponding to the two gain sections of the cavity, which is typical for a Mamyshev laser. In the main manuscript this distinction is not shown for simplicity. The spectral filters convert the spectral shaping induced by nonlinear propagation in each arm into amplitude transmission functions, $\mathscr{F}_b$ and $\mathscr{F}_r$, which closely follow the spectral profiles (Fig. 3). The full energy map is then given by $\mathscr{F} = g_b(n)\mathscr{F}_b\left(g_r(n)\mathscr{F}_r\left(E(n), \Delta\lambda\right), \Delta\lambda\right)$, which is well approximated by the energy map function discussed above. The energy map curves in Fig. 4e were generated by fitting the generic functional form above to the experimental transmission data in Fig. 3c to determine a representative parameter set, and subsequently varying a single control parameter to reproduce the experimentally observed shift shown in Fig. 4f. Because the available data do not uniquely constrain all parameters of the phenomenological model, these curves are not intended as quantitative fits. Rather, they provide a physically motivated reconstruction of the experimentally relevant energy map, whose qualitative structure is robust against moderate parameter variations.

**Hierarchical Level 3**

The gain evolves as it gets depleted by each pulse and recovers slowly via pumping, as described by

$$\dot{g}(t) = \frac{g(t)}{E_{\mathrm{sat}}}\left(\epsilon P_{\mathrm{p}} - \sum_{n=1}^{\infty}\sum_{j=1}^{N}\frac{E_j(n)}{T_{\mathrm{c}}}\delta\left(\frac{t}{T_{\mathrm{c}}} - \frac{\tau_j}{T_{\mathrm{c}}} - n\right)\right) \tag{11}$$

where we approximate each pulse by a Dirac delta, $t$ denotes time in the laboratory frame, $P_{\mathrm{p}}$ is the pump power, $T_{\mathrm{c}}$ is the cavity roundtrip time, $E_j(n)$ is energy of the $j^{\mathrm{th}}$ pulse for the $n^{\mathrm{th}}$ roundtrip, $E_{\mathrm{sat}}$ is the saturation energy, and $\epsilon$ is the pump-to-signal conversion efficiency. This dynamics naturally separates into a slowly evolving baseline gain $g_{\mathrm{o}}$, and an intra-roundtrip modulation $g_{\mathrm{m}}(\tau)$, $g(t) = g_{\mathrm{o}}(t) + g_{\mathrm{m}}(\tau)$, with $g_{\mathrm{m}} \ll g_{\mathrm{o}}$. The baseline gain evolves according to the roundtrip-averaged equation $\dot{g}_{\mathrm{o}} \simeq g_{\mathrm{o}}\left(\epsilon P_{\mathrm{p}} - \sum_{j=1}^{N} E_j^{*,m}(g_{\mathrm{o}})/T_{\mathrm{c}}\right)/E_{\mathrm{sat}}$. In the breathing-like pulse state, $N = 1$, and $m$ switched periodically between 2 and 4 because the pump power was set such that $g_{\mathrm{o}}$ values corresponding to $m = 2$ and $m = 4$ fall outside the respective ranges of $E^{*,2}(g_{\mathrm{o}})$ and $E^{*,4}(g_{\mathrm{o}})$.

**Hierarchical Level 4**

The laser cavity supports interactions that are both gain and acoustic in origin, which are detailed in Supplementary Section 8 and 9, respectively. The gain-mediated interactions show up in

$$\frac{g_{\mathrm{m}}}{g_{\mathrm{o}}} = \frac{E^{*,2}}{E_{\mathrm{sat}}}\left(\frac{N\tau}{T_{\mathrm{c}}} - \sum_{j=1}^{N}\Theta\left(\tau - \tau_j\right)\right), \tag{12}$$

where $\Theta$ is the Heaviside step function, assuming all pulses have identical pulse energies, $E^{*,2}$. The gain modulation experienced by the $i^{\mathrm{th}}$ pulse is then

$$\frac{g_{\mathrm{m}}\left(\tau_i\right)}{g_{\mathrm{o}}} = \frac{E^{*,2}}{E_{\mathrm{sat}}}\left(\frac{N\tau_i}{T_{\mathrm{c}}} - \sum_{j=1}^{N}\Theta\left(\tau_i - \tau_j\right)\right) = \frac{E^{*,2}}{E_{\mathrm{sat}}}\left(\frac{\tau_i}{T_{\mathrm{R}}} - i\right). \tag{13}$$

Here, $i$ is the index of the pulse in temporal order within the cavity roundtrip. Because $g_{\mathrm{m}} \ll g_{\mathrm{o}}$, we can assume the pulse speed depends linearly on it,

$$v_{\mathrm{i,gain}} = \frac{\partial v_i}{\partial g_{\mathrm{m}}\left(\tau_i\right)} \frac{E^{*,2}}{E_{\mathrm{sat}}} \left(\frac{\tau_i}{T_{\mathrm{R}}} - i\right) \equiv \Gamma\left(\frac{\tau_i}{T_{\mathrm{R}}} - i\right), \tag{14}$$

where $v_{\mathrm{i,gain}}$ is the contribution of the gain-mediated interaction to the speed of the $i^{\mathrm{th}}$ pulse.

There are two contributions from the acoustic waves to the relative pulse speeds in Equation (4). The first is a direct contribution given by

$$v_{i,n_{\mathrm{I,m}}} = \alpha_6 \tilde{n}_6 + \alpha_{10} \tilde{n}_{10} = \sum_{h=-\infty}^{+\infty} \left(\alpha_6 T_6(\omega_h) + \alpha_{10} T_{10}(\omega_h)\right) \sum_{j=1}^{N} \mathrm{e}^{i\omega_h\left(\tau - \tau_j\right)}, \tag{15}$$

where $\tilde{n}_6$ and $\tilde{n}_{10}$ are the index modulations per unit energy in the 6-μm and 10-μm fibres, the coefficients $\alpha_6$, $T_6$, and $\alpha_{10}$, $T_{10}$, are their acoustic interaction coefficients and transfer functions, and $\omega_h = h \cdot 2\pi / T_{\mathrm{c}}$ are the harmonic frequencies of the laser cavity. The second acoustic contribution to the pulse speed is indirect, arising from a shift of the pulse's central wavelength, given by

$$v_{i,\frac{d}{d\tau}\tilde{n}_{\mathrm{I,m}}} = \sum_{h=-\infty}^{+\infty} i\omega_h \left(\beta_6 T_6(\omega_h) + \beta_{10} T_{10}(\omega_h)\right) \sum_{j=1}^{N} \mathrm{e}^{i\omega_h\left(\tau_i - \tau_j\right)}, \tag{16}$$

where $\beta_6$ and $\beta_{10}$ are the coefficients for the two fibre types, determined from simulations.

**Numerical model for pulse propagation and gain dynamics**

The simulation of pulse propagation is based on numerically integrating the Nonlinear Schrödinger Equation (NLSE), generalised to include gain, third-order dispersion, and spontaneous emission noise using the Runge-Kutta for Interaction Picture algorithm[49]. Output coupling and parasitic losses, including fibre splice losses, were also included. The gain was modelled via the effective transition cross-sections of Yb-doped germanosilicate fibre[50]. The concentration of excited gain ions, $N_{\mathrm{e}}$, was taken as constant over the simulated time window (tens of picoseconds), but allowed to vary between propagation steps along the gain fibres and between roundtrips. Different procedures were used to calculate $N_{\mathrm{e}}$. In preparatory simulations, $N_{\mathrm{e}}$ was quickly obtained by setting it in each propagation step to a value that balances the numbers of absorbed and emitted photons. In this case, $N_e$ varied strongly between roundtrips until convergence of the pulse energy. In the main simulations, we either fixed $N_e$ or updated it dynamically according to the net photon flux in each propagation step. The former was used to model effects that differ from pulse to pulse or fluctuate rapidly in time, averaging out their influence on the gain, while the latter was used to simulate the gain evolution. Simulations involving spontaneous emission noise included an additional term in each gain propagation step. This was implemented in the Fourier domain with random phase and amplitude of $\sqrt{dz N_e \sigma_e(\omega)\hbar\omega}$, where $dz$ is the propagation step size. This term was disabled in simulations of purely deterministic effects. The coefficients for pulse pattern dynamics and the simulations are discussed in Supplementary Section 10 and 11, respectively.

**Experimental setup**

The setup (Supplementary Fig. 3) is a dual-arm Mamyshev laser cavity. All fibres are polarisation-maintaining (PM) and single-mode for environmental stability and long-term robustness. Each arm consists of a narrowband spectral filter followed by nonlinear amplification in Yb-doped fibre, forming the characteristic Mamyshev configuration. In the blue arm, a reflective grating (600 lines/mm) and collimator define the filter, followed by a pump–signal combiner and a 0.63 m Yb-doped 10/125 gain fibre pumped at 976 nm. Residual pump light is removed by a second pump–signal combiner. Signal exits through a collimator after travelling an additional 2.11 m of PM980 fibre. After passing a second grating–collimator pair forming the red filter, and a free-space isolator, it enters the red-arm filter. Both gratings are mounted on rotational stages for wavelength tuning. The red arm follows a similar layout, with a longer 10/125 Yb-doped fibre (2.45 m) and a total of 2.76 m of PM980 fibre. The signal exits through a collimator with a 0.98 m 10-µm pigtail, then passes a free-space isolator, an adjustable output coupler (half-wave plate and polarising beam splitter), and a 30/70 beam splitter before reentering to the blue filter. The total fibre lengths in the blue and red arms are 5.1 m and 7.1 m, respectively. The group-velocity dispersion of the fibres is 21 $fs^2/mm$, corresponding to a net cavity dispersion of ~0.25 $ps^2$. The fundamental repetition rate is ~15 MHz, corresponding to a cavity roundtrip time of $T_c \sim 67$ ns. For pulse injection we built a separate 43-MHz mode-locked fibre laser. The injected pulses are amplified and gated by an acousto-optic modulator driven by an arbitrary waveform generator setting the discrete state $\mathbb{S}$.

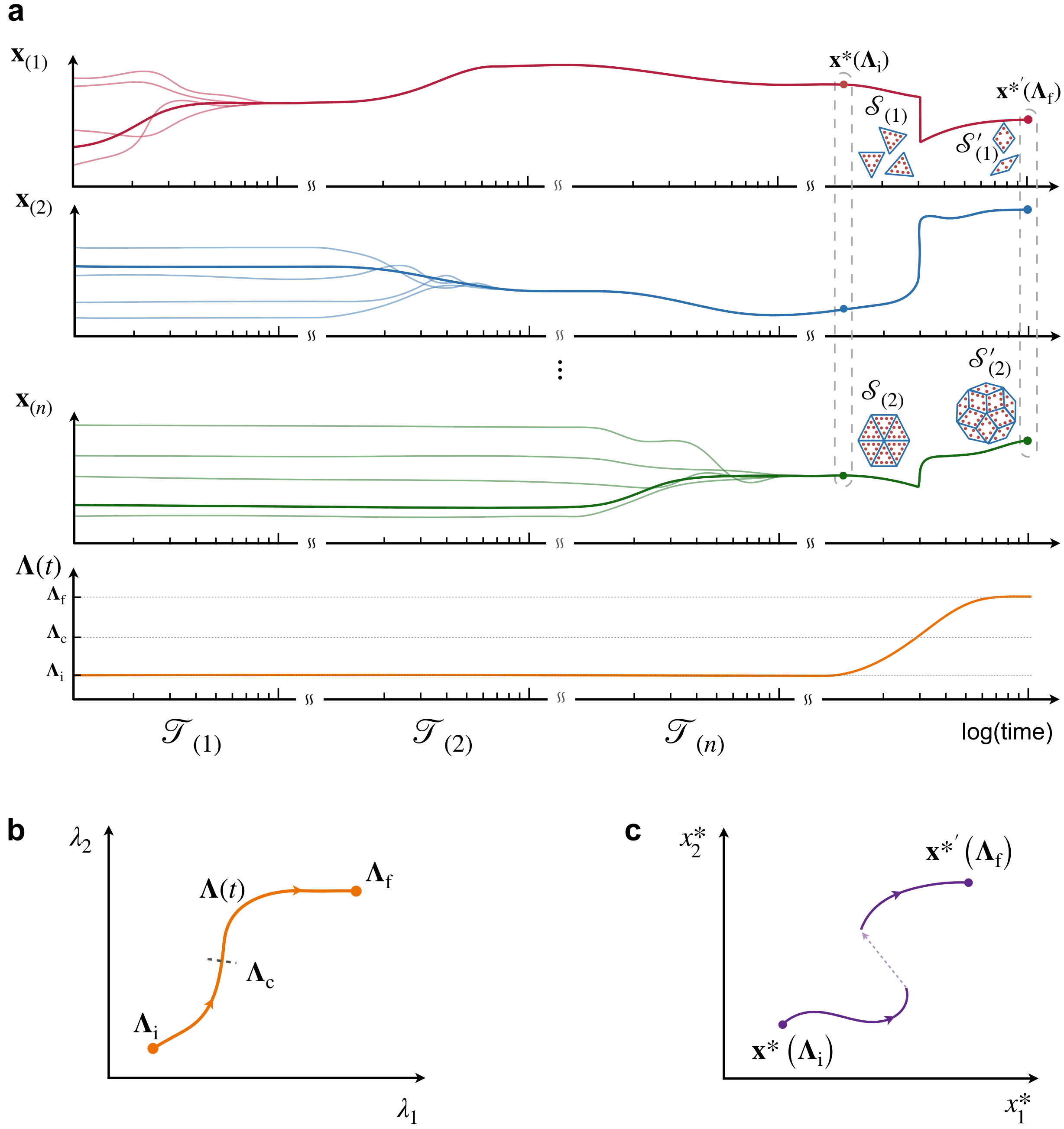

**Figure 1. Programmable emergence through hierarchical self-organisation. a,** Schematic time-domain view: Coordinates are grouped from fast microscopic variables $\mathbf{x}_{(1)}$ to progressively slower $\mathbf{x}_{(2)}, \ldots, \mathbf{x}_{(n)}$. Fast coordinates relax and lock to slower collective coordinates at the next-higher level, iteratively reducing the effective dynamics. Once locked, the system can be steered quasi-statically via external control parameters. Symbols indicate hierarchical organisation: first-level structures $\mathcal{S}_{(1)}$ form building blocks for $\mathcal{S}_{(2)}$, which may transform to $\mathcal{S}'_{(1)}$ and $\mathcal{S}'_{(2)}$ under parameter variation. **b,** Control-parameter space showing a continuous trajectory $\mathbf{\Lambda}(t)$ connecting initial and final values $\mathbf{\Lambda}_\mathrm{i}$ and $\mathbf{\Lambda}_\mathrm{f}$. Crossing a bifurcation at $\mathbf{\Lambda}_\mathrm{c}$ induces qualitative changes of the emergent state. **c,** Reduced state-space representation of the corresponding emergent state $\mathbf{x}^*$.

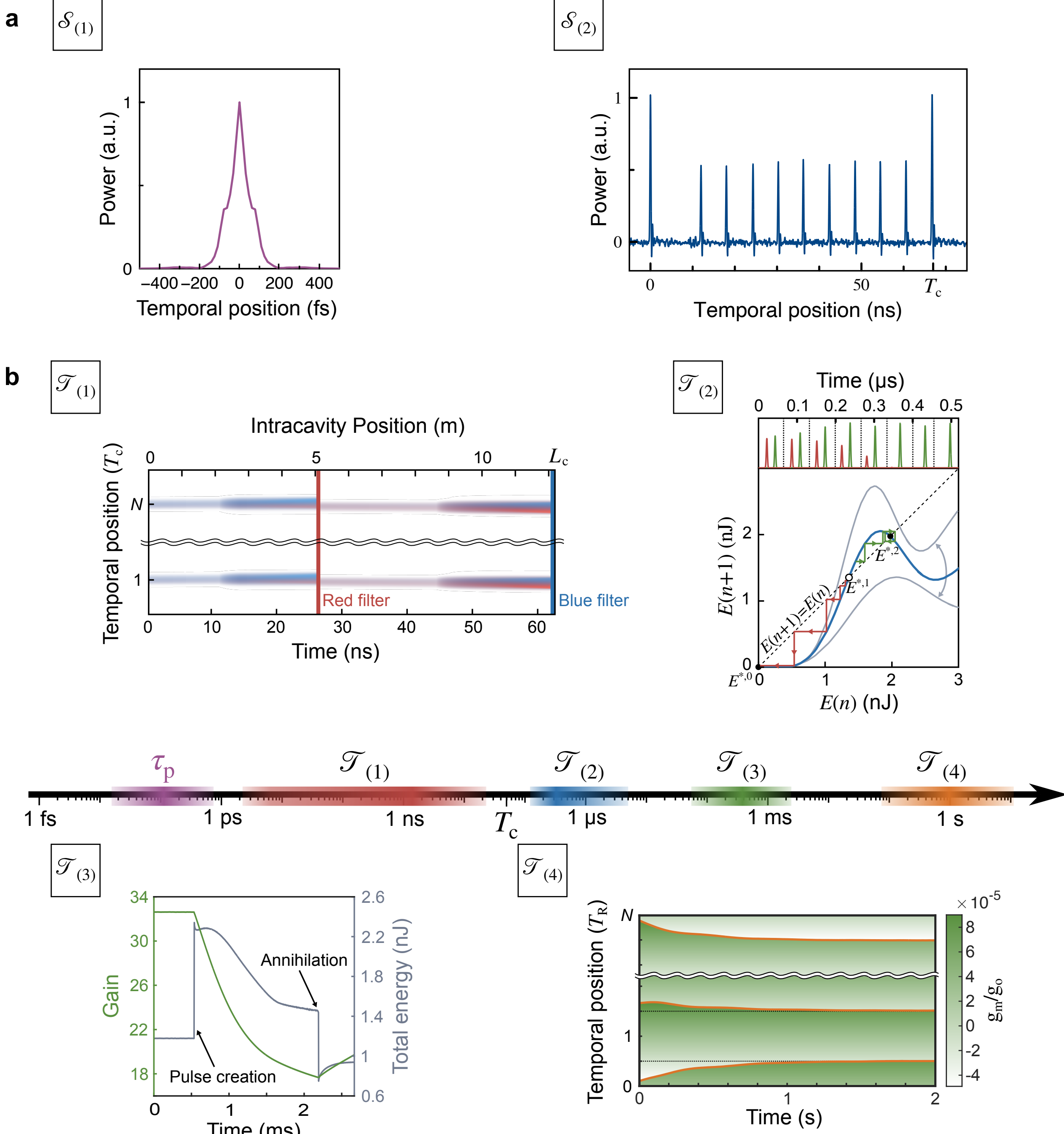

**Figure 2. Hierarchical self-organisation in a mode-locked laser. a,** The first organisational structure, $\mathcal{S}_{(1)}$, is the individual pulse, whose femtosecond-scale temporal profile defines the shortest timescale, $\tau_p$ (left). The second organisational structure, $\mathcal{S}_{(2)}$, is the pattern formed by all pulses (right), shown here in an intentionally anharmonic configuration. The pulse that appears to have higher energy corresponds to two bound pulses separated by less than 50 ps, which are not resolved. **b,** Hierarchical dynamical levels organised by timescale (panels labelled by $\mathcal{T}_{(1)} - \mathcal{T}_{(4)}$). The central axis highlights the temporal depth of the laser dynamics, spanning more than ten orders of magnitude from femtoseconds to seconds. $\mathcal{T}_{(1)}$: nanosecond-scale intracavity pulse-shape evolution (colour denotes instantaneous frequency shift). $\mathcal{T}_{(2)}$: microsecond-scale pulse-energy dynamics, shown as pulse creation (green) and annihilation (red) time series (top) and their corresponding trajectories on an iterative energy map (bottom). $\mathcal{T}_{(3)}$: millisecond-scale gain depletion and recovery following pulse creation. $\mathcal{T}_{(4)}$: seconds-scale evolution of pulse positions toward harmonic order (orange lines indicate simulated pulse trajectories; green shading denotes gain modulation).

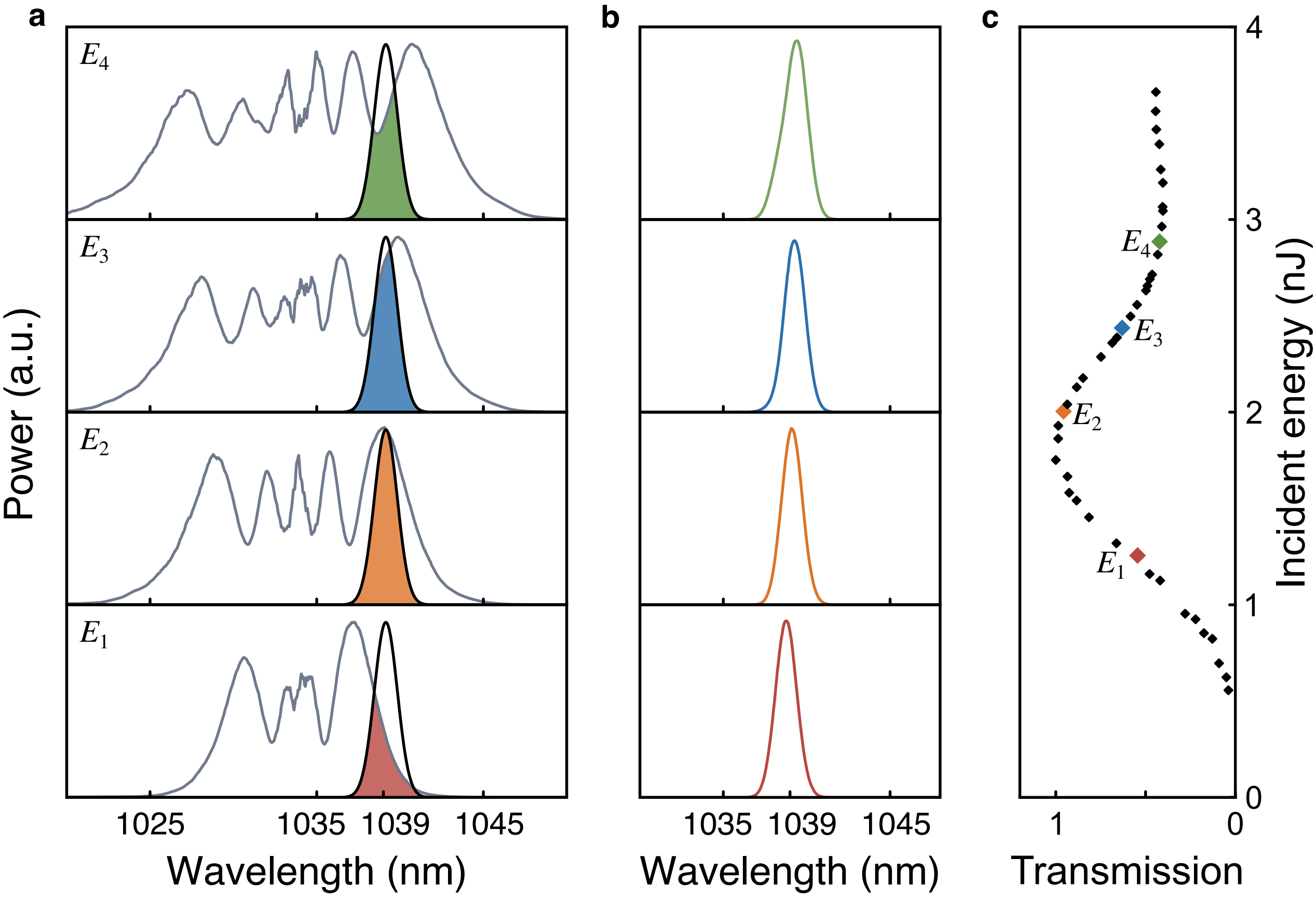

**Figure 3. Pulse shape parametrised by pulse energy. a,** Measured spectra before and after filtering for different energies. As energy increases, the outermost spectral lobe red-shifts, altering its overlap with the filter. **b**, Filtered spectral shapes are nearly identical. **c**, Energy transmission as a function of incident ($E(n)$) closely tracks the spectral intensity at the filter wavelength, enabling direct inference of the energy map.

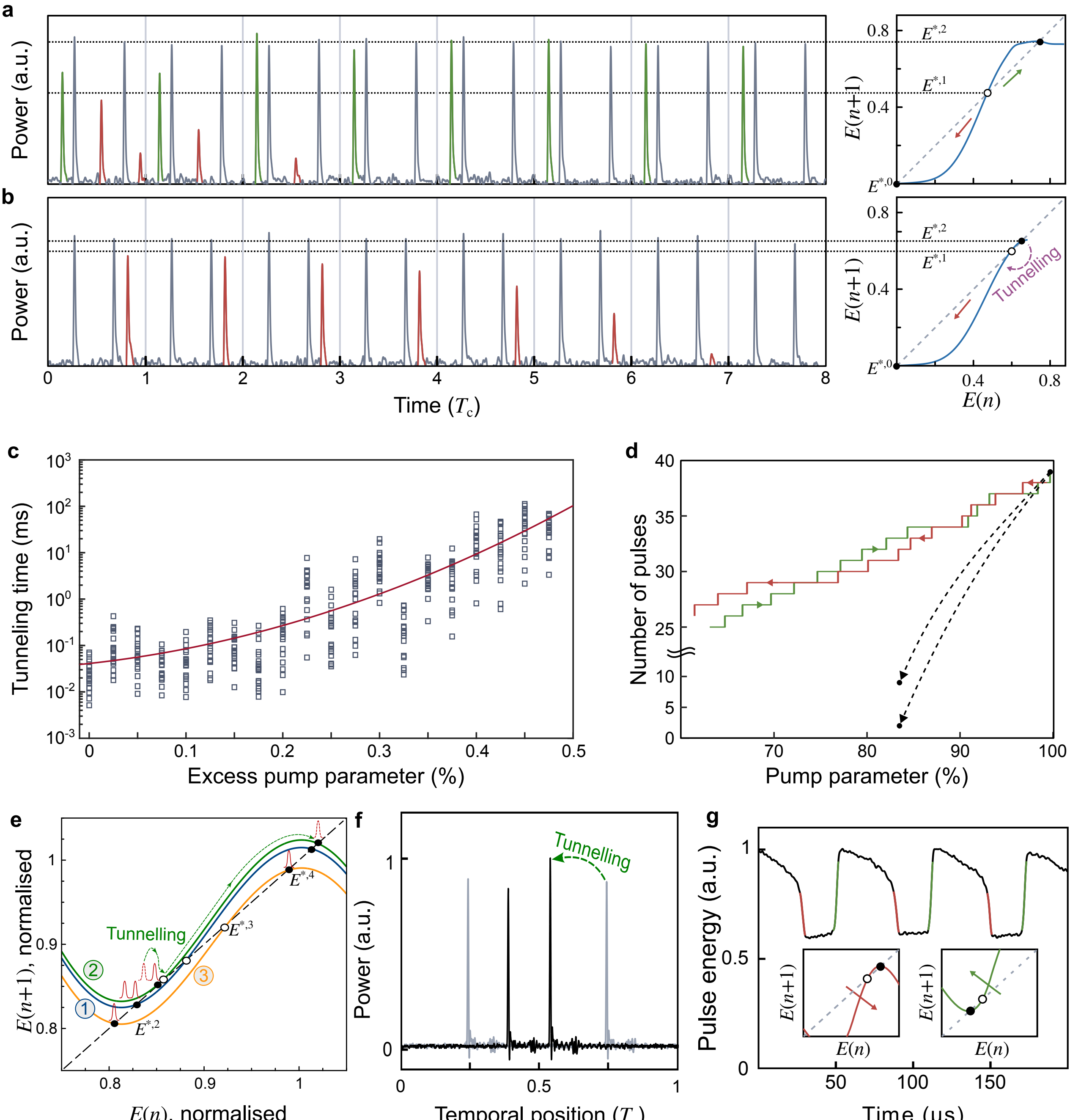

**Figure 4. Programmable control of pulse number and energy states. a,** Pulse injection experiments showing deterministic nucleation and decay for pulse energy above (green) and below (red) the threshold (right, empty circle), respectively. **b,** Measured single-pulse annihilation through tunnelling across the threshold induced by noise near the saddle-node bifurcation. **c,** Simulated tunnelling times versus pump power, showing super-exponential scaling near the bifurcation. **d,** Experimental steering of number of pulses via controlled pump-power trajectories: step-wise creation (green) and annihilation (red), demonstrating path dependence. Black arrows indicate uncontrolled multi-pulse loss due to rapid pump reduction. **e,** Energy-map engineering enabling two coexisting stable pulse-energy fixed points. **f,** Oscilloscope traces before (grey) and after (black) transition to non-identical pulses. **h,** Breathing-pulse-like state with reduced power.

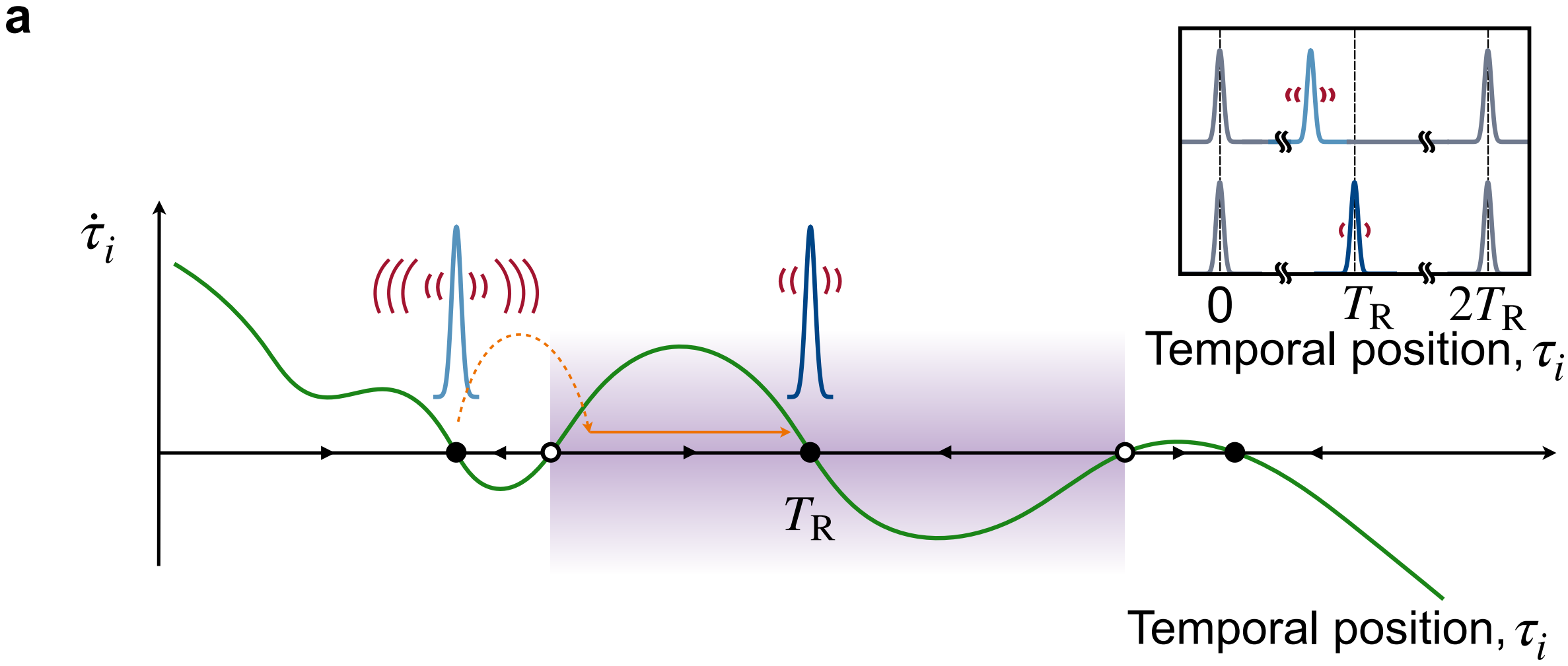

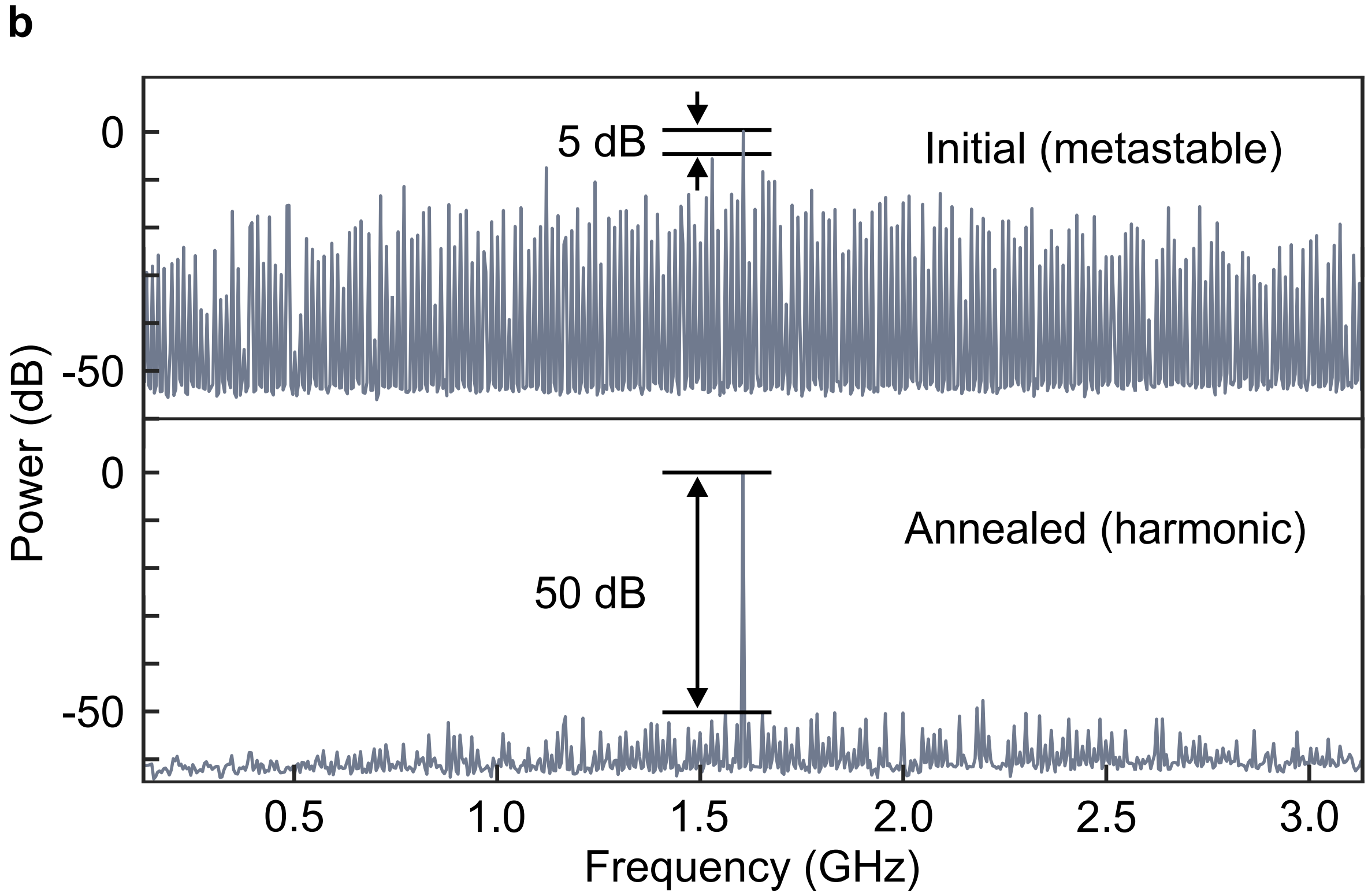

**Figure 5. Annealing toward harmonic pulse patterns. a,** Schematic of the annealing mechanism: externally injected subthreshold pulses stochastically perturb the gain landscape, allowing an anharmonically trapped intracavity pulse (light blue) to escape a shallow local maximum and relax into the global harmonic configuration (dark blue). **b,** Radio-frequency spectra for the initial and harmonised states, demonstrating the removal of metastable configurations and the emergence of high-purity harmonic order with 106 pulses inside the laser cavity.

## Author Contributions

A.Ş. built the experimental setup, performed the experiments, simulations, and developed the laser-specific theory. F.Ö.I. developed the concept of programmable emergence through hierarchical steering via its general theoretical framework, supervised the study, and co-wrote the manuscript with A.Ş.

## Acknowledgements:

This work was supported by the Alexander von Humboldt Foundation through an Alexander von Humboldt Professorship awarded to F. Ö. Ilday and the European Research Council (ERC) under the European Union's Horizon 2022 research and innovation programme (grant agreement no. 101055055, ERC Advanced Grant UNILASE). The authors thank Mesut Laçin for help with harmonic state characterisation and the experimental setup, Amirhossein Maghsoudi for building an amplifier for the injected pulses, and Serim Ilday for inspiring the programmable emergence through hierarchical self-organisation perspective.

## SUPPLEMENTARY INFORMATION

### 1. A minimal illustration of hierarchical steering in a two-variable system

To build intuition for hierarchical steering in its simplest possible form, we begin with a deliberately minimal dynamical system comprising only two coupled variables evolving on widely separated timescales. This system is not intended to model self-organisation or emergence, which generally require many degrees of freedom at each level. Instead, it serves as a pedagogical limit case that isolates the essential mechanism underlying hierarchical steering: the slaving of fast dynamics to slow collective variables, and the ability to steer the system by varying external parameters on the slowest timescale.

The purpose of this section is therefore conceptual rather than physical. It demonstrates how timescale separation alone allows a hierarchy of dynamics to be predictively steered, even in the absence of collective behaviour or pattern formation. The same mechanism is later realised in higher-dimensional systems that do support emergent structures (Supplementary Section 2 and the main text).

We consider the fictitious dynamical system

$$\mathcal{T}_x \dot{x} = y^2 - x, \quad \dot{y} = \sin\left(xy\right) - \lambda\left(t\right)y, \tag{1}$$

where $x(t)$ is the fast variable evolving on the timescale of $\mathcal{T}_x$=0.01 and $y(t)$ is a slow variable, with the timescale of unity, and $\lambda(t)$ is a slowly varying control parameter (Supplementary Figure 1).

Because $x$ evolves much faster than $y$, the dynamics naturally organise into two hierarchical levels. On the timescale relevant for $x$, these trajectories appear nearly horizontal because $x$ evolves much faster than $y$; the slow variable $y$ and the control parameter $\lambda(t)$ are effectively frozen. Treating $y$ as a constant, the fast subsystem has a single stable fixed point $x = y^2$. As a result, for any initial condition, $x(t)$ rapidly relaxes onto it. This fast convergence occurs well before any appreciable evolution of $y$ or $\lambda(t)$, and constitutes the locking of the fast variable to the slow one. Solving for the steady state of $x$ in terms of $y$, either analytically or graphically, then substituting this steady state relation into the evolution equation for $y$, reduces the system to the one-dimensional slow subsystem $\dot{y} = \sin\left(y^3\right) - \lambda(t)y$, which can again be analysed analytically or graphically. Crucially, this reduced equation depends only on the slow variable y and the externally controlled parameter $\lambda(t)$. The fast dynamics no longer appear explicitly, yet they remain implicitly present through the locking relation.

As long as $\lambda(t)$ is varied quasi-statically, the system continuously tracks the instantaneous attractor of the reduced equation. Steering applied at the slowest level propagates deterministically through the hierarchy without disrupting the fast dynamics. Time-domain simulations (Supplementary Figure 1) starting from a randomly picked range of initial conditions make this hierarchy explicit: $x(t)$ converges first, followed by the slower convergence of $y(t)$, with both ultimately steered by the externally prescribed trajectory $\lambda(t)$.

Although this system is intentionally two-dimensional and non-emergent, it illustrates the core logical structure of hierarchical steering used throughout the paper. A more representative toy model is discussed in Supplementary Section 2 and the main text extends it to demonstrate programmable emergence in the full-fledged laser system that supports complex pattern formation and collective behaviour at multiple hierarchical levels.

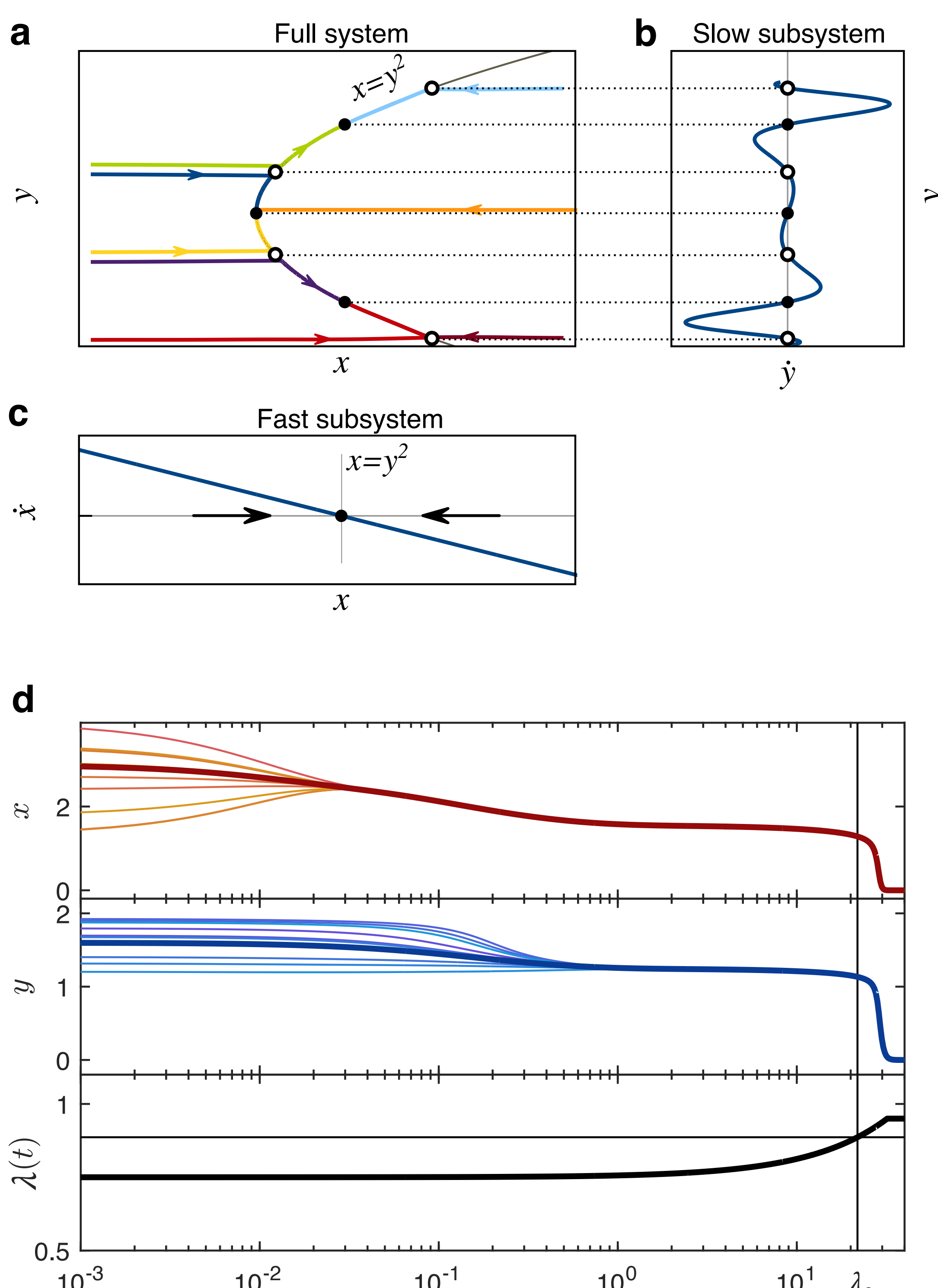

**Supplementary Figure 1**: **Minimal illustration of hierarchical steering in a two-variable system. a,** Phase-plane trajectories $(x(t),y(t))$ showing rapid convergence of the fast variable $x$ onto $x=y^2$ (black curve), followed by slow evolution along it. **b,** Fast subsystem showing evolution of $x$ with the order parameter $y$ held constant. **c,** Slow subsystem showing evolution of $y$ with the slaved variable, $x$, in its steady state ($x = y^2$), correctly predicting stable (filled circles) and unstable (open circles) fixed points. **d,** Time evolution of $x(t)$ and $y(t)$ plotted on a logarithmic time axis for a range of initial conditions, illustrating the clear separation of timescales, and steering through the external control parameter $\lambda(t)$. $\lambda_c$ indicates to a bifurcation point.

## 2. Hierarchical pattern formation in a generalised Swift–Hohenberg toy model

We consider a minimal theoretical model that illustrates hierarchical self-organisation and predictive steering in a driven–dissipative system with well-separated dynamical timescales. The purpose of this model is not to represent a specific physical platform, but to provide an intuitive and generic realisation of the hierarchical steering mechanism developed in the main text. Accordingly, the model is deliberately minimal and intended for conceptual clarity rather than quantitative realism. Its structure is inspired by multiscale pattern formation in driven–dissipative systems, where fast pattern-forming dynamics are shaped by slower, cumulative fields.

As a broadly applicable starting point, we build on the one-dimensional Swift–Hohenberg equation, a canonical model for pattern formation across fluid dynamics, reaction–diffusion systems, nonlinear optics, and soft-matter physics. We introduce a controlled generalisation that incorporates multiple dynamical levels with explicit timescale separation, allowing the emergence of nested structures and their steering via a single slow control parameter.

We begin with a system comprising three dynamical levels; a fast pattern-forming field $u$, an intermediate coarse-graining field $\theta$, and a slow field $v$.

Level 1 has the fastest dynamics and supports short-wavelength pattern formation, described by

$$\partial_t u(x,t) = \left(r_u(t) + v(x,t)\right) u - \left(1 + \partial_x^2\right)^2 u - u^3 + \xi_u(x,t), \tag{2}$$

where $r_u(t)$ is the first externally controlled quasi-statically varied parameter, and $\xi_u(x,t)$ is a white noise term with zero mean. The effective growth rate $r_{\text{eff}}(x,t) \equiv r_u(t) + v(x,t)$, implies that pattern formation in $u(x,t)$ is locally enabled or suppressed by the slow field $v(x,t)$.

Level 2 comprises the intermediate field $\theta(x,t)$, which effectively couples the fast field $u(x,t)$ to the slow field $v(x,t)$:

$$\mathcal{T}_\theta\, \partial_t \theta(x,t) = \ell_\theta^2 \partial_x^2 \theta - \theta + u(x,t)^2, \tag{3}$$

with $\ell_\theta \gg 1$ and $\mathcal{T}_\theta \gg 1$ are large characteristic spatial and time scales compared to the fast field $u$. This equation has the form of a heat equation with the intensity or square of the fast field $u^2$ acting as a heat source, constituting a diffusive, weakly nonlinear field driven by the fast pattern. Because $\ell_\theta \gg 1$, spatial variations of $u$ on its intrinsic scale are strongly attenuated, $\theta$, therefore, follows a spatially smoothed version of $u$.

Level 3 is the slowest, and governs long-wavelength pattern formation,

$$\mathcal{T}_v \partial_t v(x,t) = \left(r_v(t) + \theta(x,t)\right) v - \left(1 + \ell_v^2 \partial_x^2\right)^2 v - v^3 + \xi_v(x,t), \tag{4}$$

where $\ell_v \simeq \ell_\theta \gg 1$ and $\mathcal{T}_v \gg 1$ sets a characteristic wavelength much larger than that of $u$, $r_v(t)$ is an externally controlled parameter that is varied quasi-statically in time, and $\xi_v(x,t)$ is a white noise term with zero mean.

At the fastest Level 1, $u$ will rapidly converge to either of two different attractors, depending on the quasi-static value of $\theta(x)$, which acts as an order parameter at this timescale. We consider a standard trial solution of form $A(x)\cos(x)$, where $A(x)$ is a slowly varying amplitude so that its high-order derivatives can be neglected. Further assuming $A(x)$ to be small in magnitude enough that generation of higher Fourier components are slower than the low-pass filtering, we conclude that the two attractor states are the uniform solution, $u^* = 0$, where $r_u + v(x) < 0$, and a periodic pattern with non-zero $A(x)$ elsewhere. To identify the latter, we substitute the trial solution into Equation (2), and look for a stationary solution,

$$\dot{A}\cos(x) = \left(r_u(t) + v(x,t)\right) A\cos(x) - \frac{3}{4}A^3\cos(x) + O\left(\cos(3x)\right) = 0. \tag{5}$$

Ignoring higher harmonic terms, $u^* \approx \sqrt{4\left(r_u + v(x)\right)/3}\,\cos(x)$.

At Level 2, $\theta$ relaxes much faster than $v$. On the slowest timescale $\mathcal{T}_v$, Equation (3) can be approximated, to first order, by

$$(1 - \ell_\theta^2 \partial_x^2)\theta \simeq u^2, \tag{6}$$

using the adiabatic elimination. This reveals that $\theta$ acts as a low-pass spatial filter on $u^2$, spatially averaging out its short-wavelength structure and retaining only its coarse-grained envelope. Therefore, $\theta^* = \langle u^{*2} \rangle$, where $\langle \cdot \rangle$ denotes spatial average over the scale $\ell_\theta$. Then, $\theta^* \simeq \left(r_u + v\right)/3$, wherever $r_u + v(x) > 0$ and $\theta^* \simeq 0$, elsewhere.

Finally, the slowest dynamics of Level 3 allow replacing $\theta(x,t)$ with $\theta^*(x)$, yielding the following generalised Swift-Hohenberg equation,

$$\mathcal{T}_v \partial_t v(x,t) = \left(r_v(t) + H(r_u(t) + v)\frac{2r_u(t)}{3}\right) v - \left(1 + \partial_x^2\right)^2 v + H(r_u(t) + v)\frac{2}{3}v^2 - v^3 + \eta_v(x,t), \tag{7}$$

where $H(r_u + v)$ is the Heaviside function, reflecting whether Level 1 is in a uniform or patterned state. The former reduces to a decoupled (isolated) Swift-Hohenberg equation, and the latter to a non-symmetrical Swift–Hohenberg equation, both versions supporting two distinct states, either uniform or patterned but with different and path-dependent conditions on the external control parameters $r_u(t)$ and $r_v(t)$.

Altogether this system supports four distinct attractor states, uniform $u$ and $v$, represented by $\mathbf{x}^{*,1}$; patterned $u$ with uniform $v$, represented by $\mathbf{x}^{*,2}$; uniform $u$ with patterned $v$, represented by $\mathbf{x}^{*,3}$; hierarchical state with patterned $u$ and $v$, represented by $\mathbf{x}^{*,4}$. Their stability regions are separated by a set of bifurcations along with regions of multiple stability, several of which are determined analytically, and others numerically. These introduce the richness of path-dependence, and hysteresis. In other words, the final state is not uniquely determined by the control parameters, but the order in which they are arrived. These features are illustrated through numerical simulations of this system in Supplementary Figure 2).

This toy model illustrates, with minimal complexity, how slow control parameters can steer a hierarchy of self-organised structures by first shaping the slow attractor and thereby locking the fast dynamics. Although deliberately simplified, the model is reminiscent of the multiscale laser-induced periodic patterns discussed in Supplementary Section 14. Formulated as a generalisation of the Swift-Hohenberg equation, a widely used paradigm for pattern formation across physics, it highlights that the framework is not tied to laser dynamics but applies generically to driven–dissipative systems with well-separated timescales.

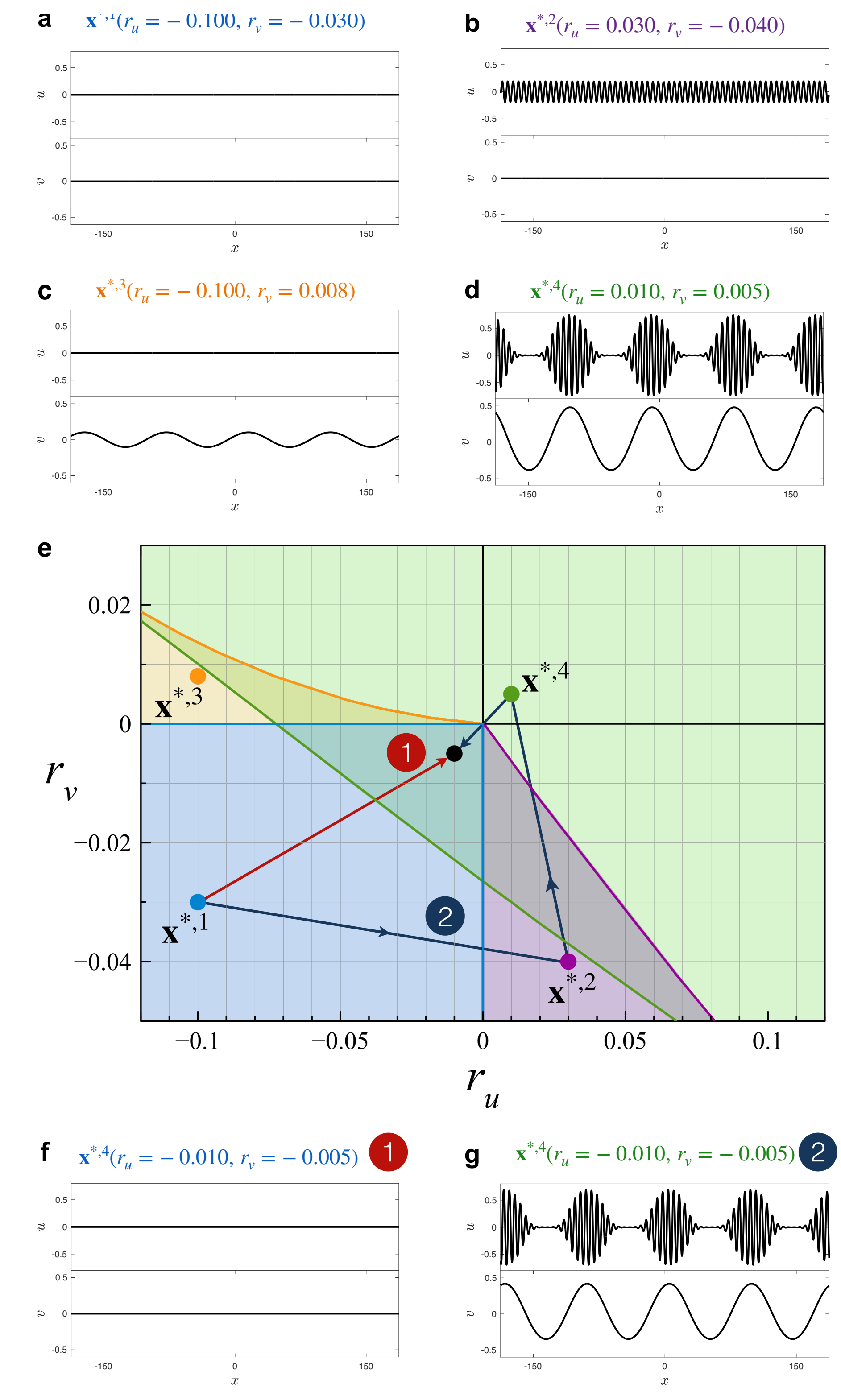

**Supplementary Figure 2**: **Predictive steering in a hierarchical Swift-Hohenberg Toy Model. a-d** Four distinct attractor states of the 3-level system: **a,** uniform $u$ and $v$; **b,** patterned $u$ with uniform $v$; **c,** uniform $u$ with patterned $v$; **d,** hierarchical state with patterned $u$ and $v$. **e,** Two control-parameter trajectories with identical initial and final values but different paths in parameter space. **f, g,** Final states reached following Trajectory 1 and Trajectory 2, respectively. The former remains trapped in a uniform state, whereas the latter yields hierarchical patterns.

## 3. Schematic of the laser system

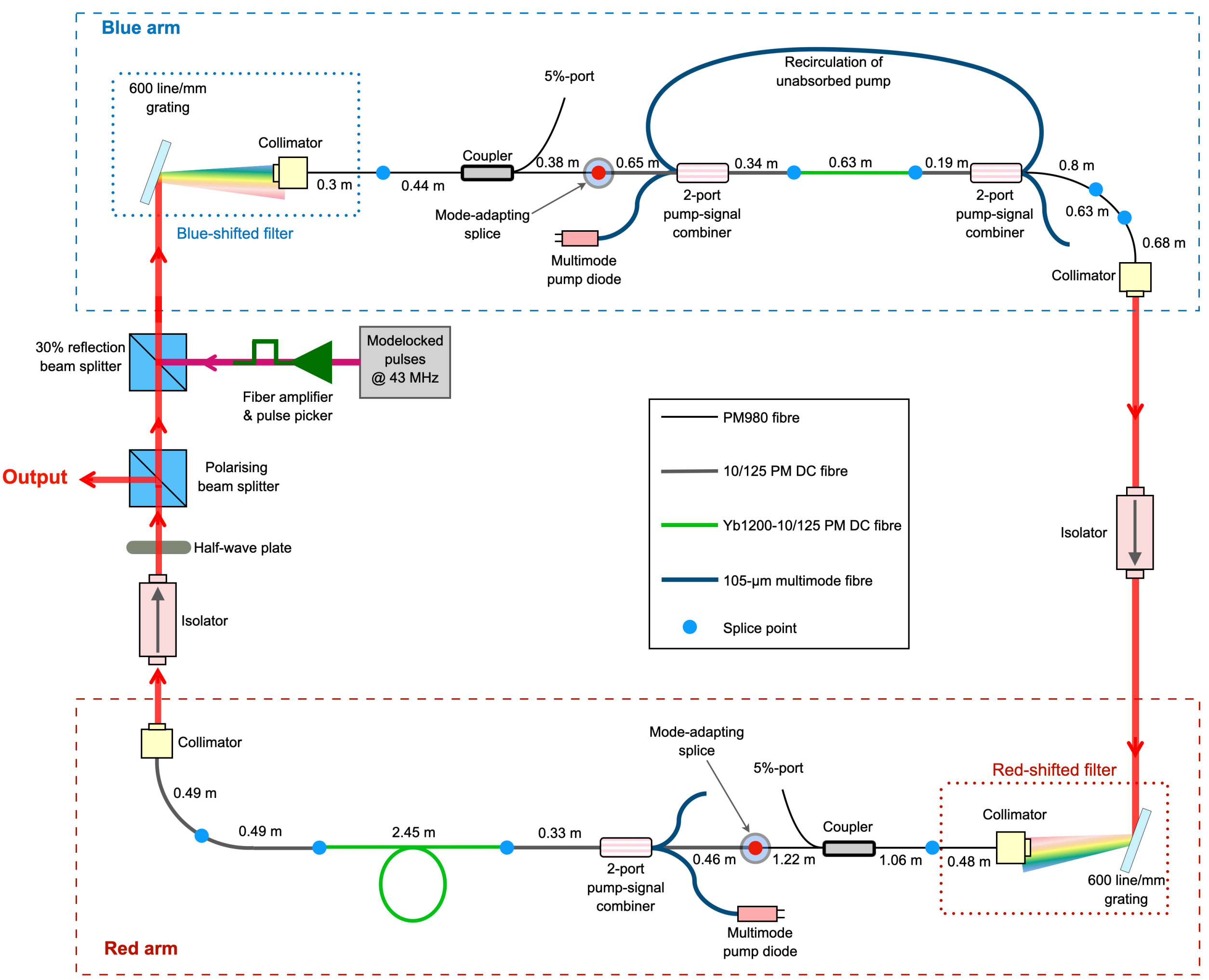

**Supplementary Figure 3**: **Schematic of the laser system.** Legend indicates different fibre types used. Further system details are provided in Methods.

## 4. Numerical demonstration of parameterisation of pulse shapes by narrow-band filtering

The effect of passing two very different pulses through a sufficiently narrow filter is illustrated in Supplementary Figure 4, showing both temporal shapes and chirps. Because the filter is narrow, the filtered pulses are nearly transform-limited, making them almost identical in the temporal domain. However, they are delayed differently. This temporal offset and its role in pulse repositioning are discussed in Supplementary Section 6.

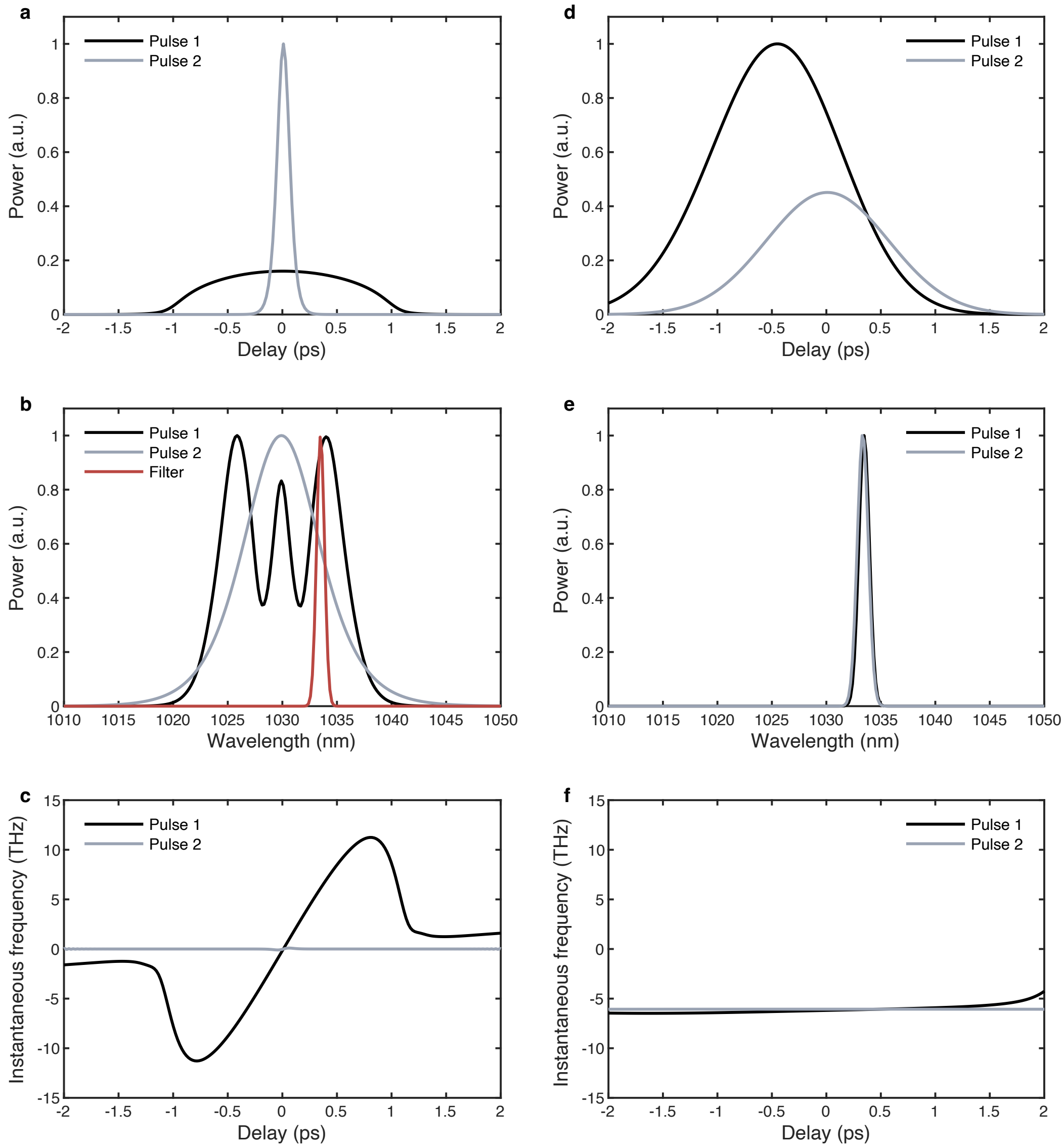

**Supplementary Figure 4**: **Effect of narrow-band filtering. a-c**, Temporal profile, spectrum and instantaneous frequency chirps of two very different pulses incident on a narrow filter. **d-f**, Corresponding temporal, spectral, and instantaneous frequency profiles after filtering. The spectra are nearly identical, instantaneous frequencies are approximately flat, and temporal profiles have the same shape. Differences are limited to pulse energies and temporal positions, which serve as order parameters. The temporal offset depends on the filter's central wavelength and is crucial for subsequent pulse repositioning.

## 5. Energy map engineering

As shown in Figure 3 of the main manuscript, the transmitted pulse energy through a filter in our laser follows the spectral shape near the filter. If the filter lies near a steep slope that descends outwards, then a small increase in the incident energy will result in a large increase in the transmitted energy. Thus, the slope of the energy map can be manipulated with guidance from the spectra. For the period-doubling experiment, the blue-most spectral lobe in the red arm was steep enough to produce an energy map slope below −1. This slope was achieved by positioning the blue filter at the inner side of the blue-most spectral lobe in the red arm, while the red filter was placed at the red edge of the blue spectrum, as illustrated in Supplementary Figure 5c,d.

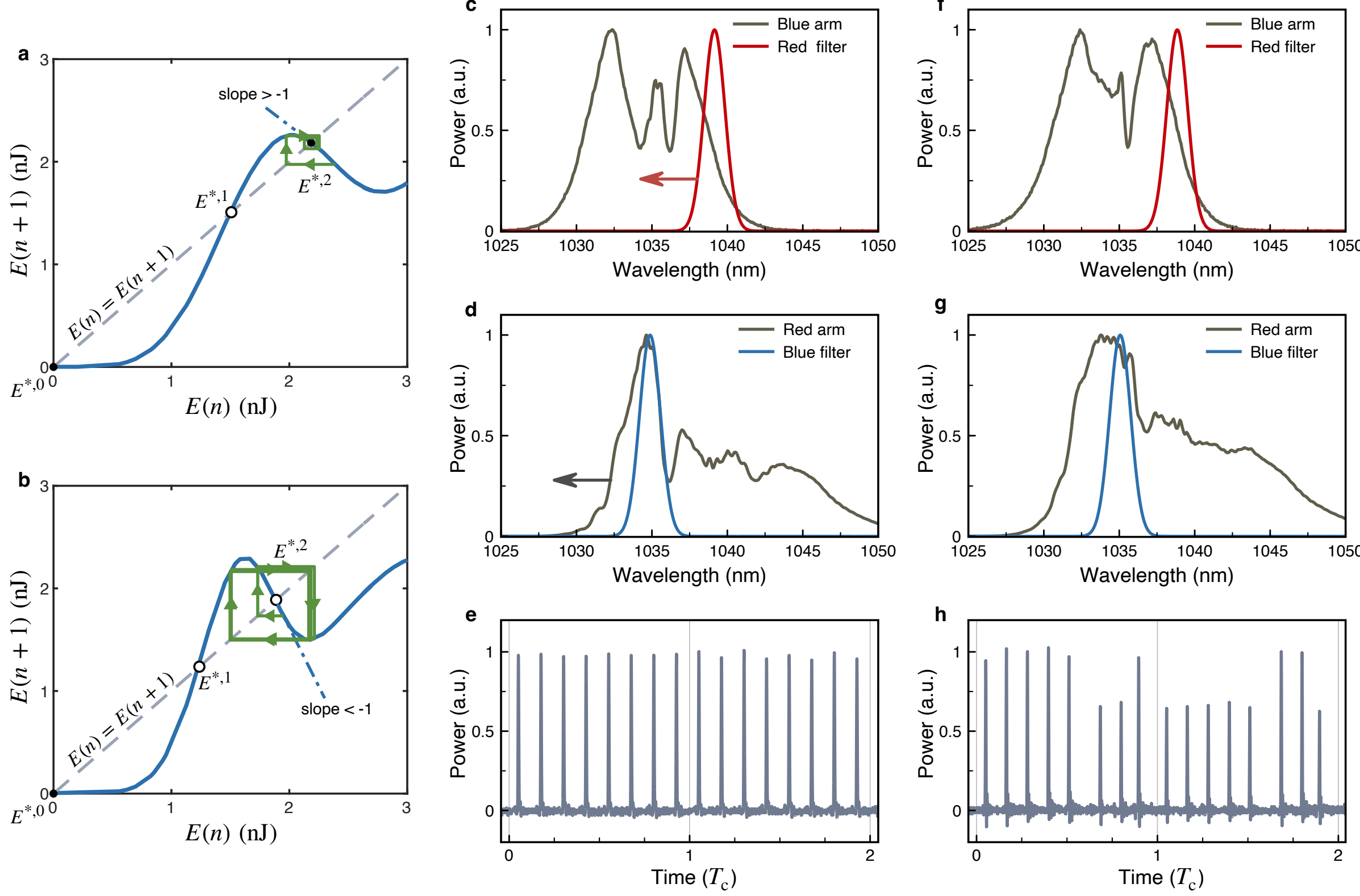

**Supplementary Figure 5**: **Period-doubled states. a-b**, Energy map before and after crossing the period-doubling bifurcation. The trajectory converges to a stable alternating pattern once the slope decreases below −1. **c-d**, Spectra before period-doubling. The energy map is modified by shifting the red filter, and consequently, the red spectral lobe, as indicated by the arrows, producing steep slopes of opposite signs. **e**, Pulse train before shifting the red filter, showing eight stable pulses. **f-g**, Spectra after shifting the red filter (averaged over the low- and high-energy pulses shown in **h**). **h**, Pulse train after the filter shift, showing period-doubled pulses, with high-energy pulses in one roundtrip becoming low-energy in the following roundtrip, and vice versa.

Following this adjustment, the pulses exhibited alternating energies on consecutive roundtrips. The sequence of the high- and low-energy pulses is random because the period-doubling of different pulses grow from independent random deviations from the now unstable fixed energy, $E^{*,2}$ (Supp. Fig. 5h).

By correlating the spectral slopes with the energy map slope, one identifies minima and maxima in the energy map from spectral minima and maxima. This is the basis of the non-identical pulses and breathing pulse experiment shown in Fig. 4e-g, the optical spectra for which are shown in Supp. Fig. 6. We generated these spectra by increasing the pump power in the red arm relative to the blue arm and adjusting the filter offset so the red filter lies at the edge of the blue arm spectrum, while the red arm developed two spectral lobes surrounding the blue filter. From the correlation between the spectral and energy map slopes, one concludes that such a spectrum modifies the energy map, introducing a new pair of fixed points: a second stable one at $E^{*,4}$, separated from $E^{*,2}$ by an unstable point at $E^{*,3}$. $E^{*,2}$ is the energy at which the first local minimum of the red spectrum lies near the blue filter, and $E^{*,4}$ corresponds to the following local maximum. When the gain increases, it broadens the spectrum for the same input energy, delivering a steep slope to the blue filter wavelength, which implies an increased energy map slope. The saddle-node bifurcations, like the one shown in the insets Fig. 4g, are reached when the energy map slope reaches unity.

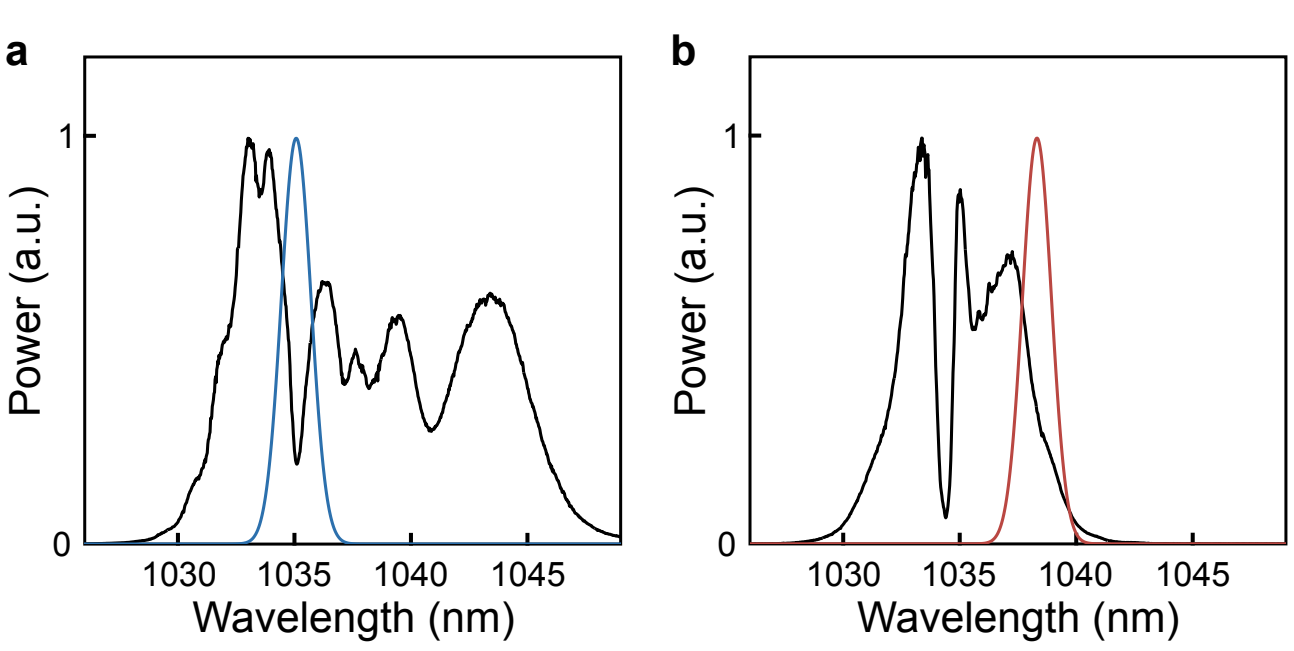

**Supplementary Figure 6. Laser spectra for the deliberate creation of non-identical pulse states.** Measured intracavity spectra and corresponding spectral filtering used to during the experiments described in Fig. 4f and 4f of the main manuscript, **a,** spectrum measured after the blue arm (black trace) of the cavity and before the red filter (red trace), **b,** spectrum measured after the red arm (black trace) and before the blue filter (blue trace).

## 6. Dependence of pulse-repositioning coefficients on control parameters

In Methods, we summarised the effects of gain modulation and acoustic oscillations to the pulse speed, yielding an explicit equation for pulse repositioning. Here, we show representative examples, then explain qualitatively how these coefficients depend on the spectral settings. Supplementary Figure 7 and Supplementary Table 1 present simulation results linking five canonical spectra to the corresponding values of $\Gamma$, $\alpha_6$, $\alpha_{10}$, $\beta_6$, $\beta_{10}$, and $\langle \eta_\tau^2 \rangle$. We label the canonical spectral settings shown in Supplementary Figure 7a-e as *a*–*e* in what follows.

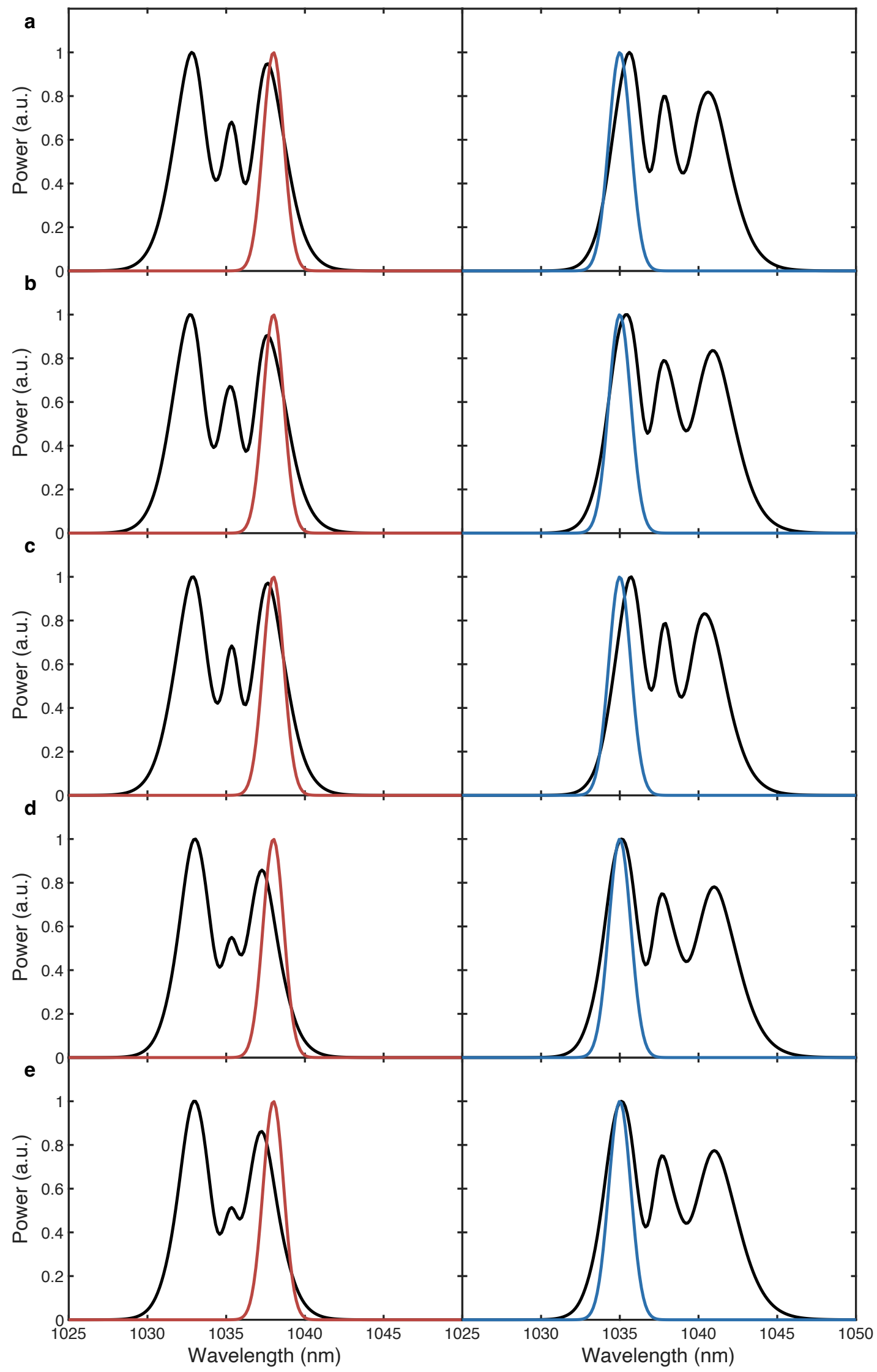

**Supplementary Figure 7**: **Canonical spectra used in pulse-speed simulations.** Simulation results corresponding to each spectrum are summarised in Supplementary Table 1. Only the pump powers were changed, except in **e**, which has a 10 dB higher output coupling loss and a compensating increase in the blue-arm pump power relative to **d**, yielding the same $E_r^*$ and $E_b^*$ as in **d**.

| | *a* | *b* | *c* | *d* | *e* |
|---|---|---|---|---|---|
| $(1-\mathcal{F}')^{-1}$ | 4.2 | 1.4 | 17.7 | 2.6 | 3.3 |
| $D_{\mathrm{E}}$ | 4 fs | 65 fs | -41 fs | 269 fs | 270 fs |
| $\Gamma T_{\mathrm{c}}$ | 0.003 fs | 0.017 fs | -0.130 fs | 0.122 fs | 0.164 fs |
| $\left\langle \eta_\tau^2 \right\rangle T_{\mathrm{c}}$ | 0.039 $\mathrm{fs}^2$ | 0.034 $\mathrm{fs}^2$ | 0.27 $\mathrm{fs}^2$ | 0.20 $\mathrm{fs}^2$ | 1.20 $\mathrm{fs}^2$ |
| $\alpha_6 T_{\mathrm{c}}$ | -6.2 ns nJ | -6.4 ns nJ | -6 ns nJ | -5.4 ns nJ | -5.3 ns nJ |
| $\alpha_{10} T_{\mathrm{c}}$ | -21.5 ns nJ | -23.4 ns nJ | -20.2 ns nJ | -23.6 ns nJ | -23.3 ns nJ |
| $\beta_6 T_{\mathrm{c}}$ | -0.88 $\mathrm{ns}^2$ nJ | -1 $\mathrm{ns}^2$ nJ | -1.34 $\mathrm{ns}^2$ nJ | 1.46 $\mathrm{ns}^2$ nJ | 2.25 $\mathrm{ns}^2$ nJ |
| $\beta_{10} T_{\mathrm{c}}$ | -3.44 $\mathrm{ns}^2$ nJ | -4.67 $\mathrm{ns}^2$ nJ | 2.85 $\mathrm{ns}^2$ nJ | -2.36 $\mathrm{ns}^2$ nJ | -1.64 $\mathrm{ns}^2$ nJ |

**Supplementary Table 1**: **Pulse-repositioning coefficients from simulations.** Quantities characterising the pulse-speed dynamics are shown for each of the canonical spectral settings in Supplementary Figure 7a-e.

We begin with pulse-repositioning coefficient, $\Gamma$, which depends strongly on the relative placement of the spectral filters. Placing the blue filter near the peak of the red spectrum and the red filter at the tail of the blue spectrum produces a large positive $\Gamma$ (spectral settings *d* and *e*), promoting harmonic modelocking. Conversely, the opposite arrangement yields a large negative $\Gamma$ (spectral setting *c*). Positioning both filters near spectral peaks results in a small $\Gamma$ (spectral setting *b*), while intermediate configurations give moderate values (spectral setting *a*). These numerical trends reveal two competing forces by which the gain modulates the pulse speed, both enhanced by the spectral slopes. We next derive these effects analytically by relating the pulse energy at the output of both amplifiers to the gain modulation, and subsequently connecting the pulse speed to the pulse energy. This framework also provides insight into the contributions of noise and acoustically induced wavelength shifts.

The parametrisation of the gain by the pulse positions translates into a parametrisation of the energy map, which reads

$$E_i(n+1) = \mathscr{F}\left(E_i(n), g_{\mathrm{o,b}} + g_{\mathrm{m,b}}\left(\tau_i\right), g_{\mathrm{o,r}} + g_{\mathrm{m,r}}\left(\tau_i\right), \Delta\lambda\right), \tag{8}$$

where $i$ enumerates the sequentially ordered pulses in the laser, as in the main text. Linearising with respect to the gain modulations, the evolution of the deviation of the pulse energy at the output of the blue arm, $\delta E_{\mathrm{b}}$, becomes

$$\delta E_{\mathrm{b}}(n+1) = \mathscr{F}'\delta E_{\mathrm{b}}(n) + \frac{g_{\mathrm{m,b}}}{g_{\mathrm{o,b}}}E_{\mathrm{b}}^{*} + g_{\mathrm{o,b}}\frac{\partial\mathscr{F}_{\mathrm{b}}}{\partial E_{\mathrm{r}}}\frac{g_{\mathrm{m,r}}}{g_{\mathrm{o,r}}}E_{\mathrm{r}}^{*}, \tag{9}$$

where $\mathscr{F}'$ denotes $d\mathscr{F}/dE_{\mathrm{b}}$. Here, the energy map is taken for the output of the blue arm, as in the main text, so that $E_{\mathrm{b}}^{*} = E^{*,2}$. Because of the last two terms, the deviation stabilises at a non-zero value, $\delta E_{\mathrm{b}}^{*}$, representing the shift of the stable fixed point in the energy map due to the $g_{\mathrm{m}}$'s,

$$\frac{\delta E_{\mathrm{b}}^{*}}{E_{\mathrm{b}}^{*}} = \frac{E_{\mathrm{r}}^{*}/E_{\mathrm{sat}}}{1-\mathscr{F}'}\left(\frac{E_{\mathrm{b}}^{*}}{E_{\mathrm{r}}^{*}} + g_{\mathrm{o,b}}\frac{\partial\mathscr{F}_{\mathrm{b}}}{\partial E_{\mathrm{r}}}\frac{E_{\mathrm{r}}^{*}}{E_{\mathrm{b}}^{*}}\right)\left(\tau_i/T_R - i\right), \tag{10}$$

where we have substituted the parametrisation of the gain by the pulse position. Note that the division by $1-\mathscr{F}'$ allows the energy perturbation to accumulate to more than an order of magnitude larger than the amount delivered in a single roundtrip. This is an important diagnostic, and we quantify it in Supplementary Table 1 (measured by simulating a few roundtrips with a small initial energy perturbation and noting its decay rate).

Applying the same linearisation to the red arm gives

$$\delta E_{\mathrm{r}}(n+1) = g_{\mathrm{o,r}}\frac{\partial\mathscr{F}_{\mathrm{r}}}{\partial E_{\mathrm{b}}}\delta E_{\mathrm{b}}(n) + \frac{g_{\mathrm{m,r}}}{g_{\mathrm{o,r}}}E_{\mathrm{r}}^{*}, \tag{11}$$

which leads to similar form as for the blue arm,

$$\delta E_{\mathrm{r}}(n+1) = \mathscr{F}'\delta E_{\mathrm{r}}(n) + \frac{g_{\mathrm{m,r}}}{g_{\mathrm{o,r}}}E_{\mathrm{r}}^{*} + g_{\mathrm{o,b}}\frac{\partial\mathscr{F}_{\mathrm{r}}}{\partial E_{\mathrm{b}}}\frac{g_{\mathrm{m,b}}}{g_{\mathrm{o,b}}}E_{\mathrm{b}}^{*}, \tag{12}$$

and consequently to

$$\frac{\delta E_{\mathrm{r}}^{*}}{E_{\mathrm{r}}^{*}} = \frac{E_{\mathrm{b}}^{*}/E_{\mathrm{sat}}}{1-\mathscr{F}'}\left(\frac{E_{\mathrm{r}}^{*}}{E_{\mathrm{b}}^{*}} + g_{\mathrm{o,r}}\frac{\partial \mathscr{F}_{\mathrm{r}}}{\partial E_{\mathrm{b}}}\frac{E_{\mathrm{b}}^{*}}{E_{r}^{*}}\right)\left(\tau_i/T_R - i\right). \tag{13}$$

We now clarify how these accumulated energy deviations affect the pulse speeds. The effect arises locally at the offset filters. As seen in Supplementary Figure 4d, the filtered pulse is temporally displaced because the incident pulse is chirped, with its spectrum spread across its temporal width. When the passband of a filter coincides with, say, the red edge of the incident spectrum, the transmitted pulse is positioned close to the leading edge of the incident pulse. We model the resulting position offsets at the two filters as

$$\tau_{\mathrm{r}} = -\Delta\lambda\frac{\tau_{\mathrm{p,b}}}{\lambda_{\mathrm{p,b}}}, \qquad \tau_{\mathrm{b}} = \Delta\lambda\frac{\tau_{\mathrm{p,r}}}{\lambda_{\mathrm{p,r}}}, \tag{14}$$

where $\tau_{\mathrm{r}}$ and $\tau_{\mathrm{b}}$ denote the temporal shifts at the red and blue filters, respectively; $\tau_{\mathrm{p,b}}$ and $\lambda_{\mathrm{p,b}}$ are the temporal and spectral widths at the output of the blue arm (incident on the red filter), and $\tau_{\mathrm{p,r}}$ and $\lambda_{\mathrm{p,r}}$ are their counterparts in the red arm.

An energy deviation at either arm increases the spectral and temporal widths by different amounts, thereby modifying the position shifts at the filters. This often produces a substantial net delay that recurs every roundtrip. Linearising with respect to the pulse energies in the two arms yields the delay rate,

$$\dot{\tau}_i = -\frac{\Delta\lambda}{T_{\mathrm{c}}}\left(\frac{\partial}{\partial E_{\mathrm{b}}}\left(\frac{\tau_{\mathrm{p,b}}}{\lambda_{\mathrm{p,b}}}\right)\cdot\delta E_{\mathrm{b}} - \frac{\partial}{\partial E_{\mathrm{r}}}\left(\frac{\tau_{\mathrm{p,r}}}{\lambda_{\mathrm{p,r}}}\right)\cdot\delta E_{\mathrm{r}}\right). \tag{15}$$

There is a competition between the contributions from the two filters. By adjusting the spectral settings, one can tune this balance, enhancing the energy deviations or their prefactors asymmetrically between the arms. To quantify this effect, we used the same simulation employed for $\Gamma$ to evaluate the ratio of the relative pulse speed to the relative energy deviation in one arm. We chose the red arm, as its contribution generally dominates Equation (15) but the results for the blue arm would be similar. We denote this ratio by $D_{\mathrm{E}}$ and list its values for each canonical spectral setting in Supplementary Table 1. $D_{\mathrm{E}}$ may be regarded as the energy analogue of chromatic dispersion and is used in Supplementary Section 7 to extend the trapped Brownian motion theory[1] to Mamyshev oscillators. In terms of this energy dispersion, the coefficient of the gain-mediated pulse-repositioning coefficient becomes,

$$\Gamma = \frac{D_{\mathrm{E}}}{T_{\mathrm{c}}} \frac{\delta E_{\mathrm{r}}^{*}(\tau_i)}{E_{\mathrm{r}}^{*}\left(\tau_i/T_R - i\right)}. \tag{16}$$

We are now in a position to explain the $\Gamma$ simulation results qualitatively. Spectral setting *a* features a large $(1-\mathcal{F}')^{-1}$ factor, allowing large energy deviations (Equations 10, 13). However, the temporal shifts at the two filters nearly cancel, yielding $D_{\mathrm{E}}$ and $\Gamma$ values close to zero. If the Mamyshev arms were symmetric, this cancellation would have been achieved by symmetric spectra. In our laser, however, the longer fibre after amplification in the blue arm causes stronger temporal broadening for the same relative energy deviation, weakening its prefactor in Equation (15) compared with the red arm. This necessitates increasing the blue energy deviation through a steeper spectral slope at the blue filter. Spectral setting *b* is nearly symmetric between the arms, so $D_{\mathrm{E}}$ is large and positive due to the stronger prefactor for the red term in Equation (15). The energy deviations are small, however, because the spectral slopes are mild, giving only a modest $(1-\mathcal{F}')^{-1}$ factor (Equations 10, 13). As a result, $\Gamma$ remains relatively small. Spectral setting *c* is strongly asymmetric in favour of the blue term in Equation (15), producing a negative $D_{\mathrm{E}}$. The slopes here are so steep that the $(1-\mathcal{F}')^{-1}$ factor boosts the energy deviation by more than an order of magnitude, yielding a large negative $\Gamma$. Finally, spectral settings *d* and *e*, which are very similar, exaggerate the asymmetry in favour of the red term in Equation (15), resulting in large positive values of both $D_{\mathrm{E}}$ and $\Gamma$.

Next, we briefly discuss the pathways by which the wavelength shifts affect pulse speed, explaining the $\beta$ values in Supplementary Table 1. The dispersive pathway, which is usually dominant, is weak in our laser because it requires the wavelength shift to accumulate over many roundtrips, a process suppressed by the narrow filters. Instead, the wavelength shifts modify the filter-induced delay. This can occur directly, by changing the offset between the pulse's central wavelength and the filter transmission, thereby shifting the temporal shifting the transmitted pulse. An acoustically induced red shift is equivalent to decreasing $\Delta\lambda$ at the red filter and increasing it at the blue filter in Equation (14), leading to a net decrease in pulse speed.

Wavelength shifts also contribute indirectly to pulse speed by repeatedly inducing energy perturbations, as in the gain-mediated interaction. These perturbations arise mainly because the shifted spectrum alters the filter overlap, and to a lesser extent from the spectral dependence of the gain. Together, these direct and indirect contributions explain the calculated $\beta$ values.

A positive derivative of the index modulation produces a red wavelength shift. Since the direct effect translates red shift to reduced pulse speed, it contributes negatively to both $\beta$ values. This negative contribution dominates when the energy perturbations cancel at the two filters ($D_{\mathrm{E}} \approx 0$, as in setting *a*) or when the energy deviations remain small ($(1-\mathscr{F}')^{-1} \approx 1$, as in setting *b*).

When $D_{\mathrm{E}}(1-\mathscr{F}')^{-1}$ is large, the indirect contribution must be considered. In the blue arm (predominantly 6-µm fibres), a positive wavelength shift causes a positive energy perturbation, proportional to the spectral slope at the red filter. The resulting indirect contribution to $\beta_6$ scales with $D_{\mathrm{E}}(1-\mathscr{F}')^{-1}$, explaining the $\beta_6$ values at spectral settings *c*-*e*. In the red arm (predominantly 10-µm fibres), the red shift instead causes a negative energy perturbation when the blue filter slope is steep, explaining the negative $\beta_{10}$ at setting *c*. However, when the blue filter coincides with a spectral peak, the spectral dependence of the gain becomes relevant, turning the red shift into a weak positive energy perturbation. This weak positive perturbation is responsible for weakening the negative values at settings *d* and *e*, compared to *a* and *b*.

Unlike gain modulation and acoustically induced wavelength shift, the acoustically induced index modulation does not couple to the energy map. Consequently, $\alpha_6$ and $\alpha_{10}$ are largely independent of the spectral settings, with variations reflecting only their linear dependence on pulse energy.

Finally, we explain the behaviour of the noise variance $\langle \eta_\tau^2 \rangle$. Spontaneous emission noise directly perturbs the pulse energy, temporal position, and central wavelength, each with variance proportional to$h\nu/E_{\mathrm{in}}$, where $E_{\mathrm{in}}$ is the pulse energy entering the amplifier[1,2]. This immediately accounts for the difference between settings *d* and *e*, which are otherwise nearly identical and have nearly equal pulse-repositioning coefficients: setting *e* was obtained from setting *d* by reducing the output coupler transmission tenfold, thereby lowering the input energy to the blue amplifier and increasing the noise variance. If the blue amplifier were the sole source of spontaneous emission, the variance would have increased by a factor of ten.

The differences between spectral settings *a*-*d* can be understood by recalling that the energy perturbations can accumulate a Mamyshev oscillator. This accumulation produces the roundtrip-to-roundtrip correlations. The correlation time thus matches the energy-evolution timescale, which lengthens as the energy-map slope approaches unity. As a result, the energy contribution to the noise dominates when $D_{\mathrm{E}}(1-\mathcal{F}')^{-1}$ is large (settings *c*-*e*). In fact, we show below that this factor is the inverse of an effective viscosity in the Brownian model, replacing the role of spectral filtering and chromatic dispersion from our earlier paper[1]. Consequently, the noise variance scales quadratically with it.

## 7. A trapped Brownian particle model for the energy and position deviations

In this section we extend the trapped Brownian-particle model to the Mamyshev laser and determine the pulse energy and temporal position deviations. Focusing on the gain-mediated interaction and neglecting the acoustic contributions, the pulse repositioning reduces to a linear equation in the relative deviation from the harmonic pattern,

$$\dot{\delta}_{\tau i} = -\frac{\Gamma}{T_\mathrm{R}}\delta_{\tau i} + \frac{\eta_{\tau i}(t)}{T_\mathrm{R}}, \tag{17}$$

where $\delta_{\tau i} \equiv \left(\tau_i/T_\mathrm{R} - i\right)$ is the relative position deviation. Equation (17) is formally identical to the Langevin equation for a Brownian particle in a harmonic potential, with the restoring force proportional to Γ. The corresponding variance is,

$$\left\langle \delta_{\tau i}^2 \right\rangle = \frac{\left\langle \eta_\tau^2 \right\rangle}{2\Gamma T_\mathrm{R}}. \tag{18}$$

The discussion so far on the noise and the gain-mediated interaction highlights the central role of the pulse energy deviations. When these deviations are allowed to accumulate, both the random fluctuations and the deterministic interaction strengths increase, raising the question of how to optimally suppress the fluctuations of the pulse positions. In our previous work[1], we demonstrated the equivalence of these fluctuations to trapped overdamped Brownian motion, where the pulse speed was proportional to a wavelength deviation damped by spectral filtering. In the present laser, the energy deviation plays this role, with an effective viscosity parameter arising from the energy map slope, $\mathscr{F}'$. This mapping clarifies the factors that control fluctuation suppression in the Mamyshev laser.

We start by rewriting Equation (12) using the relative energy deviation $\delta_{E\mathrm{r}} \equiv \delta E_r / E_r^*$ and substituting the gain modulation explicitly,

$$\delta_{Er}(n+1) = \mathscr{F}'\delta_{Er}(n) + \frac{1}{E_{\text{sat}}}\left(E_{\text{r}}^{*} + g_{\text{o,r}}\frac{\partial \mathscr{F}_{\text{r}}}{\partial E_{\text{b}}}\frac{E_{\text{b}}^{*}}{E_{\text{r}}^{*}}E_{\text{b}}^{*}\right)\left(\frac{\tau_i}{T_{\text{R}}} - i\right). \tag{19}$$

This discrete equation can be recast into a differential form,

$$\dot{\delta}_{Er} = \frac{\delta_{Er}(n+1) - \delta_{Er}(n)}{T_{\text{c}}} = -\frac{1-\mathscr{F}'}{T_{\text{c}}}\delta_{Er} + \kappa\delta_{\tau i}T_{\text{R}} + \eta_E'(t), \tag{20}$$

where we have added an energy noise term, $\eta_E'$, and introduced the stiffness coefficient,

$$\kappa \equiv \frac{1}{E_{\text{sat}}T_{\text{c}}T_{\text{R}}}\left(E_{\text{r}}^{*} + g_{\text{o,r}}\frac{\partial \mathscr{F}_{\text{r}}}{\partial E_{\text{b}}}\frac{E_{\text{b}}^{*}}{E_{\text{r}}^{*}}E_{\text{b}}^{*}\right). \tag{21}$$

Equation (20) has the same form as Equation (3) in the report[1], where the fluctuating variable was the wavelength deviation. In principle, one should treat the energy deviations in both arms separately, each with its own noise term and energy-dispersion coefficient. However, this generalisation produces algebraically longer but qualitatively equivalent expressions. For clarity, we therefore express the pulse speed in terms of the red-arm energy deviation alone:

$$T_{\text{R}}\dot{\delta}_{\tau i} = \frac{\partial \dot{\tau}_i}{\partial \delta_{Er}}\delta_{Er} \equiv -\frac{D_{\text{E}}}{T_{\text{c}}}\delta_{Er}. \tag{22}$$

We can therefore recast the dynamics into the Brownian framework introduced previously. This puts the evolution of the position deviation in the same form as in our earlier Brownian paper. Rewriting the energy deviation equation in terms of the position deviation,

$$\frac{T_{\text{c}}}{D_E}\ddot{\delta}_{\tau i} = -\gamma\dot{\delta}_{\tau i} - \kappa\delta_{\tau i} - \frac{\eta_E'}{T_{\text{R}}}, \tag{23}$$

where $\gamma \equiv \left(1-\mathscr{F}'\right)/D_E$ is the effective viscosity. By making the equations dimensionless and evaluating the coefficients of the deterministic terms[1], one finds that the inertial term is negligible for typical parameters. A simpler but equivalent argument is to note that the energy deviation is parametrised by the pulse position and use its steady state by taking $\dot{\delta}_E = 0$. Either argument leads to,

$$\dot{\delta}_{\tau i} = -\frac{\kappa}{\gamma}\delta_{\tau i} + \frac{\eta'_E}{\gamma T_\mathrm{R}}, \tag{24}$$

which is of the same form as Equation (17), with $\Gamma = T_\mathrm{R}\kappa/\gamma$ and $\eta_\tau = \eta'_E/\gamma$. This shows that the noise variance scales like $1/\gamma^2$. Supplementary Figure 8 plots simulated $\Gamma$ and $\langle \eta_\tau^2 \rangle$ against $1/\gamma$ up to exceedingly small $\gamma$. These simulations used the same parameters as Supplementary Figure 7d, but with a varying pump power in the red arm. $\Gamma$ and $\langle \eta_\tau^2 \rangle$ fit reasonably well to a linear and a parabolic function of $1/\gamma$, respectively, as expected from this simple theory. A better fit can be obtained by incorporating the dependence of the spontaneous emission noise on the input pulse energy to each amplification arm and tracking how the energy deviations transfer from one arm to the other and affect the position shift at each filter.

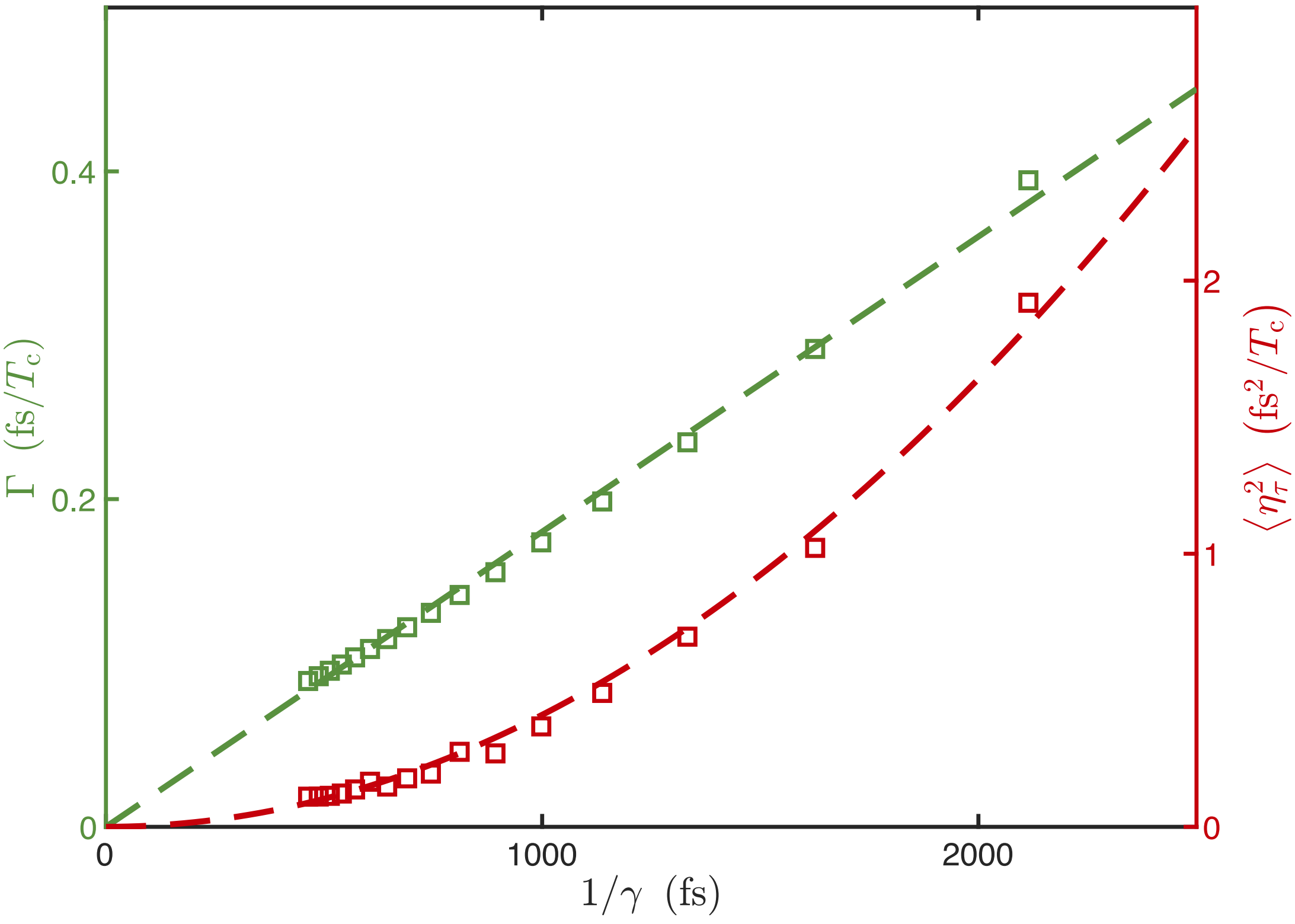

**Supplementary Figure 8**: **Simulated gain-mediated pulse-repositioning coefficient and noise variance.** The linear and parabolic curve fits correspond to the trapped Brownian theory.

From Equation (24), the variance of the position fluctuations is,

$$\left\langle \delta_{\tau i}^2 \right\rangle = \frac{\left\langle \eta_E'^2 \right\rangle}{2\gamma\kappa T_\mathrm{R}^2}. \tag{25}$$

In the Brownian-only picture, this expression suggests that position fluctuations are minimised by maximising $\gamma$, *i.e.*, by reducing $\mathscr{F}'$ and $D_\mathrm{E}$ towards zero. However, in this limit, the acoustic contribution to the pulse speed becomes increasingly important, shifting the optimum towards steeper spectral slopes and larger $\gamma$.

This conclusion changes when the Brownian model is extended to include a non-inertial noise source: interference between the spontaneously emitted light and the signal pulses can shift the pulse positions directly, without first accumulating energy and speed deviations. At large $\gamma$, this direct noise dominates and explains why the values of $\left\langle \eta_\tau^2 \right\rangle$ deviate from the $1/\gamma^2$ scaling by the Brownian-only model at settings *a* and *b*. To incorporate this effect, we add a direct, viscosity-independent noise term to Equation (22),

$$T_\mathrm{R}\dot{\delta}_{\tau i} = \frac{D_E}{T_\mathrm{c}}\delta_{E\mathrm{r}} + \eta_\mathrm{direct}(t). \tag{26}$$

Then, using Equation (20) and again taking $\dot{\delta}_E = 0$, Equation (24) is updated to include a direct viscosity-independent noise term,

$$\dot{\delta}_{\tau i} = -\frac{\kappa}{\gamma}\delta_{\tau i} + \frac{\eta_E'}{\gamma T_\mathrm{R}} + \frac{\eta_\mathrm{direct}}{T_\mathrm{R}}. \tag{27}$$

These two noise contributions are uncorrelated and white[2]. Combining them and comparing with Equation (17) yields,

$$\eta_\tau(t) = \frac{1}{\gamma}\sqrt{\left\langle \eta_E'^2 \right\rangle + \gamma^2 \left\langle \eta_\mathrm{direct}^2 \right\rangle}\,\xi(t), \tag{28}$$

where $\xi(t)$ is white noise with a variance of unity (in inverse square-root time units). The variance of the pulse position fluctuations then becomes,

$$\left\langle \delta_{\tau i}^{2} \right\rangle = \frac{\left\langle \eta_{E}^{\prime 2} \right\rangle + \gamma^{2} \left\langle \eta_{\mathrm{direct}}^{2} \right\rangle}{2\gamma \kappa T_{\mathrm{R}}^{2}}, \tag{29}$$

which finds a minimum at,

$$\gamma = \sqrt{\frac{\left\langle \eta_{E}^{\prime 2} \right\rangle}{\left\langle \eta_{\mathrm{direct}}^{2} \right\rangle}}. \tag{30}$$

Thus, unlike the origin model's prediction of monotonous improvement by indefinitely increasing $\gamma$, the extended Brownian model shows that the best suppression occurs at a finite $\gamma$, requiring the red filter to be placed close to the edge of the blue spectrum, as in Supplementary Figure 7b.

## 8. Gain-mediated pulse-to-pulse interactions

The amplification or absorption as light propagates through the gain fibre can be modelled via the effective emission and absorption cross-sections of Yb-doped germanosilicate fibre[3]. The stimulated emission is proportional to the concentration of excited gain atoms, $N_\mathrm{e}$, and the absorption to the concentration of the lower-level gain atoms, $N_0 - N_\mathrm{e}$, where $N_0$ is the doping concentration. For the signal power, which we denote $P_\mathrm{s}(t, z)$, this model reads,

$$\frac{\partial P_\mathrm{s}}{\partial z} = \gamma_\mathrm{s}\left(\left(\sigma_\mathrm{e}\left(\lambda_\mathrm{s}\right) + \sigma_\mathrm{a}\left(\lambda_\mathrm{s}\right)\right) N_\mathrm{e} - \sigma_\mathrm{a}\left(\lambda_\mathrm{s}\right) N_0\right) P_\mathrm{s}, \tag{31}$$

and for the pump,

$$\frac{dP_\mathrm{p}}{dz} = \gamma_\mathrm{p}\left(\left(\sigma_\mathrm{e}\left(\lambda_\mathrm{p}\right) + \sigma_\mathrm{a}\left(\lambda_\mathrm{p}\right)\right) N_\mathrm{e} - \sigma_\mathrm{a}\left(\lambda_\mathrm{p}\right) N_0\right) P_\mathrm{p}. \tag{32}$$

Here, $\gamma_\mathrm{s}$ and $\gamma_\mathrm{p}$ denote the ratio of the light contained in the doped core: $\gamma_\mathrm{s} \approx 1$ for the signal, and $\gamma_\mathrm{p}$ equals the cladding-to-core area ratio which is roughly 1 for the signal and the ratio of the cladding to core areas for the pump (~1/150 in our 10/125 gain fibres). The emission and absorption cross-sections, $\sigma_\mathrm{e}$ and $\sigma_\mathrm{a}$, depend strongly on wavelength. Their approximate values are 25, 25, 7, and 1 for $\sigma_\mathrm{e}\left(\lambda_\mathrm{p}\right)$, $\sigma_\mathrm{a}\left(\lambda_\mathrm{p}\right)$, $\sigma_\mathrm{e}\left(\lambda_\mathrm{s}\right)$, and $\sigma_\mathrm{a}\left(\lambda_\mathrm{s}\right)$, respectively, all in units of $10^{-25}\mathrm{m}^2$.

Integrating over the gain fibre length yields the signal gain,

$$g = \frac{P_\mathrm{s}\left(z = z_\mathrm{gain}\right)}{P_\mathrm{s}\left(z = z_\mathrm{gain} - L_\mathrm{gain}\right)} = \exp\left(\gamma_\mathrm{s}\left(\sigma_\mathrm{e}\left(\lambda_\mathrm{s}\right) + \sigma_\mathrm{a}\left(\lambda_\mathrm{s}\right)\right)\mathcal{N}_\mathrm{e}\right)\exp\left(-\gamma_\mathrm{s}\sigma_\mathrm{a}\left(\lambda_\mathrm{s}\right)\mathcal{N}_0\right), \tag{33}$$

where $z_\mathrm{gain} - L_\mathrm{gain}$ and $z_\mathrm{gain}$ are the fibre entrance and exit points, and $\mathcal{N}_\mathrm{e}$ and $\mathcal{N}_0$ are respective integrals of $N_\mathrm{e}$ and $N_0$ along the fibre. A similar integration can be used to calculate the pump absorption, $\alpha$, but in practice this dependence on the gain is negligibly weak; therefore, we treat it as a constant in the analytical model.

Finally, the dynamic evolution of the gain can be expressed in terms of the excited ion population:

$$\dot{g} = \frac{\mathrm{d}g}{\mathrm{d}\mathcal{N}_\mathrm{e}}\dot{\mathcal{N}}_\mathrm{e} = g\gamma_\mathrm{s}\left(\sigma_\mathrm{e}\left(\lambda_\mathrm{s}\right) + \sigma_\mathrm{a}\left(\lambda_\mathrm{s}\right)\right)\dot{\mathcal{N}}_\mathrm{e}. \tag{34}$$

Neglecting spontaneous emission, nonradiative transitions, and excited state absorption, the difference between the number of signal and pump photons entering and leaving the fibre equals the change in the number of excited gain atoms,

$$\dot{\mathcal{N}}_\mathrm{e} = \frac{1}{A_\mathrm{core} h\nu_\mathrm{s}}\left(\frac{\alpha\nu_\mathrm{s}}{\nu_\mathrm{p}}P_\mathrm{p} - \Delta P_\mathrm{s}\right), \tag{35}$$

where $A_\mathrm{core}$ is the core area, $h\nu_\mathrm{p}$ and $h\nu_\mathrm{s}$ are the pump and signal photon energies, and $\Delta P_\mathrm{s}$ is the net signal power output across the gain fibre. The signal input into the amplifier is much smaller than the output, it can be neglected, so that $\Delta P_\mathrm{s}(t) \simeq P_\mathrm{s}\left(t, z_\mathrm{gain}\right)$. The signal consists of ultrashort pulses repeating every cavity roundtrip, $T_\mathrm{c}$, each centred at $\tau_j$,

$$\dot{\mathcal{N}}_\mathrm{e} = \frac{1}{A_\mathrm{core} h\nu_\mathrm{s}}\left(\frac{\alpha\nu_\mathrm{s}}{\nu_\mathrm{p}}P_\mathrm{p} - \sum_{n=1}^{\infty}\sum_{j=1}^{N}\left|a_j\left(n; z_\mathrm{gain}, t - nT_\mathrm{c} - \tau_j\right)\right|^2\right). \tag{36}$$

Combining with Equation (34) and approximating the ultrashort pulses by a Dirac delta, the gain evolution reads,

$$\dot{g} = g\frac{\gamma_\mathrm{s}\left(\sigma_\mathrm{e}\left(\lambda_\mathrm{s}\right) + \sigma_\mathrm{a}\left(\lambda_\mathrm{s}\right)\right)}{A_\mathrm{core} h\nu_\mathrm{s}}\left(\frac{\alpha\nu_\mathrm{s}}{\nu_\mathrm{p}}P_\mathrm{p} - \sum_{n=1}^{\infty}\sum_{j=1}^{N}\frac{E_j(n)}{T_\mathrm{c}}\delta\left(\frac{t - nT_\mathrm{c} - \tau_j}{T_\mathrm{c}}\right)\right), \tag{37}$$

where $E_j(n)$ is the pulse energy at the output of the gain fibre. This simplifies to,

$$\dot{g}(t) = \frac{g(t)}{E_{\mathrm{sat}}}\left(\epsilon P_{\mathrm{p}} - \sum_{n=1}^{\infty}\sum_{j=1}^{N}\frac{E_j(n)}{T_{\mathrm{c}}}\delta\left(\frac{t - nT_{\mathrm{c}} - \tau_j}{T_{\mathrm{c}}}\right)\right), \tag{38}$$

where $\epsilon = \dfrac{\alpha\nu_{\mathrm{s}}}{\nu_{\mathrm{p}}}$ is the pump-to-signal conversion efficiency, and $E_{\mathrm{sat}} = \dfrac{A_{\mathrm{core}}h\nu_{\mathrm{s}}}{\gamma_{\mathrm{s}}\left(\sigma_{\mathrm{e}}\left(\lambda_{\mathrm{s}}\right) + \sigma_{\mathrm{a}}\left(\lambda_{\mathrm{s}}\right)\right)}$ is the saturation energy of the amplifier. For our parameters, this evaluates to ~20 μJ.

Next, we invoke the adiabatic elimination (slaving) approximation, replacing the pulse energies with their stable fixed points parametrised by the gain, $E_j^{*,m}(g)$, and neglecting the short time required to reach them. Because pulse adaptation occurs on the timescale of a few roundtrips, *i.e.*, much faster than gain recovery, this approximation is justified. Consistent with this separation of timescales, we take a time average over one roundtrip, $T_{\mathrm{c}}$. This captures the slow evolution of the baseline of the gain by accounting for the aggregate depletion or recovery, while neglecting the intra-roundtrip modulation. The error of this averaging, *i.e.*, the intra-roundtrip gain modulation, can be collected in a separate term, $g_{\mathrm{m}}$, so that $\dot{g} = \dot{g}_{\mathrm{o}} + \dot{g}_{\mathrm{m}}$. Thus, the evolution of the baseline gain reads,

$$\dot{g}_{\mathrm{o}} = \frac{g}{E_{\mathrm{sat}}}\left(\epsilon P_{\mathrm{p}} - \frac{1}{T_{\mathrm{c}}}\sum_{j=1}^{N} E_j^{*,m}\left(g\right)\right). \tag{39}$$

To find $\dot{g}_{\mathrm{m}}$, we compare the full gain Equation (11) of Methods with that of the baseline gain, leading to,

$$\dot{g}_{\mathrm{m}} = \frac{g}{T_{\mathrm{c}}E_{\mathrm{sat}}}\sum_{j=1}^{N} E_j^{*,m}\left(g\right)\left(1 - \sum_{n=1}^{\infty}\delta\left(\frac{t - nT_{\mathrm{c}} - \tau_j}{T_{\mathrm{c}}}\right)\right). \tag{40}$$

The terms in the parentheses average out to zero, so they do not induce any net change, only intra-roundtrip modulation. Because $E_j^{*,m} \ll E\mathrm{sat}$, we have $|g_{\mathrm{m}}| \ll g_{\mathrm{o}}$. This allows replacing $g$ in the right-hand sides of Equations (39, 40) with $g_{\mathrm{o}}$. Furthermore, the minute change of $g_{\mathrm{o}}$ within a single roundtrip allows regarding it as a constant in the $\dot{g}_{\mathrm{m}}$ equation above, making $g_{\mathrm{m}}$ periodic,

$$\dot{g}_{\mathrm{m}}(\tau;\tau_1,\cdots,\tau_N)=\frac{g_{\mathrm{o}}}{T_{\mathrm{c}}E_{\mathrm{sat}}}\sum_{j=1}^{N}E_j^{*,m}\left(g_{\mathrm{o}}\right)\left|1-\delta\left(\frac{\tau-\tau_j}{T_{\mathrm{c}}}\right)\right|. \tag{41}$$

This approximation becomes exact in the steady state, where $\dot{g}_{\mathrm{o}}=0$. Integrating Equation (41), we get,

$$g_{\mathrm{m}}(\tau;\tau_1,\cdots,\tau_N)=\frac{g_{\mathrm{o}}}{E_{\mathrm{sat}}}\sum_{j=1}^{N}E_j^{*,m}\left(g_{\mathrm{o}}\right)\left(\frac{\tau}{T_{\mathrm{c}}}-\Theta\left(\tau-\tau_j\right)\right), \tag{42}$$

where $\Theta$ is the Heaviside step function. For harmonic modelocking, we are interested in $g_{\mathrm{m}}$ when the pulses are all identical, say at $E^{*,2}$. Then,

$$\frac{g_{\mathrm{m}}}{g_{\mathrm{o}}}=\frac{E^{*,2}}{E_{\mathrm{sat}}}\left(\frac{N\tau}{T_{\mathrm{c}}}-\sum_{j=1}^{N}\Theta\left(\tau-\tau_j\right)\right), \tag{43}$$

as stated in the main text. The gain modulation experienced by the $i^{\mathrm{th}}$ pulse is then

$$\frac{g_{\mathrm{m}}\left(\tau_i\right)}{g_{\mathrm{o}}}=\frac{E^{*,2}}{E_{\mathrm{sat}}}\left(\frac{N\tau_i}{T_{\mathrm{c}}}-\sum_{j=1}^{N}\Theta\left(\tau_i-\tau_j\right)\right)=\frac{E^{*,2}}{E_{\mathrm{sat}}}\left(\frac{\tau_i}{T_{\mathrm{R}}}-i\right). \tag{44}$$

Here, $i$ is the index of the pulse in temporal order within the cavity roundtrip. Because $g_{\mathrm{m}}\ll g_{\mathrm{o}}$, we can assume the pulse speed depends linearly on it,

$$v_{\mathrm{i,gain}}=\frac{\partial v_i}{\partial g_{\mathrm{m}}\left(\tau_i\right)}\frac{E^{*,2}}{E_{\mathrm{sat}}}\left(\frac{\tau_i}{T_{\mathrm{R}}}-i\right)\equiv\Gamma\left(\frac{\tau_i}{T_{\mathrm{R}}}-i\right), \tag{45}$$

where $v_{\mathrm{i,gain}}$ is the contribution of the gain-mediated interaction to the speed of the $i^{\mathrm{th}}$ pulse. This mechanism is explained qualitatively via an analytical model in Supplementary Section 6, which leverages the parametrisation of the pulse energy and shape by the gain.

## 9. Acoustic pulse-to-pulse interactions

As pulses propagate in the fibre, they attract the fused silica molecules of the fibre towards the core through electrostriction (attraction of induced dipoles to the electric field). This drives longitudinal acoustic waves which travel through the fibre cross-section. At any given point along the fibre, $z$, with a mode intensity radius of $w(z)$ and a signal power of $P_s(z,t)$, the acoustic waves evolve according to the following wave equation[4,5],

$$\frac{\partial^2 \rho}{\partial t^2} - v_{\text{sound}}^2 \nabla_\perp^2 \rho - 2A \frac{\partial}{\partial t} \nabla_\perp^2 \rho = - \frac{\gamma_e n_I}{2\pi \epsilon_r c} \frac{P_s(z,t)}{w^2} \underbrace{\nabla_\perp^2 \exp\left[ -(r/w)^2 \right]}_{\text{mode-overlap factor}}, \tag{46}$$

where $\rho$ is the material density perturbation. The first term represents inertia, the second the elastic restoring force, the third the viscous attenuation, and the right-hand side is the driving force. $v_{\text{sound}}$ is the speed of longitudinal sound waves in fused silica ($5.9 \times 10^3$ m/s[5]), $A_{\text{vis}}$ is the viscous attenuation coefficient, and $\gamma_e \equiv \rho_0 \partial \epsilon_r / \partial \rho$ is the electrostriction coefficient, with $\rho_0$ the density of fused silica. $\gamma_e$ can be estimated using the Lorentz-Lorenz law as[4] $\gamma_e = (n_I^2 - 1)(n_I^2 + 2)/3$, where $n_I$ is the refractive index, which is related to the dielectric constant through $\epsilon_r = n_I^2$. In the driving term, $n_I$ has been equated to the group velocity index.

By the adiabatic elimination (slaving) approximation, the pulse pattern is effectively constant on the slower timescale of acoustic wave evolution. This makes the signal power, $P_s$ periodic, allowing it to be expressed as a function of the delay coordinate and the pulse positions:

$$P_s(\tau; \tau_1, \cdots, \tau_N) = \frac{E^{*,2}}{T_c} \sum_{j=1}^{N} \delta\left( \frac{\tau - \tau_j}{T_c} \right), \tag{47}$$

where the pulse shape has been approximated by a Dirac delta, since its duration is negligible compared to the acoustic oscillations. The pulse energy has likewise been approximated by its steady-state value.

Since the acoustic waves are linear, they can be divided into multiple acoustic modes, each with a characteristic spatial distribution and natural frequency. The oscillations of these modes can, in turn, be analysed in terms of the frequency components of the driving force. Consequently, every radio-frequency component in $P_s$ contributes to the oscillation of each acoustic mode. The resulting material density perturbation, $\rho$, can therefore be expressed as a sum over all optical frequency components and all acoustic modes:

$$\rho\left(r,\tau\right)=\sum_{M=1}^{\infty}J\left(K_M r\right)\sum_{h=-\infty}^{+\infty}\mathrm{e}^{i\omega_h\tau}\,\tilde{\rho}_M\left(\omega_h\right), \tag{48}$$

where $\omega_h = h\cdot 2\pi/T_c$ are the harmonic angular frequencies of the laser cavity, $T_c$ is the cavity period, $\tilde{\rho}_M\left(\omega_h\right)$ is the Fourier transform of the oscillation of the $M^{\mathrm{th}}$ acoustic mode, and $J\left(K_M r\right)$ is the zeroth-order Bessel function of the first kind, which describes the cylindrically symmetric spatial profile of each mode. This assumes perfect cylindrical symmetry and neglects the stress rods in the polarisation-maintaining fibre.

For any real positive acoustic wave number $K_M$, this solution is valid. Imposing the boundary condition at the cladding–polymer interface discretises $K_M$ into values corresponding to the allowed acoustic modes.

To determine the amplitudes $\tilde{\rho}_M\left(\omega_h\right)$ and, ultimately, the full density perturbation, the driving term must be expressed in the frequency domain via the Fourier transform of the pulse train,

$$P_s(\tau)=\frac{E^{*,2}}{T_c}\sum_{j=1}^{N}\delta\left(\frac{\tau-\tau_j}{T_c}\right)=\frac{E^{*,2}}{T_c}\sum_{h=-\infty}^{+\infty}\sum_{j=1}^{N}\mathrm{e}^{i\omega_h(\tau-\tau_j)}, \tag{49}$$

and as a linear combination of the acoustic modes to describe the Gaussian optical mode:

$$\frac{1}{\pi w^2}\mathrm{e}^{-\left(\frac{r}{w}\right)^2}=\sum_{M=1}^{\infty}S_M J(K_M r), \tag{50}$$

where $S_m$ is the optoacoustic overlap coefficient, calculated as the spatial inner product of the optical and the acoustic modes, *i.e.*, by integrating over their spatial overlap,

$$S_M = \int_0^{r_{\text{clad}}} J(K_M r) \frac{1}{\pi w^2} \mathrm{e}^{-\left(\frac{r}{w}\right)^2} 2\pi r dr, \tag{51}$$

with $r_{\text{clad}}$ the cladding radius. Low-$K_M$ modes have a broad maximum at the fibre core, while higher-order modes localise their oscillations more tightly, improving overlap with the optical mode. For sufficiently high $M$, the acoustic oscillations are confined within the core and the overlap averages to near zero. Consequently, only a limited band of acoustic modes is efficiently excited by the pulses. With the mode-intensity radii corresponding to our experiments, this band falls roughly to 1 GHz, which can be estimated as the speed of sound divided by the mode-intensity diameter.

Substituting Equations (48-50) into the wave Equation (46) and rearranging the terms verifies the solution, yielding the density perturbation:

$$\rho(r,\tau) = \frac{\gamma_{\text{e}}}{2cn_{\text{I}}} \sum_{M=1}^{\infty} J(K_M r) \sum_{h=-\infty}^{+\infty} \frac{S_M K_M^2}{v_{\text{sound}}^2 K_M^2 - \omega_h^2 + 2iA\omega_h K_M^2} \frac{E^{*,2}}{T_{\text{c}}} \sum_{j=1}^{N} \mathrm{e}^{i\omega_h(\tau-\tau_j)}. \tag{52}$$

The denominator in the $h$-sum is the resonance factor of a damped-driven harmonic oscillator with $v_{\text{sound}} K_M$ as the natural frequency. The factor $S_M K_M^2$ plays the role of an effective reciprocal mass of this harmonic oscillator.

The acoustically induced refractive index modulation is proportional to the density perturbation, and thus inherits its spatial dependence. However, the optical pulses only interact within the spatial overlap between the acoustic and optical modes, introducing an additional factor of $S_M$. Using $\gamma_{\text{e}} = \rho_0 \partial n_{\text{I}}^2 / \partial \rho$ to relate index and density perturbations, we obtain the index modulation experienced by the laser light:

$$n_{\text{I,m}}(\tau; \tau_1, \ldots, \tau_N) = \frac{\gamma_{\text{e}}^2}{4cn_{\text{I}}^2 \rho_0} \sum_{h=-\infty}^{+\infty} \sum_{M=1}^{\infty} \frac{S_M^2 K_M^2}{v_{\text{sound}}^2 K_M^2 - \omega_h^2 + 2iA\omega_h K_M^2} \frac{E^{*,2}}{T_{\text{c}}} \sum_{j=1}^{N} \mathrm{e}^{i\omega_h\left(\tau - \tau_j\right)}. \tag{53}$$

For convenience, we introduce a transfer function $T(\omega)$ such that

$$n_{\mathrm{I,m}}(\tau;\tau_1,\ldots,\tau_N) \equiv \sum_{h=-\infty}^{+\infty} T\left(\omega_h\right) E^{*,2} \sum_{j=1}^{N} \mathrm{e}^{i\omega_h\left(\tau-\tau_j\right)} \equiv \tilde{n}_{\mathrm{I,m}} E^{*,2}, \tag{54}$$

where $\tilde{n}_{\mathrm{I,m}}$ expresses the index modulation in an energy-independent way, which will be convenient for analysing the pulse speeds.

There are two contributions from the acoustic waves to the relative pulse speeds. The first is a direct contribution whereby the $i^{\text{th}}$ pulse slows down proportionally to $n_{\mathrm{I,m}}(\tau_i;\tau_1,\ldots,\tau_N) \times L$, where $L$ is the fibre length where the index modulation is experienced. However, the index modulation is not the same throughout the fibres of our laser. This is partly because the steady-state pulse energy depends on the position in the cavity, so the resulting pulse speed, $v_{i,n_{\mathrm{I,m}}}$, is an integral,

$$v_{i,n_{\mathrm{I,m}}} = -\frac{1}{T_{\mathrm{c}}} \int dz \frac{E^*(z)}{\mathrm{c}} \tilde{n}_{\mathrm{I,m}} \equiv \alpha \tilde{n}_{\mathrm{I,m}}, \tag{55}$$

where $c$ is the speed of light, and $E^*(z)$ is the steady-state pulse energy as a function of the position along the cavity. At the reference point where the energy map is defined, this equals $E^{*,2}$.

The acoustic response differs between fibre segments in our laser, since some have a core diameter of 6 μm and others 10 μm and the index modulation varies mode-intensity radius $w$, which results in different opto-acoustic overlap factor $S_m$, acoustic transfer function $T(\omega)$, and index modulations $\tilde{n}_{\mathrm{I,m}}$. We label these quantities with a subscript 6 or 10, corresponding to the fibre types with 6 μm and 10 μm core diameters, respectively.

Thus, the direct contribution of this index modulation to the pulse speed, $v_{i,n_{\mathrm{I,m}}}$, reads,

$$v_{i,n_{\mathrm{I,m}}} = \alpha_6 \tilde{n}_6 + \alpha_{10} \tilde{n}_{10} = \sum_{h=-\infty}^{+\infty} \left(\alpha_6 T_6 + \alpha_{10} T_{10}\right) \sum_{j=1}^{N} \mathrm{e}^{i\omega_h\left(\tau-\tau_j\right)}, \tag{56}$$

where $\tilde{n}_6$ and $\tilde{n}_{10}$ are the index modulations per unit energy in the two fibre types and the coefficients $\alpha_6$ and $\alpha_{10}$ are each calculated by taking the integral in Equation (55) over the corresponding fibre.

The second acoustic contribution to the pulse speed is an indirect one, arising from a shift of the pulse's central wavelength that is proportional to the derivative of the index modulation[5],

$$\frac{d\lambda}{dz} = \frac{\lambda}{\mathrm{c}} \sum_{h=-\infty}^{+\infty} T\left(\omega_h\right) \cdot i\omega_h \cdot E^* \sum_{j=1}^{N} \mathrm{e}^{i\omega_h\left(\tau - \tau_j\right)} = \frac{\lambda}{\mathrm{c}} E^* \frac{d\tilde{n}_{\mathrm{I,m}}}{d\tau}. \tag{57}$$

Because this wavelength shift is small, we can take its contribution to the pulse speed as linear in the index modulation. Taking the different fibre types into account, this contribution reads,

$$v_{i,\frac{d}{d\tau}\tilde{n}_{\mathrm{I,m}}} = \beta_6 \frac{d\tilde{n}_6}{d\tau}\Big|_{\tau_i} + \beta_{10} \frac{d\tilde{n}_{10}}{d\tau}\Big|_{\tau_i} = \sum_{h=-\infty}^{+\infty} i\omega_h \left(\beta_6 T_6 + \beta_{10} T_{10}\right) \sum_{j=1}^{N} \mathrm{e}^{i\omega_h\left(\tau_i - \tau_j\right)}, \tag{58}$$

where $\beta_6$ and $\beta_{10}$ are the coefficients for the two fibre types, to be determined from simulations (see below).

For the calculation of the transfer functions, we took mode-intensity radii as $w_6 = 2.7$ µm and $w_{10} = 3.8$ µm. The acoustic wavenumbers were calculated by assuming a zero boundary condition at the cladding-polymer interface with a cladding diameter of 125 µm for all fibres, corresponding to perfect reflection of the acoustic waves. The viscous attenuation coefficient, $A_{\mathrm{vis}}$, was calculated as 1.24 µm$^2$ MHz by inserting an attenuated plane wave into Equation (46) with no driving force and using the attenuation rate reported in[6].

These assumptions substantially overestimate the acoustic interactions, particularly near resonance. In practice, significant reflection loss is expected at the cladding–polymer interface, and the resonance frequency will vary along the fibre due to geometric tolerances in the cladding diameter. These limitations should be kept in mind when interpreting the pulse pattern simulations.

## 10. Pulse pattern simulations

We calculate the coefficients governing the top-level pulse pattern dynamics, the gain restoring force coefficient $\Gamma$, the acoustic interaction coefficients $\alpha_6$, $\alpha_{10}$, $\beta_6$, $\beta_{10}$, and the variance of the white Gaussian noise, $\left\langle \eta_\tau^2 \right\rangle$, depend on the spectral settings, *i.e.*, on $P_{\mathrm{p,\ blue}}/N$, $P_{\mathrm{r,\ red}}/N$, and $\Delta\lambda$. They were obtained from single-pulse simulations, which yielded the values $\Gamma T_\mathrm{c} = 0.120\ \mathrm{fs}$, $\left\langle \eta_\tau^2 \right\rangle T_\mathrm{c} = 0.2\ \mathrm{fs}^2$, $\alpha_6 T_\mathrm{c} = -5.4\ \mathrm{ns\ nJ}$, $\alpha_{10} T_\mathrm{c} = -24\ \mathrm{ns\ nJ}$, $\beta_6 T_\mathrm{c} = 1.5\ \mathrm{ns}^2\ \mathrm{nJ}$, and $\beta_{10} T_\mathrm{c} = -2.4\ \mathrm{ns}^2\ \mathrm{nJ}$, for the spectral settings shown in Supplementary Fig. 7d. For the simulation of the pulse pattern, Euler's method was used with step size $dt = 4$ ms, with the noise term evaluated from Gaussian random numbers with variance of unity and scaled by $\sqrt{\left\langle \eta_\tau^2 \right\rangle dt}$.

The calculation for $\Gamma$ starts with a preparatory simulation to quickly obtain $N_\mathrm{e}$. The pulse naturally drifts within the simulated temporal window. We record the rate of this drift after convergence and take it as the reference against which the relative pulse speed will be calculated. Then, we simulate over a single roundtrip with no pump using the dynamic gain model, where we deplete $N_\mathrm{e}$ in each propagation step according to the number of emitted photons. The resulting $N_e$ corresponds to the gain right after a pulse depletes it, which is equivalent to the gain a pulse experiences at $\left(\tau_i/T_\mathrm{R} - i\right) = -1$. We simulate with this $N_\mathrm{e}$ until convergence and measure the change in the drift rate compared to the preparatory simulation. This change equals $\Gamma$. The calculations for $\alpha_6$, $\alpha_{10}$, $\beta_6$, and $\beta_{10}$ are discussed in Supplementary Section 9. We then calculate the optical frequency shifts by multiplying the pulse with $\exp(i\delta\omega\,\tau)$ in each propagation step, with $\delta\omega$ calculated from,

$$\delta\omega(z) = -dz\frac{\omega}{\mathrm{c}}E(z)\frac{d\tilde{n}_{\mathrm{I,m}}}{d\tau}, \tag{59}$$

where $dz$ is the propagation step size, $\omega$ is the central angular frequency, and $E(z)$ is the pulse energy at the propagation step. We run these simulation until the pulse converges to a new, slightly different state, measure the change in the pulse speed (using its drift within the temporal window), and divide it by the inserted test value of $d\tilde{n}_6/d\tau$ or $d\tilde{n}_{10}/d\tau$ to calculate $\beta_6$ or $\beta_{10}$, respectively.

The noise variance $\langle \eta_\tau^2 \rangle$ was calculated by introducing spontaneous emission noise after preparatory simulations with a fixed gain for thousands of roundtrips and noting the random displacements of the pulse within the simulated temporal window in each roundtrip. Then, we statistically analyse these displacements. To isolate the random contribution of the noise, we subtract the mean of these displacements. Assuming the noise is white, its variance is the mean of the square of these displacements. However, the noise is not exactly white; it is correlated with a short correlation time spanning up to tens of roundtrips. This correlation is negligible when simulating the pulse pattern evolution, but must be taken into account when calculating the noise variance, for which uncorrelated samples are required. To produce these uncorrelated samples, we add up the random displacements of a sufficient number of consecutive roundtrips before squaring. Then, taking their mean and dividing by the number of roundtrips per sample gives $\langle \eta_\tau^2 \rangle$.

We simulated the full pulse repositioning equation terms for all pulse numbers ranging from 2 to 200, starting near harmonic modelocking. Using the parameters corresponding to spectral setting *d* of Supplementary Figure 7, where $\Gamma$ is large and positive, harmonic modelocking remained stable at most harmonics. Slight detuning of the cladding diameter or the fundamental repetition rate shifted which harmonics were unstable, demonstrating the stringent resonance requirement for the acoustic interactions to destabilise harmonic modelocking at these spectral settings.

At spectral setting *c*, where $\Gamma$ is large and negative, harmonic modelocking was unstable at all harmonics except those with 2, 3, or 5 pulses. Settings *a*, *b*, and *e* also prohibited most harmonic states but stabilised more of them than setting *c*. Among these, setting *e* stabilised a relatively large number of harmonics, though not the majority, followed by setting *b* and then setting *a*. The latter two result in much slower pulse repositioning compared to setting *d*. This is illustrated in Supplementary Figure 8, where simulated pulse repositioning at setting $b$ takes over a minute to converge to a stable anharmonic pattern. This contrast underscores the tunability of pulse repositioning predicted by our theory.

These results carry three implications. First, harmonic mode-locking is stabilised by the gain-mediated interaction whenever $\Gamma$ is positive, regardless of the repetition rate. Second, the acoustic terms selectively stabilise some harmonics but destabilise most others, depending sensitively on the repetition rate. Third, at spectral settings such as Supplementary Figure 7d, the gain-mediated term tends to dominate over the acoustic ones. This dominance is in practice even stronger, since the simplified acoustic model exaggerates their resonance.

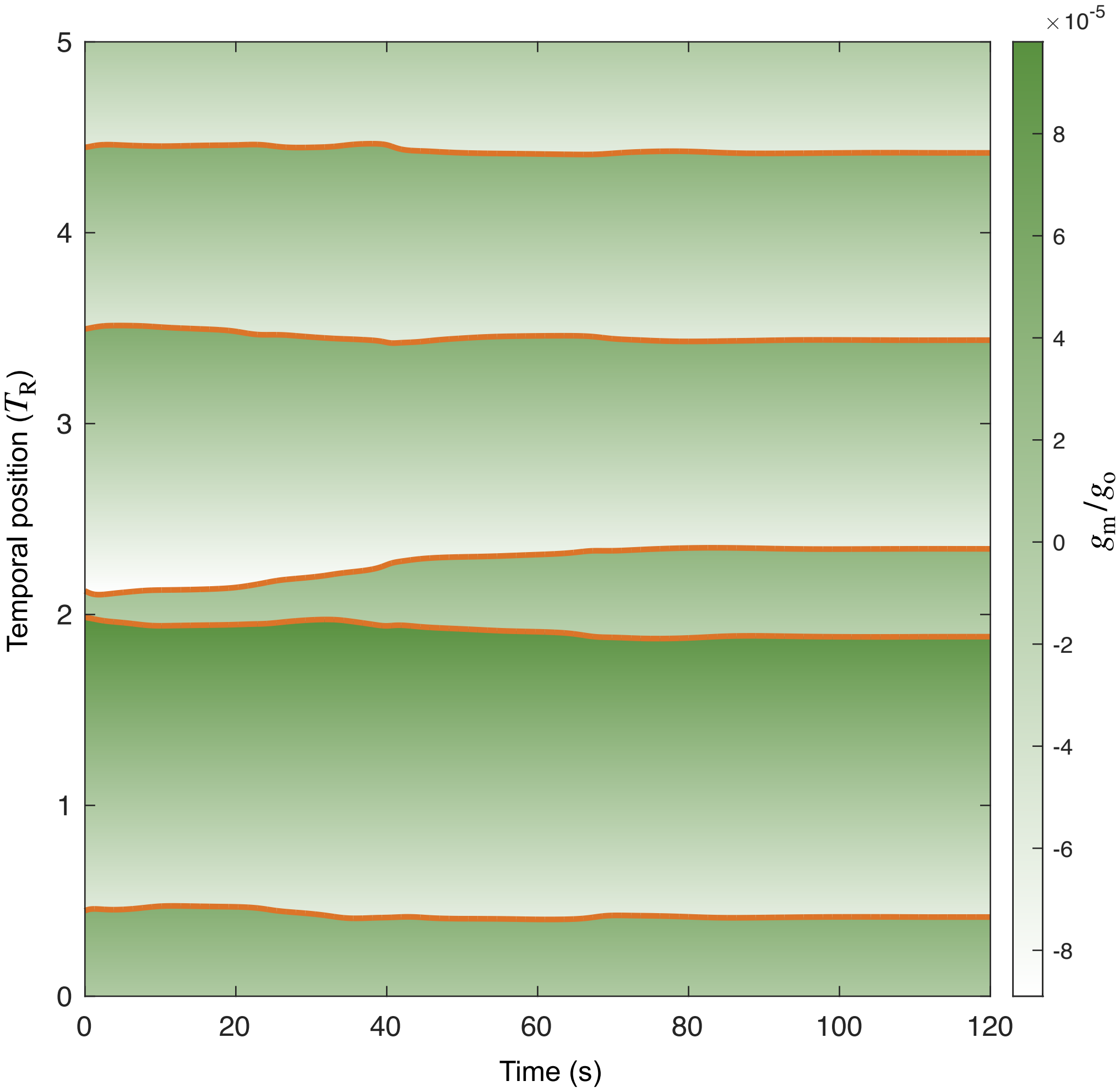

**Supplementary Figure 9**: **Slow pulse repositioning at setting *b*.** Because the gain-mediated pulse-repositioning coefficient is small (Supplementary Table 1b), the pulses take more than a minute to reposition and settle into an anharmonic pattern, where they experience significant gain disparity.

### 11. Specific Simulations

A set of specific simulations were made to obtain results in certain figures. The colour plots for Level 1 in Figure 2b are identical copies of a converged single pulse from a preparatory simulation.

For Level 2 dynamics depicted in Figure 2b, the preparatory single-pulse simulation was followed by a dynamic-gain simulation, updating $N_e$ each roundtrip according to the net photon flux. This simulation was then interrupted to add a second pulse and resumed to cover the remaining duration shown in the plot, producing the gain and total energy traces for the blue arm.

The energy maps in Figure 4a,b were obtained by running a preparatory simulation for each. The settings were chosen to produce similar spectra to the experiments. Then, we fixed the gain and ran many 2-roundtrip simulations initialised with the converged red-filtered pulse shape but varying initial energies and collected $E(2)$ versus $E(1)$ at the output of the blue arm. We checked and confirmed that this plot is nearly identical to $E(3)$ versus $E(2)$, confirming its accuracy.

For the tunnelling times (Figure 4c), we simulated a single pulse subject to spontaneous emission noise at pump powers near the saddle-node bifurcation and noted the random times when the pulse died. At each pump power, we ran a preparatory simulation as described above to obtain $N_e$ in each gain propagation step. We then ran 20 simulations with this $N_e$ (fixed), waited until the noise triggered the pulse annihilation in each, and noted the times it did. Close to the bifurcation, spontaneous emission increases the mean pulse energy at the same gain, preventing annihilation. In reality, this average effect lowers the gain, allowing annihilation. To account for this, we added noise to the preparatory simulation as well. Because the photon-balance calculation there amplifies the noise effect on $N_e$, we repeated this preparatory simulation 1000 times and averaged the resulting $N_e$ values at each pump power before the 20 fixed-gain simulations. The random relative pulse-energy change per roundtrip is typically $\sim 10^{-4}$, as estimated analytically from the number of photons in the pulse as it enters the gain fibres. This implies that the noise introduces a disparity on the order of $10^{-4}$ between the pulses in each roundtrip, which is comparable to the gain disparity in anharmonic patterns. Therefore, the simulated tunnelling times refer to pulses in harmonic patterns, where this gain disparity vanishes.

The colour map for Level 4 dynamics in Figure 2b was calculated using the analytical expression for the intra-roundtrip gain modulation with $E^{*,2}/E_{\mathrm{sat}} = 10^{-4}$. Pulse positions were simulated for 100 pulses with random initial perturbations from the harmonic positions, on the order of 0.3 $T_{\mathrm{R}}$, using the explicit form of Equation (4),

$$\dot{\tau}_i = -\Gamma\left(\frac{\tau_i}{T_{\mathrm{R}}} - i\right) + \eta_{\tau i}(t) \\ - \sum_{h=-\infty}^{+\infty}\left(T_6(\omega_h)(\alpha_6 + i\omega_h\beta_6) + T_{10}(\omega_h)(\alpha_{10} + i\omega_h\beta_{10})\right)\sum_{j=1}^{N}\exp\left(i\omega_h\left(\tau_i - \tau_j\right)\right), \tag{60}$$

which results from writing $v_i$ as a sum of the gain and acoustic contributions derived in Supplementary Sections 8, 9, using the coefficients listed in Supplementary Table 1 for settings *d*. Euler's method was used for these simulations with step size $dt = 4$ ms, with the noise term evaluated from Gaussian random numbers with standard deviation of unity and scaled by $\sqrt{\langle\eta_\tau^2\rangle\, dt}$.

## 12. Laser Performance

In addition to validating the theoretical framework, the laser exhibited excellent long-term robustness. The maximum recorded repetition rate, average power, pulse energy, and supermode suppression ratio were 1.73 GHz, 3.5 W, 20 nJ and 60 dB, respectively, with a minimum dechirped pulse duration of 100 fs. With pulse durations down to 100 fs and repositioning dynamics taking several seconds or longer (Supplementary Figure 8), the framework captures system dynamics spanning up to 14 orders of magnitude in timescales. Laser component limitations on average power prevented sustained operation with all these characteristics simultaneously. In particular, after observing partial damage to a critical component, we increased the output coupling from ~50% to ~90%, which is expected to increase both the laser noise and the supermode amplitudes[1] (Supplementary Section 6). Higher pulse energies produce shorter pulses, but the energy had to be reduced at high repetition rates due to power limitations. Even so, at 1.6 GHz, we simultaneously achieved (Supplementary Figure 10) the highest average power and supermode suppression as well as one of the shortest pulse durations and highest pulse energies reported for harmonically modelocked lasers above 1 GHz using standard fibres[7-11]. Across all experimentally explored initial conditions and control sequences, any targeted harmonic state could be reliably reached through appropriate seeding and annihilation operations, although the stochastic nature of the dynamics occasionally required repetition of intermediate steps.

The maximum repetition rate we have recorded is 1.73 GHz. We have not characterised this state fully. The highest repetition rate state with full characterisation is shown in Supplementary Figure 10. It has 110 pulses with a pulse energy of 2 nJ and a dechirped pulse duration of 125 fs. Decreasing the number of pulses allows increasing the pulse energy, broadening the spectrum, and allowing shorter dechirped durations. This is shown in the 1 GHz state in Supplementary Figure 11 with 3 nJ and 100 fs dechirped duration. Decreasing the repetition rate and increasing the pulse energy further provide diminishing returns due to decreasing compressibility. Supplementary Fig. 12 shows a supermode suppression of ~60 dB, the highest that we recorded, while also highlighting its dependence on the spectral settings. These states were obtained before we noticed any thermal degradation in any components. After some time of high average power operation, we noticed a drop in the average power and responded by lowering the pump powers, increasing the passive fibre length after the gain in the red arm (to achieve the required spectral broadening with lower pulse energies) and increasing the output coupling loss to decrease the power falling on the blue filter. The data in Figure 5 in the main text was taken after these changes. In Supplementary Figure 13, we show the optical spectra associated with Figure 5 of the main text as well as radio frequency spectra of two intermediate pulse patterns. Lastly, the state with the highest pulse energy of 20 nJ that we recorded is shown in Supplementary Figure 14. This state, too, was obtained after this partial damage.

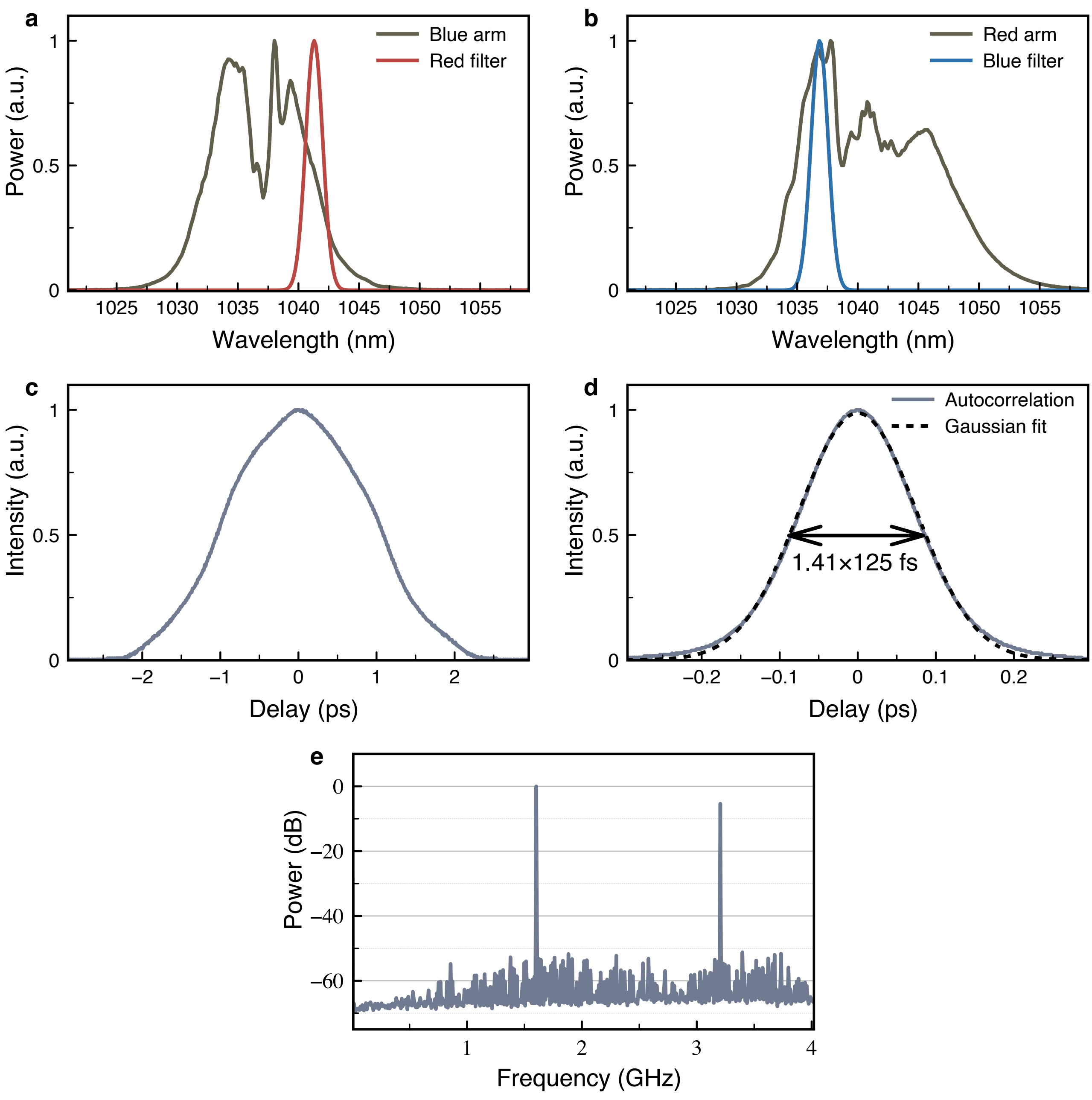

**Supplementary Figure 10**: **A 1.6 GHz harmonic state. a**,**b**, Optical spectra measured at the two arms. **c**,**d**, Autocorrelation of direct and dechirped output (red arm), respectively. A Gaussian fit is drawn on the autocorrelation of the dechirped pulses, indicating a full width at half maximum duration of ~125 fs. **e**, Radio frequency spectrum of the pulse pattern showing harmonic modelocking at ~1.6 GHz with over 50 dB supermode suppression and 106 intra-cavity pulses.

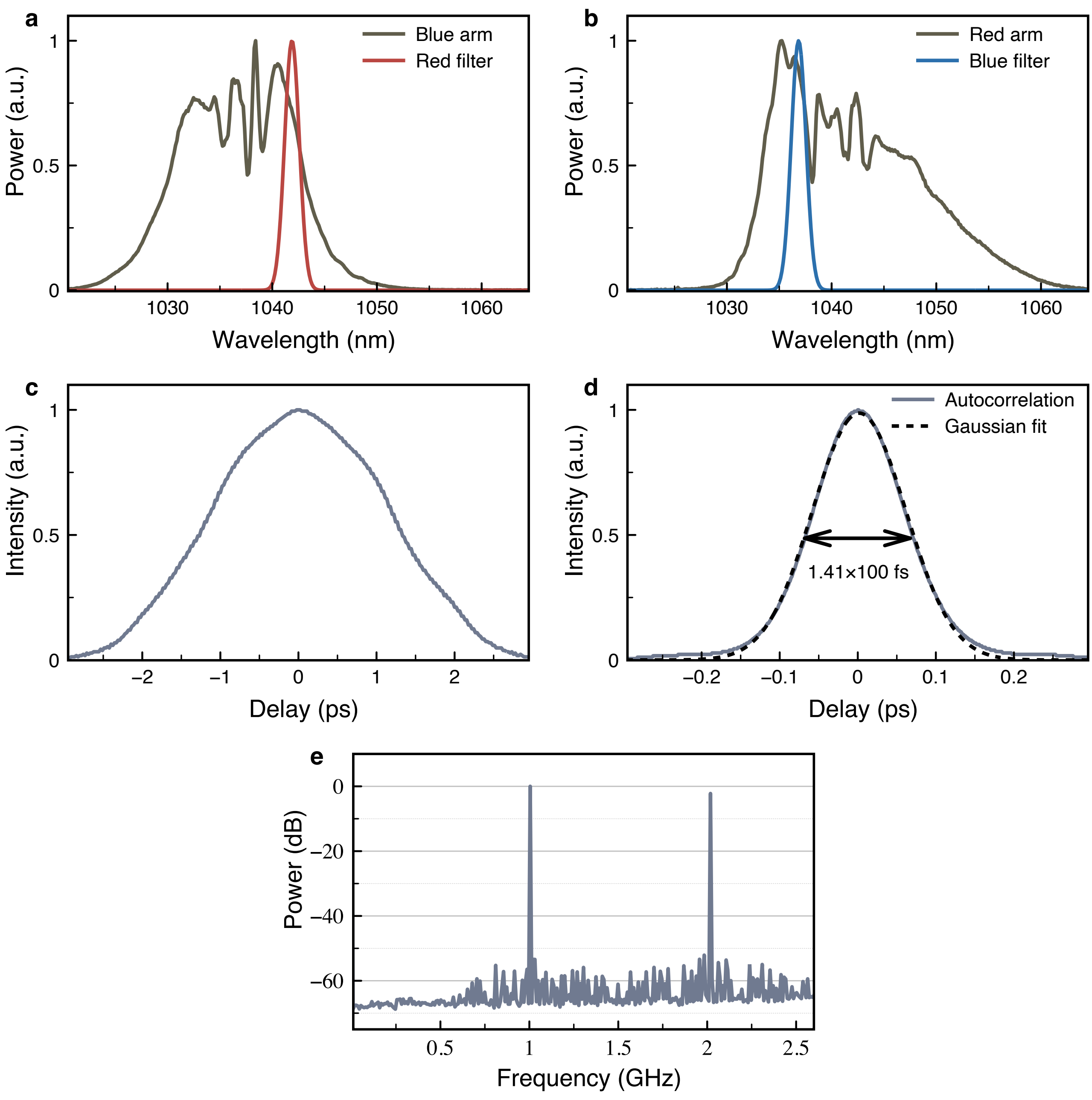

**Supplementary Figure 11**: **A short-pulsed 1 GHz harmonic state. a**,**b**, Spectral settings. **c**,**d**, Autocorrelation of direct and dechirped output (red arm), respectively. A Gaussian fit to the dechirped pulses indicates a full width at half maximum of ~100 fs. **e**, Radio frequency spectrum of the pulse pattern showing harmonic modelocking at ~1 GHz with over 50 dB supermode suppression.

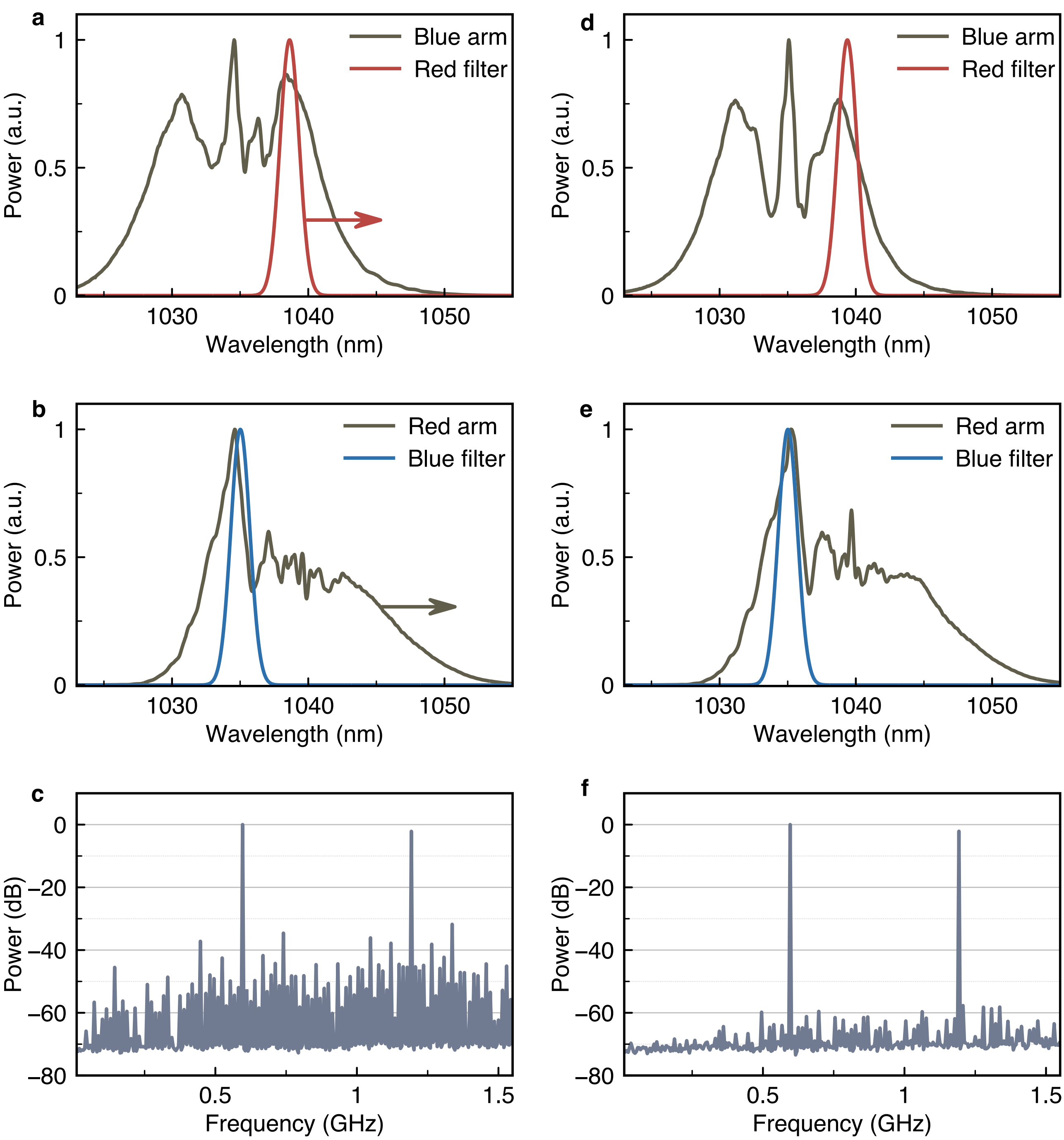

**Supplementary Figure 12**: **Optimised supermode suppression ratio. a-c**, Measured optical and radio frequency spectra of a ~0.6 GHz harmonic state with a poor supermode suppression before optimisation. Arrows indicate the direction of shifting the red filter and spectrum to suppress supermodes. **d-e**, Measured optical and radio frequency spectra after optimisation. Placing the blue filter at a spectral peak and the red filter at a steep slope improved supermode suppression to ~60 dB.

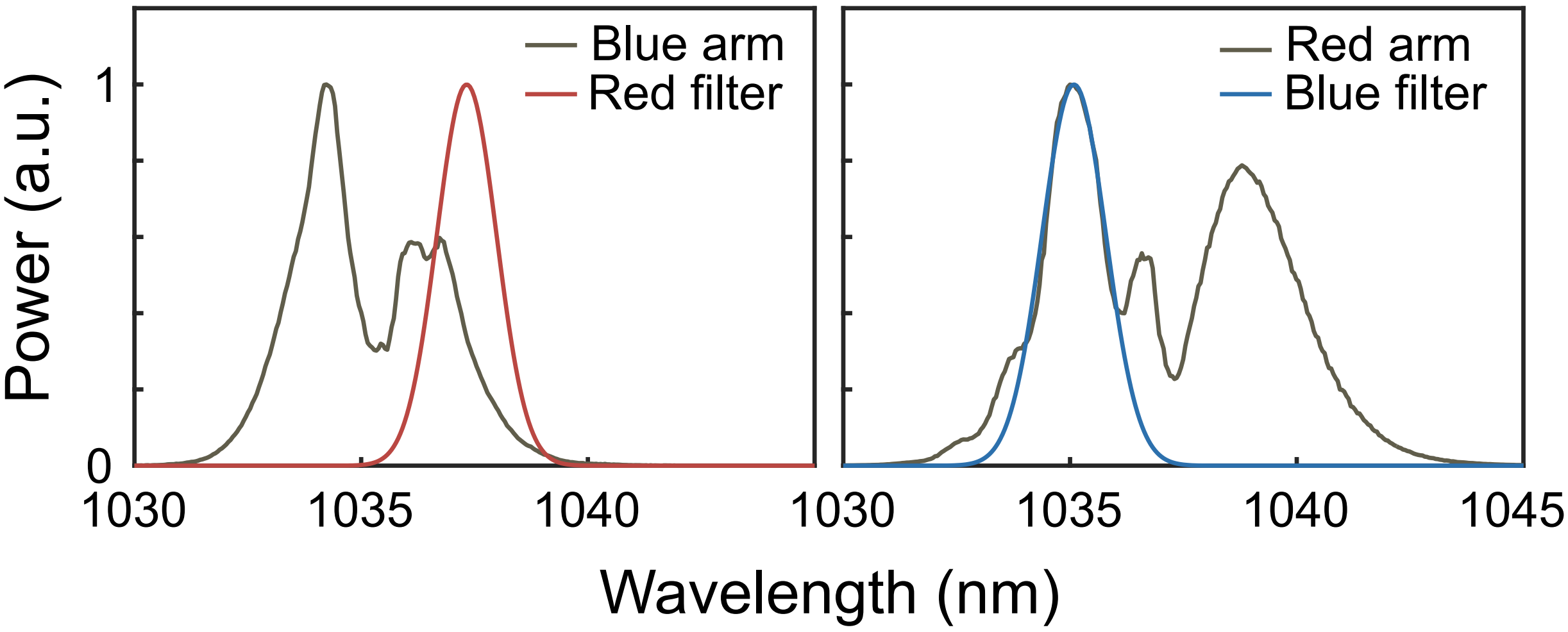

**Supplementary Figure 13**: **Spectra associated with the annealing experiment.** **a**, Measured radio frequency spectra for the initial pulse pattern after the last pulse creation, and the pulse patterns after annealing, absrupt harmonisation and optimisation of the noise suppression via pump power tuning. **b**, Optical spectra measured after optimisation.

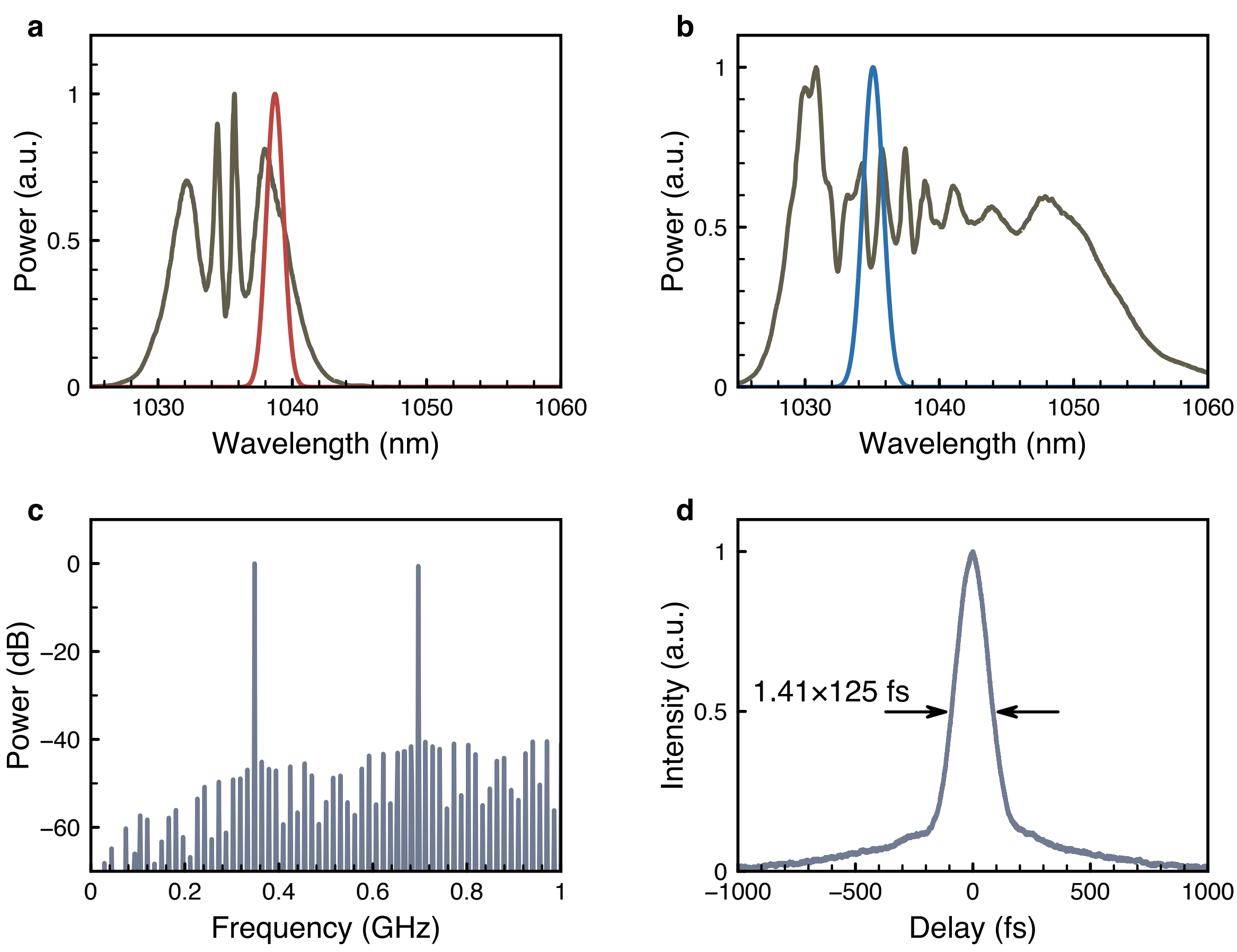

**Supplementary Figure 14**: **A harmonic state with 20-nJ pulses.** **a**,**b**, Measured optical spectra. **c**, Radio frequency spectrum of the pulse pattern showing harmonic modelocking at ~350 MHz with over ~45 dB supermode suppression. **d**, autocorrelation of dechirped output (red arm).

## 13. Analogy between multi-pulse configurations and multi-particle states in field theory

The structure of multi-pulse configurations and the transitions between them are intuitively described by an informal but instructive analogy to multi-particle states in quantum field theory[12], or more precisely, second quantisation as applied to classical many-body systems[13]. The multi-pulse ensemble may be represented as a discrete set of excitation number states, $|N\rangle$, where $N$ denotes the number of pulses circulating in the cavity. Each pulse corresponds to a quasiparticle-like excitation[14]. The $|0\rangle$ state, while containing no pulses, is not truly empty but exhibits fluctuating background power due to amplified spontaneous emission. Deterministic injection transitions the system from $|N\rangle$ to $|N+1\rangle$, adding one pulse, and increasing the intracavity energy by $E^{*,2}$. Conversely, quantum-noise-driven tunnelling removes a pulse, reducing the system from $|N\rangle$ to $|N-1\rangle$. There is a non-commutation between creation and annihilation operations. Annihilating a pulse after creation typically removes any of the $N+1$ pulses, not necessarily the most recently added one, producing a more anharmonic pulse pattern. In contrast, annihilation followed by creation is highly likely to refill the vacated temporal position if the pulse injection is properly timed, largely preserving the original pattern. This asymmetry reflects the collective, strongly nonlinear interactions that govern pulse formation and repositioning. It also clarifies why assembling a fully harmonic pulse pattern is inherently challenging, and motivates the annealing mechanism discussed below. While entirely classical and lacking a formal operator algebra or Hilbert space, our framework reveals how pulse number can be systematically controlled without requiring explicit solutions of the governing equations (Fig. 4d).

## 14. Candidate systems for hierarchical steering of permanent self-organised structures

The quantitative application of the hierarchical framework to modelocked lasers in this work builds on years of theoretical and experimental development. A comparable reduction for other platforms is beyond the scope of the present study. It is, however, possible to make an informed assessment of the generality of the hierarchical framework. There are two main conditions for its applicability, namely, well-separated timescales, and the existence of attractor states at each hierarchical level.

In many such systems, hierarchical steering produces permanent self-organised structures, even though steering itself does not imply global reversibility of the system state under parameter variation. Instead, control parameters act on a spatially and temporally localised active region, within which the hierarchy evolves continuously, while previously formed structures remain fixed and influence subsequent evolution through boundary conditions or seeding. When viewed in the local frame of the driven interaction zone, the dynamics are formally equivalent to those of adaptive, reversible systems, even though irreversibility accumulates outside that frame. This perspective naturally applies to laser–matter interactions, materials self-assembly, and related driven–dissipative processes.

There appears to be many driven–dissipative systems with strong separations of timescales[15-19]. They also exhibit various hierarchical structures suggesting the existence of attracting sets or low-dimensional manifolds, such that fast degrees of freedom rapidly collapse onto effective order parameters governing progressively slower layers. When mainly these two critical conditions are met, our framework is applicable. The framework is not claimed to apply generically to all multiscale systems, but specifically to driven–dissipative systems in which fast dynamics collapse onto attractors at each level.

Several laser-based and laser-driven systems already exhibit these structural features in a form that closely mirrors the hierarchy identified here. The examples below are therefore intended as qualitatively compatible but we emphasise that they are yet unvalidated testbeds for the framework. Furthermore, they are selected from laser-material interactions only due to our familiarity with these applications, but beyond that the possibilities do not appear to be limited to laser excitation.

***(i) Laser-induced self-organised surface patterns***

Laser-induced periodic surface structures (LIPSS) provide a representative example of multiscale pattern formation driven by ultrafast excitation. In various material surfaces, fine ripples with sub-100-nm periodicity form on top of larger surface modulations with periods of several hundred nanometres and heights of order 100 nm. These structures emerge under excitation by trains of femtosecond pulses and arise from the interaction of ultrafast optical excitation, transient melting, and cumulative morphological feedback[16].

A natural hierarchical decomposition can be identified. At the fastest level (femtoseconds to picoseconds), each pulse generates an inhomogeneous energy-deposition landscape through near-field enhancement and possible excitation of surface electromagnetic waves. These fast electronic processes rapidly reach equilibrium with the atoms of the lattice, triggering ultrafast ablation processes. At the intermediate level (nanoseconds to microseconds), transient melt-layer hydrodynamics and capillary flows reshape the surface. At the slowest level (microseconds to milliseconds, from pulse to pulse), modification of the surface feeds back on the optical near field, progressively reinforcing selected spatial frequencies in the surface topography.

Within this structure, slow morphological parameters such as ripple amplitude and periodicity act as attracting order parameters for the faster optical and hydrodynamic processes. Steering could, in principle, be achieved by controlling slow external parameters such as pulse fluence, polarisation, repetition rate, or pre-patterned seed structures, which would steer the collective degrees of freedom through the hierarchy. Although a full predictive reduced model is beyond the scope of this study, the coexistence of strong timescale separation and stable attractors makes LIPSS a viable candidate system.

***(ii) Ultrafast laser-driven zeolite synthesis***

Laser-driven synthesis of hierarchical zeolites provides another example of multiscale organisation in a strongly driven, dissipative system[15]. Tightly focused femtosecond pulses in aqueous precursor solutions can initiate nucleation and growth by multiphoton absorption, producing transient micron-scale hot zones that repeatedly process small parcels of fluid, as the fluid is recirculated by the Marangoni-like flows induced by the local heating by the laser.

At the fastest level (femtoseconds to nanoseconds), multiphoton absorption creates rapid, highly localised temperature excursions, forms a plasma and triggers chemical activation within the focal volume referred to as a tiny ultrafast reactor in the original paper[15]. The flow of any given section of the fluid containing the reacting precursor material takes microseconds, forming the intermediate timescale, determined by microfluidic flow speed. At slower times (milliseconds to longer), the section of fluid first cools down, while nanoscale structures grow and assemble into larger crystalline domains, in some cases exhibiting hierarchical structures. On the longest timescales (minutes to hours), the same section of fluid revisits the focal spot, i.e., the ultrafast reactor and the process is advanced, growing larger zeolite structures.

In hierarchical terms, the fast and highly non-equilibrium conditions created by each pulse are strongly slaved to the pulse and beam parameters, while microsecond-scale duration of the accelerated reaction at high temperatures are slaved to much slower microfluidic dynamics, determined by the position and power of the laser beam. The final structure is governed by the total number of passage of any section of the fluid from the focal point, which is determined by macroscopic properties, namely, the ratio of the volume of the container to the volume of the focal spot (the ultrafast reactor). Although the quantitative chemistry is system-specific, qualitatively this system parallels those demonstrated in the laser system studied here.

*(iii) Burst-laser-driven glass drilling*

In burst-mode femtosecond laser drilling of glass[20,21], trains of closely spaced pulses generate extremely long, narrow capillaries whose length can exceed the Rayleigh range by more than an order of magnitude. Each individual pulse produces a local ablation crater with approximately spherical symmetry on a femtosecond to picosecond timescale, followed by electron–lattice equilibration and transient melting on picosecond to nanosecond scales. Within a single burst, successive pulses interact with the evolving crater through melt expulsion and recoil pressure on nanosecond to microsecond timescales.

The final capillary, extending far beyond the focal depth, emerges only through the cumulative action of many such elementary spherical modification events. The elementary pulse-scale structures therefore constitute the lowest hierarchical level, while the burst-scale and multi-burst-scale growth of the capillary form higher-level organisational structures. External parameters such as burst repetition rate, intra-burst spacing, and average power act at the slowest level and determine the slaving of the faster ablation and melt dynamics. The resulting long-aspect-ratio structures thus represent an emergent product of multi-level hierarchical self-organisation.

*(iv) Sub-surface self-organised structures in silicon*

Nonlinear laser lithography deep inside silicon produces elongated buried microstructures with radii of order one micrometre and lengths approaching a millimetre, far exceeding the Rayleigh range[8]. Each nanosecond pulse generates a highly localised modification volume with approximately spherical symmetry. Over hundreds of to one thousand pulses at repetition rates around 150 kHz, these elementary structures accumulate into extended rod-like geometries within milliseconds.

Here again, a clear hierarchy can be identified: ultrafast nonlinear absorption and plasma formation at the fastest level, transient thermal and mechanical relaxation resulting in changes in the local index of refraction at the timescale of the temporal separation of the pulses, and cumulative structural growth at millisecond timescales as the index change cause by each pulse moves the focal point of the next pulse. The final extended structure emerges as a higher-level organisational product of many locally generated elementary units caused by ~100 pulses, slaved to slow external controls such as pulse energy and repetition rate.